\documentclass[10pt]{article}
\usepackage{amsmath,amsthm,latexsym,amssymb,amsfonts,epsfig, psfrag,color}
\usepackage{subfigure}
\usepackage{graphicx}
\usepackage{caption}
\usepackage{sidecap}
\usepackage{wrapfig}
\makeatletter \@addtoreset{equation}{section} \makeatother

\newtheorem*{Example}{Example}%[section]
\def\be{\begin{equation}}
\def\ee{\end{equation}}
\def\ba{\begin{eqnarray}}
\def\ea{\end{eqnarray}}

\newcommand\nn{\nonumber}
\newcommand\q{\quad}
\def\Nl{{\mathchoice
{\setbox0=\hbox{$\displaystyle\rm N$}\hbox{\hbox to0pt
{\kern0.4\wd0\vrule height0.9\ht0\hss}\box0}}
{\setbox0=\hbox{$\textstyle\rm N$}\hbox{\hbox to0pt
{\kern0.4\wd0\vrule height0.9\ht0\hss}\box0}}
{\setbox0=\hbox{$\scriptstyle\rm N$}\hbox{\hbox to0pt
{\kern0.4\wd0\vrule height0.9\ht0\hss}\box0}}
{\setbox0=\hbox{$\scriptscriptstyle\rm N$}\hbox{\hbox to0pt
{\kern0.4\wd0\vrule height0.9\ht0\hss}\box0}}}}
\def\Zl{{\mathchoice
{\setbox0=\hbox{$\displaystyle\rm Z$}\hbox{\hbox to0pt
{\kern0.4\wd0\vrule height0.9\ht0\hss}\box0}}
{\setbox0=\hbox{$\textstyle\rm Z$}\hbox{\hbox to0pt
{\kern0.4\wd0\vrule height0.9\ht0\hss}\box0}}
{\setbox0=\hbox{$\scriptstyle\rm Z$}\hbox{\hbox to0pt
{\kern0.4\wd0\vrule height0.9\ht0\hss}\box0}}
{\setbox0=\hbox{$\scriptscriptstyle\rm Z$}\hbox{\hbox to0pt
{\kern0.4\wd0\vrule height0.9\ht0\hss}\box0}}}}
\def\Ql{{\mathchoice
{\setbox0=\hbox{$\displaystyle\rm Q$}\hbox{\hbox to0pt
{\kern0.4\wd0\vrule height0.9\ht0\hss}\box0}}
{\setbox0=\hbox{$\textstyle\rm Q$}\hbox{\hbox to0pt
{\kern0.4\wd0\vrule height0.9\ht0\hss}\box0}}
{\setbox0=\hbox{$\scriptstyle\rm Q$}\hbox{\hbox to0pt
{\kern0.4\wd0\vrule height0.9\ht0\hss}\box0}}
{\setbox0=\hbox{$\scriptscriptstyle\rm Q$}\hbox{\hbox to0pt
{\kern0.4\wd0\vrule height0.9\ht0\hss}\box0}}}}
\def\Rl{{\mathchoice
{\setbox0=\hbox{$\displaystyle\rm R$}\hbox{\hbox to0pt
{\kern0.4\wd0\vrule height0.9\ht0\hss}\box0}}
{\setbox0=\hbox{$\textstyle\rm R$}\hbox{\hbox to0pt
{\kern0.4\wd0\vrule height0.9\ht0\hss}\box0}}
{\setbox0=\hbox{$\scriptstyle\rm R$}\hbox{\hbox to0pt
{\kern0.4\wd0\vrule height0.9\ht0\hss}\box0}}
{\setbox0=\hbox{$\scriptscriptstyle\rm R$}\hbox{\hbox to0pt
{\kern0.4\wd0\vrule height0.9\ht0\hss}\box0}}}}
\def\Cl{{\mathchoice
{\setbox0=\hbox{$\displaystyle\rm C$}\hbox{\hbox to0pt
{\kern0.4\wd0\vrule height0.9\ht0\hss}\box0}}
{\setbox0=\hbox{$\textstyle\rm C$}\hbox{\hbox to0pt
{\kern0.4\wd0\vrule height0.9\ht0\hss}\box0}}
{\setbox0=\hbox{$\scriptstyle\rm C$}\hbox{\hbox to0pt
{\kern0.4\wd0\vrule height0.9\ht0\hss}\box0}}
{\setbox0=\hbox{$\scriptscriptstyle\rm C$}\hbox{\hbox to0pt
{\kern0.4\wd0\vrule height0.9\ht0\hss}\box0}}}}
\def\Hl{{\mathchoice
{\setbox0=\hbox{$\displaystyle\rm H$}\hbox{\hbox to0pt
{\kern0.4\wd0\vrule height0.9\ht0\hss}\box0}}
{\setbox0=\hbox{$\textstyle\rm H$}\hbox{\hbox to0pt
{\kern0.4\wd0\vrule height0.9\ht0\hss}\box0}}
{\setbox0=\hbox{$\scriptstyle\rm H$}\hbox{\hbox to0pt
{\kern0.4\wd0\vrule height0.9\ht0\hss}\box0}}
{\setbox0=\hbox{$\scriptscriptstyle\rm H$}\hbox{\hbox to0pt
{\kern0.4\wd0\vrule height0.9\ht0\hss}\box0}}}}
\def\Ol{{\mathchoice
{\setbox0=\hbox{$\displaystyle\rm O$}\hbox{\hbox to0pt
{\kern0.4\wd0\vrule height0.9\ht0\hss}\box0}}
{\setbox0=\hbox{$\textstyle\rm O$}\hbox{\hbox to0pt
{\kern0.4\wd0\vrule height0.9\ht0\hss}\box0}}
{\setbox0=\hbox{$\scriptstyle\rm O$}\hbox{\hbox to0pt
{\kern0.4\wd0\vrule height0.9\ht0\hss}\box0}}
{\setbox0=\hbox{$\scriptscriptstyle\rm O$}\hbox{\hbox to0pt
{\kern0.4\wd0\vrule height0.9\ht0\hss}\box0}}}}
\newcommand{\cg}{\mathcal G}
\newcommand{\ch}{\mathcal H}

\newcommand{\cq}{\mathcal Q}

\def\nn{\nonumber}

\newcommand{\eqa}{\begin{eqnarray}}
\newcommand{\neqa}{\end{eqnarray}}

\newcommand{\p}{\partial}

\def\f{\frac}

\usepackage{bbm}

\def\q{{\quad}}

\definecolor{bianca}{rgb}{0,0.,0.8}

\begin{document}

{\renewcommand{\thefootnote}{\fnsymbol{footnote}}

\title{Classification of constraints and degrees of freedom for quadratic discrete actions}
\author{Philipp A H\"ohn\footnote{e-mail address: {\tt phoehn@perimeterinstitute.ca}} \\
 \small Perimeter Institute for Theoretical Physics,\\
 \small 31 Caroline Street North, Waterloo, Ontario, Canada N2L 2Y5
}

\date{}

}

\setcounter{footnote}{0}
\maketitle

\vspace{-.9cm}

\begin{abstract}
We provide a comprehensive classification of constraints and degrees of freedom for variational discrete systems governed by quadratic actions. This classification is based on the different types of null vectors of the Lagrangian two-form and employs the canonical formalism developed in \cite{Dittrich:2013jaa,Hoehn:2014fka}. The analysis is carried out in both the classical and quantum theory and applies to systems with both temporally varying or constant discretization. In particular, it is shown explicitly how changes in the discretization, e.g.\ resulting from canonical coarse graining or refining operations or an evolving background geometry, change the dynamical content of the system. It is demonstrated how, on a temporally varying discretization, constraints, Dirac observables, symmetries, reduced phase spaces and physical Hilbert spaces become spacetime region dependent. These results are relevant for free field theory on an evolving lattice and linearized discrete gravity models.
\end{abstract}

\section{Introduction}

Discrete gravitational systems generically feature a discretization or graph changing canonical dynamics \cite{Dittrich:2011ke,Hoehn:2011cm,Thiemann:1996ay,Thiemann:1996aw,Alesci:2010gb,Dittrich:2013xwa,Bonzom:2011hm}. Such a discretization changing time evolution can be interpreted as canonical coarse graining or refining operations \cite{Dittrich:2013xwa,Hoehn:2014fka,Hoehn:2014wwa}. However, a temporally varying discretization is not restricted to discrete gravity models, but also arises when putting a field onto an evolving spatial lattice \cite{Dittrich:2013jaa,Foster:2004yc,Jacobson:1999zk,Hoehn:2014fka,Hoehn:2014wwa}. The difference is that, in the gravitational case, the lattice structure itself satisfies a dynamics, while in the latter case only the field satisfies a dynamics and the evolving lattice is given as a fixed background. The evolution of a field on an evolving lattice can have the interpretation of either a dynamical coarse graining or refining if the change in the number of lattice sites results from a change in the density of lattice points in the background geometry, or of a cosmological field on a genuinely expanding or shrinking discrete universe if the change in the number of lattice sites results from the evolving background geometry.  What is common to all these cases is the temporally varying number of degrees of freedom and the ensuing question of how to formulate and understand such a coarse graining or refining dynamics in a canonical language. In particular, the notion of evolving phase and Hilbert spaces is required \cite{Dittrich:2013jaa,Hoehn:2014fka,Dittrich:2011ke,Hoehn:2014wwa,Doldan:1994yg}. Furthermore, the question arises how and in which sense such a discretization changing dynamics can be symplectic in the classical case or unitary in the quantum theory. Related to this, questions concerning the fate of observables as propagating degrees of freedom and of constraints and symmetries on temporally varying discretizations become pertinent.

In order to systematically understand such discretization changing dynamics or dynamics on evolving lattices, a general formalism for variational discrete systems has been developed in \cite{Dittrich:2011ke,Dittrich:2013jaa} for the classical case and in \cite{Hoehn:2014fka,Hoehn:2014wwa} for the quantum theory. This formalism encompasses systems on both temporally varying or constant discretizations and includes a comprehensive constraint analysis and classification of the appearing degrees of freedom. In particular, this formalism provides thorough answers to the questions alluded to above. However, the analysis in the last four references has been very general and at times formal. It is therefore worthwhile to restrict to a simple class of systems which permits us to apply the formalism and to carry out an exploration of its dynamical content explicitly.

In this manuscript, we shall therefore study the special class of discrete actions that are quadratic in the configuration variables. The reason we shall consider such quadratic discrete actions is twofold: firstly, they allow us to solve all equations of motion explicitly and thereby to classify all possible constraints and degrees of freedom, and, secondly, such actions are nevertheless interesting for many systems, e.g.\ (free) field theory on a lattice or linearized theories. A linearized theory describes perturbations around highly symmetric solutions of some more complicated discrete theory. %For such quadratic discrete systems, we can therefore explicitly test and extend the general characterization of constraints and degrees of freedom provided in \cite{Dittrich:2011ke,Dittrich:2013jaa,Hoehn:2014fka,Hoehn:2014wwa}. 
For instance, the considerations of the present work are relevant for linearized 4D Regge Calculus \cite{Dittrich:2009fb,dh4,Bahr:2009ku,Rocek:1982fr}.

More concretely, we shall restrict to discretization changes generated by so-called {\it global} time evolution moves and classify all constraints and degrees of freedom arising from the corresponding quadratic discrete actions into eight types. The subsequent discourse will cover both the classical and the quantum case. As generally explained in \cite{Dittrich:2013jaa,Hoehn:2014fka,Hoehn:2014wwa}, we shall see that the notion of observables as propagating degrees of freedom, of symmetries, of the reduced phase space and physical Hilbert spaces depends, in general, on pairs of time steps and thus, in a spacetime context, on spacetime regions. This is not a fundamental problem of the formalism but merely reflects the fact that coarse graining and refining, or growing and shrinking the lattice, change the dynamical content of the system; changes in the discretization lead to changes in the amount of dynamical information which the discretization can support. In particular, we shall see that coarse graining leads to a `propagation' of constraints and thereby to an increase in the number of constraints at a given time step and to non-unitary projections of physical Hilbert spaces in the quantum theory. In this work we shall not specifically discuss the subtleties that arise when applying this formalism to discrete gravity models as this has been done in \cite{Dittrich:2011ke,Dittrich:2013jaa,Dittrich:2013xwa,Hoehn:2014fka,Hoehn:2014wwa,dh4}.

The outline of the rest of this article is as follows: in section \ref{sec_prelim} we recall basic concepts of time evolution in variational discrete systems. In section \ref{sec_quadact} we introduce the class of quadratic discrete actions and the general symplectic structure, constraints and dynamics which they generate. Subsequently, in section \ref{sec_nullvecs}, we classify left and right vectors of the Lagrangian two-form into eight types. According to these eight types, we classify in section \ref{sec_constraintclass} constraints and equations of motion and in section \ref{sec_classdof} the occurring degrees of freedom. In section \ref{sec_unique} we consider `effective' actions (or Hamilton-Jacobi functions) arising on solutions when composing different time evolution moves, while in sections \ref{sec_varcon} and \ref{sec_classeff} we analyze which consequences `effective' actions have for the constraint structure and the classification of constraints and degrees of freedom at a fixed discrete time step. Then we shall switch to the quantum theory in section \ref{sec_QT}, in order to study physical Hilbert spaces, propagators, quantum time evolution maps, quantum constraints and degrees of freedom in an analogous manner. Lastly, in section \ref{sec_summary} we close with a brief conclusion. Along the way, and for direct illustration, we shall apply the formalism to the example of a Euclidean (massive) scalar field on an expanding two--dimensional square lattice.

\section{Preliminaries}\label{sec_prelim}

In the class of systems which we shall henceforth consider, namely variational discrete systems \cite{marsdenwest,Jaroszkiewicz:1996gr,Dittrich:2013jaa,Dittrich:2011ke,DiBartolo:2004cg,DiBartolo:2002fu}, the dynamics does {\it not} proceed by a continuous time evolution of the variables, but by a discrete time evolution of the latter on a discrete set of time steps. We shall label these time steps by $n\in\mathbb{Z}$. Time evolution of the dynamical variables can thus not be generated by a Hamiltonian through some bracket structure as in the continuum  because this would necessarily yield an infinitesimal and therefore continuous dynamics. Rather, the time evolution of the dynamical variables from a time step $n$ to the next step $n+1$ is generated by a {\it time evolution move} $n\rightarrow n+1$ and an action $S_{n+1}(x_n,x_{n+1})$ which is associated to this move. The continuous configuration variables $x_n,x_{n+1}$ are the dynamical variables in the Lagrangian picture and coordinatize the configuration manifolds $\cq_n,\cq_{n+1}$ of time steps $n,n+1$, respectively. An additional index $i=1,\ldots,\dim\cq_n$ on $x^i_n$ will sometimes be suppressed for notational purposes. Since the configuration manifolds are continuous one can vary and differentiate the action (hence, the name variational discrete systems). Examples of such variational discrete systems occur in discrete mechanics \cite{marsdenwest,Jaroszkiewicz:1996gr}, lattice field theory \cite{smitbook,Sorkin:1975jz,Dittrich:2013jaa,Hoehn:2014fka,Hoehn:2014wwa,Foster:2004yc,Jacobson:1999zk,Bahr:2010cq} and discrete gravity \cite{Regge:1961px,Williams:1996jb,Dittrich:2009fb,Dittrich:2011ke,Dittrich:2013jaa,Bahr:2011xs}.

\begin{figure}[hbt!]
\begin{center}
\psfrag{0}{$0$}
\psfrag{1}{$1$}
\psfrag{2}{$2$}
\psfrag{s1}{$S_1$}
\psfrag{s2}{$S_2$}
\psfrag{p0}{\small${}^-p^0,{}^-C^0$}
\psfrag{p1}{\small ${}^+p^1,{}^+C^1$}
\psfrag{p12}{\small${}^-p^1,{}^-C^1$}
\psfrag{p2}{\small${}^+p^2,{}^+C^2$}
\includegraphics[scale=.5]{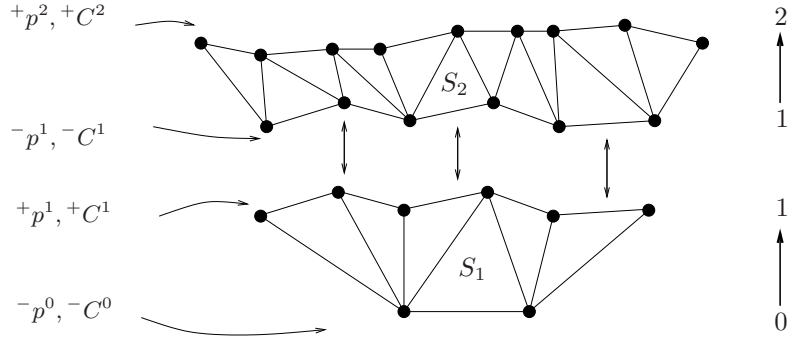}
\caption{\small In a spacetime setting, the individual evolution moves $0\rightarrow1$ and $1\rightarrow2$ correspond to two (neighbouring) spacetime regions with action contributions $S_1,S_2$, respectively. Gluing the two regions (along a common boundary hypersurface) to a new region and solving the equations of motion of the new bulk variables amounts to composing the two moves to a new move $0\rightarrow2$. To every such region or move there is associated a set of pre-- and post--momenta and pre-- and post--constraints. Solving the equations of motion at $n=1$ is equivalent to matching post-- and pre--momenta ${}^+p^1={}^-p^1$.}\label{fig1}
\end{center}
\end{figure}

In a spacetime setting, an evolution move corresponds to a region of (discrete) spacetime, while a time step corresponds to a (piece of) boundary hypersurface of that region. To each such region one associates an action with boundary (and possibly also bulk) variables. The composition of evolution moves corresponds to gluing two spacetime regions along a common boundary and an addition of the associated action contributions. A schematic illustration of a composition of moves is provided in figure \ref{fig1}. In the sequel, we shall solely focus on {\it global evolution moves} $n\rightarrow n+1$ \cite{Dittrich:2013jaa,Hoehn:2014fka} which, in a spacetime context, correspond to evolving an entire boundary hypersurface at once such that the initial hypersurface $n$ does {\it not} overlap with the final hypersurface $n+1$ (except possibly in a common boundary). Accordingly, neighbouring time steps $n,n+1$ do not share any subsets of coinciding variables. This saves us from worrying about some subtleties which are besides the main point of this article and do not affect the classification. Indeed, we emphasize that {\it local evolution moves}, such as Pachner moves or tent moves in triangulations, which involve overlapping time steps and variables that appear at various time steps, can be treated within the same formalism; this has been exhibited in detail in \cite{Dittrich:2013jaa,Dittrich:2011ke,Hoehn:2014wwa,dh4}. In particular, given that global evolution moves as discussed in this manuscript can be decomposed into local moves, the subsequent classification of constraints and degrees of freedom is general and applies to those arising in local moves as well.

In the remainder of this work, we shall restrict to variational discrete actions which are quadratic in the configuration variables. As mentioned in the introduction, such quadratic discrete actions occur in free field theory on a lattice \cite{Sorkin:1975jz,Foster:2004yc,Dittrich:2013jaa,Hoehn:2014fka,Hoehn:2014wwa} (see also the accompanying free massive scalar field example in the main body below) and in perturbative expansions to linear order (around some special solution) of more complicated interacting theories as, for instance, in linearized Regge Calculus \cite{Dittrich:2009fb,dh4,Bahr:2009ku,Rocek:1982fr}. 

Instead of reviewing the full Lagrangian and canonical formalism for general variational discrete systems---which was developed in \cite{Dittrich:2013jaa,Dittrich:2011ke} for the classical case and in \cite{Hoehn:2014fka,Hoehn:2014wwa} for the quantum case---we shall explain and illustrate the relevant concepts along the way by means of this simple class of quadratic discrete actions. For details on the general formalism we refer the reader to \cite{Dittrich:2013jaa,Dittrich:2011ke,Hoehn:2014fka,Hoehn:2014wwa}. These systematically generalize the earlier works \cite{marsdenwest,Jaroszkiewicz:1996gr,Gambini:2002wn,DiBartolo:2004cg,Gambini:2005vn,DiBartolo:2002fu,DiBartolo:2004dn,Campiglia:2006vy} to a full constraint analysis for variational discrete systems which, in particular, encompasses systems with a temporally varying discretization and thus a temporally varying number of degrees of freedom. (A related multisymplectic formulation has recently also been developed in \cite{Arjang:2013dya}.) The restriction to quadratic discrete actions permits us to explicitly solve the dynamics and thereby to provide an explicit classification of all the possible types of constraints and degrees of freedom occurring in such variational discrete systems. This serves as a complement to the general classification in \cite{Dittrich:2013jaa,Dittrich:2011ke,Hoehn:2014fka,Hoehn:2014wwa}.

\section{Quadratic discrete actions}\label{sec_quadact}

The most general form of a quadratic discrete action is 
\ba\label{Sk}
S_n(\{x_{n-1}\},\{x_n\})=\frac{1}{2}a^n_{ij}x_{n-1}^ix_{n-1}^j+\frac{1}{2}b^n_{ij}x^i_nx^j_n+c^n_{ij}x_{n-1}^ix^j_n\,,
\ea
where the (constant) coefficient matrices $a^n,b^n,c^n$ can vary with evolution step $n\in\mathbb{Z}$. Note that $a^n_{ij}=a^n_{ji}$ and $b^n_{ij}=b^n_{ji}$ because $\frac{\partial^2S_n}{\partial x^i_n\partial x^j_n}$ must be symmetric. However, generally $c^n_{ij}\neq c^n_{ji}$ since $c^n_{ij}=-\Omega^n_{ij}=\f{\p^2S_n}{\p x^i_{n-1}\p x^j_n}$ is (minus) the coordinate form of the Lagrangian two--form \cite{Dittrich:2013jaa,Dittrich:2011ke,marsdenwest}. Despite expressly allowing for varying numbers of {\it dynamical} variables from step to step, as necessary for a temporally varying discretization \cite{Dittrich:2013jaa,Dittrich:2011ke,Hoehn:2014fka,Hoehn:2014wwa}, all indices run over the same number of variables $i\in1,\ldots,Q$ because we directly work on extended configuration spaces $\cq_n$. That is, if necessary, we artificially introduce spurious non-dynamical degrees of freedom (on which the action does not depend) at each step $n$ until $\dim\cq_n=Q$ $\forall\,n$ (see \cite{Dittrich:2013jaa,Dittrich:2011ke} for details). As a result, $a^n,b^n,c^n$ are square matrices $\forall\,n$; if there are more dynamical variables $x^i_n$ at step $n$ than at step $n-1$, we artificially extend the matrices $a^n_{ij}$ and $c^n_{ij}$ to square matrices by introducing rows/columns of zeros. Conversely, if there are more dynamical variables at step $n-1$ than at step $n$, we extend the matrices $b^n_{ij}$ and $c^n_{ij}$ by columns/rows of zeros and so forth. The following discussion therefore applies both to systems with temporally constant and temporally varying discretization; it encompasses systems with temporally constant or varying numbers of dynamical degrees of freedom.

Now consider the action for three such steps $n$ or two global moves $0\rightarrow1$ and $1\rightarrow2$ as in figure \ref{fig1}, i.e.
\ba\label{2step}
S(x_0,x_1,x_2)=S_1(x_0,x_1)+S_2(x_1,x_2)\,,
\ea
where the individual $S_n$ are given by (\ref{Sk}) and we assume the $x_1$ to be `bulk' at step $n=2$, while $x_0,x_2$ then define boundary data. %In the remainder of this work, we shall solely focus on {\it global} evolution moves $n\rightarrow n+1$ \cite{Dittrich:2013jaa} where steps $n$ and $n+1$ do not overlap (except possibly in the boundary). Local moves \cite{Dittrich:2013jaa} in linearized Regge Calculus are discussed in detail in \cite{thesis,linreg}.

We shall directly apply the formalism developed in \cite{Dittrich:2013jaa}. The $S_n$ are generating functions of the first kind and thus generate the momenta canonically conjugate to the $x^i_n$. Since each $x^i_n$ occurs in the two actions $S_n,S_{n+1}$ there will be two canonically conjugate momenta ${}^-p^n_i,{}^+p^n_i$ conjugate to every $x^i_n$, which are called {\it pre--} and {\it post--momenta}, respectively. For the current problem, these are given by \cite{Dittrich:2013jaa} 
\ba\label{mom1}
{}^-p^0_i&=&-\frac{\partial S_1}{\partial x^i_0}=-a^1_{ij}x^j_0-c^1_{ij}x^j_1\,,\q\q\q {}^+p^1_i=\,\,\,\,\frac{\partial S_1}{\partial x^i_1}=b^1_{ij}x^j_1+c^1_{ji}x^j_0\,,\\
{}^+p^2_i&=&\,\,\,\,\frac{\partial S_2}{\partial x^i_2}\,=b^2_{ij}x^j_2+c^2_{ji}x^j_1   \,,\q\q\q\,\,\,\,\, {}^-p^1_i=-\frac{\partial S_2}{\partial x^i_1}=-a^2_{ij}x^j_1-c^2_{ij}x^j_2\,.\label{mom1b}
\ea 
These equations constitute the discrete Legendre transforms from $\cq_n\times\cq_{n+1}$ (the discrete analogue of the tangent bundle $T\cq$) to the phase spaces $T^*\cq_n$ and $T^*\cq_{n+1}$ at steps $n,n+1$, respectively \cite{Dittrich:2013jaa,marsdenwest}. Furthermore, (\ref{mom1}) defines implicitly the global Hamiltonian time evolution map $\mathfrak{H}_0:T^*\cq_0\rightarrow T^*\cq_1$ and (\ref{mom1b}) defines implicitly $\mathfrak{H}_1: T^*\cq_1\rightarrow T^*\cq_2$ \cite{Dittrich:2013jaa}.

Clearly, if the Lagrangian two--forms $c^{n}$ possess any null vectors, we directly obtain constraints upon contracting the momentum equations (\ref{mom1}, \ref{mom1b}) with these null vectors. If $c^{n}$ is degenerate, it must have the same number of left (or transposed) and right null vectors,\footnote{For notational ease, we suppress the index enumerating the null vectors for the moment.}
\ba
(L_{n-1})^ic^n_{ij}=0\,,\q\q\q c^n_{ij}(R_n)^j=0\,.\label{leftright}
\ea
This leads to {\it pre--constraints} \cite{Dittrich:2011ke,Dittrich:2013jaa} at $n=0$ and {\it post--constraints} \cite{Dittrich:2011ke,Dittrich:2013jaa} at $n=2$
\ba\label{kincon1}
{}^-C^0=(L_0)^i\left({}^-p^0_i+a^1_{ij}x^j_0\right)\,,\q\q\q {}^+C^2=(R_2)^i\left({}^+p^2_i-b^2_{ij}x^j_2\right)\,,
\ea
respectively, and at step $n=1$ to the post-- and pre--constraints
\ba\label{kincon}
{}^+C^1=(R_1)^i\left({}^+p^1_i-b^1_{ij}x^j_1\right)\,,\q\q\q {}^-C^1=(L_1)^i\left({}^-p^1_i
+a^2_{ij}x^j_1\right)\,,
\ea
respectively. In contrast to the continuum, we therefore have two types of constraint surfaces at each time step. The {\it pre--constraints} restrict the pre-image of the Hamiltonian time evolution map and correspond to conditions on the canonical data which must be satisfied before an evolution move can be carried out. {\it Post--constraints}, on the other hand, restrict the image of the Hamiltonian time evolution map and must be satisfied after an evolution move is performed. Pre-- and post--constraints at a given time step generally do not coincide. For a detailed discussion of pre-- and post--constraints, see \cite{Dittrich:2013jaa,Dittrich:2011ke}. Among other things, it is shown in these references that $\mathfrak{H}_1$ is a presymplectic transformation which preserves the symplectic structure restricted to the pre--constraint surface at $n=1$ and the post--constraint surface at $n=2$.

%These pre-- and post--constraints are therefore directly associated to single null vectors of the Lagrangian two--form. %In fact, they are precisely of the form (\ref{gaugen2}) which for general pre-- and post--constraints is only locally defined on the constraint surface. In the present case the above constraints define the entire constraint surface.

On-shell, pre-- and post--momenta must coincide because {\it momentum matching} \cite{marsdenwest,Dittrich:2013jaa}, ${}^+p^1_i={}^-p^1_i$, is equivalent to implementing the equations of motion for $x_1$,
\ba\label{eom}
h^{12}_{ij}x^j_1=-c^1_{ji}x^j_0-c^2_{ij}x^j_2\,,
\ea
where $h^{12}_{ij}=b^1_{ij}+a^2_{ij}$ is the Hessian of the action (\ref{2step}) for the `internal' variables $x_1$ (see also figure \ref{fig1}). Obviously, $h^{12}_{ij}=h^{12}_{ji}$. When considering (\ref{eom}) as a boundary value problem (i.e.\ attempting to solve for $x_1$, given $x_0,x_2$) it is evident that the boundary value problem admits non--uniqueness of solutions iff $\det h^{12}=0$.\footnote{Some of this non--uniqueness may arise as a consequence of artificially extending the configuration and phase spaces at all steps $n$ to equal dimension (so that $a,b,c$ are square matrices). The artificially added configuration variables will necessarily be free parameters.} On the other hand, when considering (\ref{eom}) as an initial value problem (i.e.\ attempting to solve for $x_2$, given $x_0,x_1$), it is clear that the initial value problem admits non--uniqueness of solutions iff $\det c^2=0$. In order to understand arbitrariness in the evolution, we thus need to classify the null vectors of $c^1,c^2,h^{12}$ and corresponding constraints at step $n=1$.

\begin{Example}
\emph{Before classifying the null vectors, let us introduce the toy model of a Euclidean massive scalar field on a two--dimensional expanding square lattice which we shall employ to illustrate the formalism in each of the following sections. Figure \ref{fig_exp} provides a schematic illustration of the global time evolution moves from `nothing' to a larger square, in analogy to the discrete version of the `no boundary' proposal \cite{Hartle:1983ai} discussed in \cite{Dittrich:2011ke,Dittrich:2013jaa,Hoehn:2014fka}. We shall not explicitly consider the details of the evolution to steps with $n>2$ because nothing qualitatively new happens and the evolution $0\rightarrow1\rightarrow2$ is sufficient for illustrative purposes.  }
\begin{figure}[htbp!]
\psfrag{n}{`Nothing'}
\psfrag{0}{ $n=0$}
\psfrag{1}{$1$}
\psfrag{2}{$2$}
\psfrag{3}{$3$}
\psfrag{4}{$4$}
{\includegraphics[scale=.45]{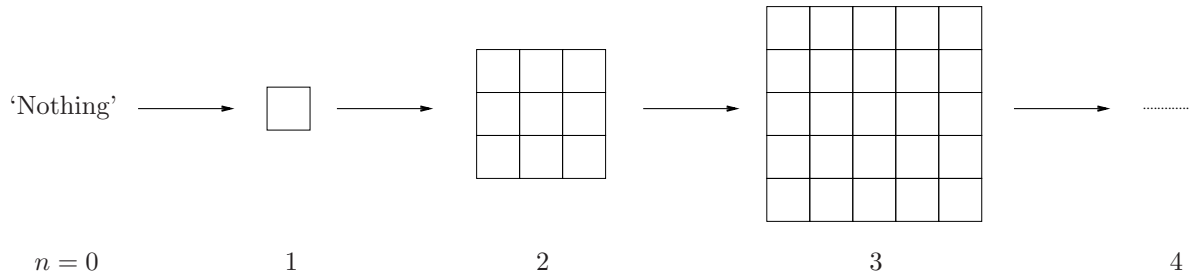}} 
\caption{\small Discrete time evolution of the expanding square lattice; the field is defined on the vertices.}\label{fig_exp}
\end{figure}

\emph{Consider the time evolution move $0\rightarrow1$ from `nothing', i.e., the empty lattice at step $n=0$, to the square with vertices $i=1,\ldots,4$ and unit edge lengths at $n=1$, as depicted in figure \ref{fig_n1square}. \begin{figure}[htbp!]
\psfrag{n1}{$n=1$}
\psfrag{n2}{$n=2$}
\psfrag{1}{\small $1$}
\psfrag{2}{\small $2$}
\psfrag{3}{\small $3$}
\psfrag{4}{\small $4$}
\psfrag{5}{\small $5$}
\psfrag{6}{\small $6$}
\psfrag{7}{\small $7$}
\psfrag{8}{\small $8$}
\psfrag{9}{\small $9$}
\psfrag{10}{\small $10$}
\psfrag{11}{\small $11$}
\psfrag{12}{\small $12$}
\psfrag{f1}{$\phi_1^1$}
\psfrag{f2}{$\phi_1^2$}
\psfrag{f3}{$\phi_1^3$}
\psfrag{f4}{$\phi_1^4$}
\psfrag{f21}{$\phi_2^1$}
\psfrag{f22}{$\phi_2^2$}
\psfrag{f23}{$\phi_2^3$}
\psfrag{f24}{$\phi_2^4$}
\psfrag{f25}{$\phi_2^5$}
\psfrag{f26}{$\phi_2^6$}
\psfrag{f27}{$\phi_2^7$}
\psfrag{f28}{$\phi_2^8$}
\psfrag{f29}{$\phi_2^9$}
\psfrag{f210}{$\phi_2^{10}$}
\psfrag{f211}{$\phi_2^{11}$}
\psfrag{f212}{$\phi_2^{12}$}
\subfigure[]{\label{fig_n1square}\includegraphics[scale=.35]{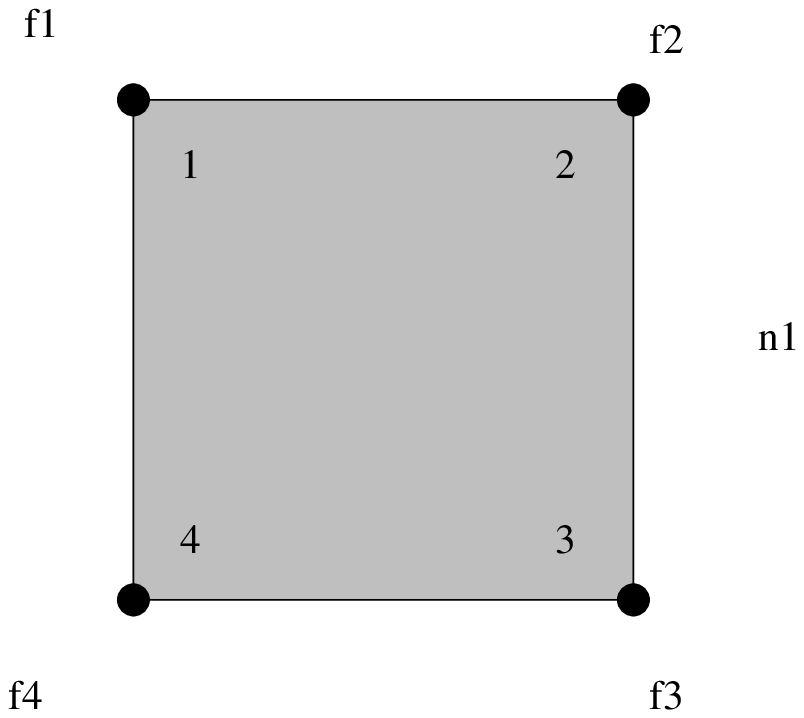}} \hspace*{2cm}
\subfigure[]{\label{fig_n2square}\includegraphics[scale=.35]{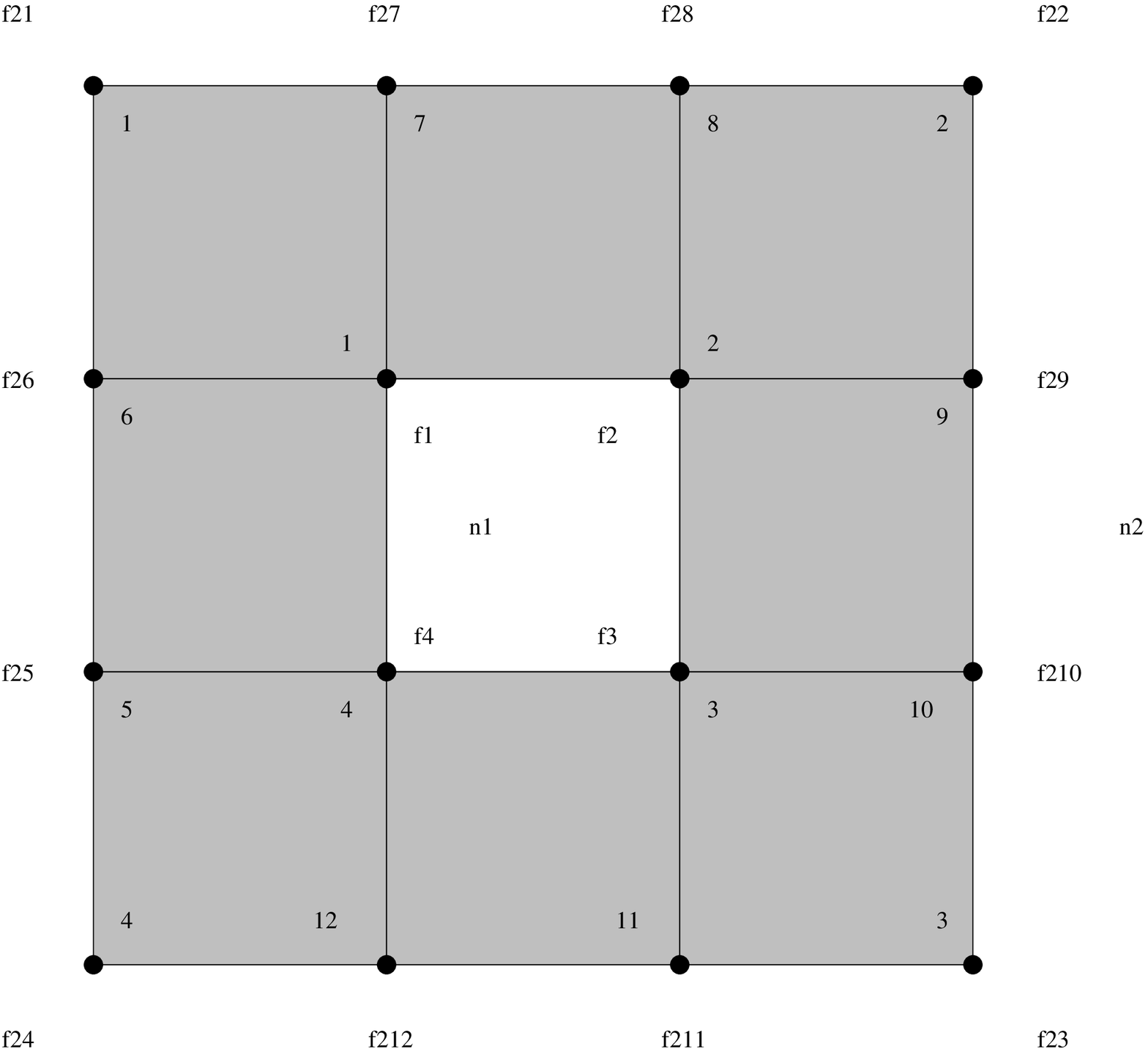}}
\caption{\small (a) The square lattice at step $n=1$. (b) The lattice piece associated to the move $1\rightarrow2$.}
\end{figure} The Euclidean action associated to such a move is that of the field on a single square and reads\footnote{It is straightforward to convince oneself that, by adding up such action contributions associated to unit squares, one obtains the standard 2D Euclidean lattice action
\ba
S_{eucl}=\sum_{i}\left(\sum_\mu\left(\phi(y^i+\vec{\mu})-\phi(y^i)\right)^2+m^2\phi(y^i)^2\right), \nn
\ea
where $i$ runs over the vertices of the lattice and ${\mu}$ runs over two unit vectors pointing in the direction of the edges.}
\ba
S_1(\{\phi_0\},\{\phi_1\})=\sum_{i=1}^4\left((1+\f{1}{4}m^2)(\phi_1^i)^2-\phi_1^i\phi_1^{i+1}\right),\nn
\ea
where $\phi^i_n:=\phi(y^i_{n})\in\mathbb{R}$ is the field at step $n$ evaluated on the vertex $i$ with coordinate $y^i_n$ and the vertex labeled by $i+1=5$ coincides with the vertex labeled by $i=1$ on account of periodicity. Clearly, this action has a trivial dependence on the `initial field values' $\{\phi_0\}$.}

\emph{In order to describe the expanding system on configuration spaces of equal dimension with $Q=12$ variables up to step $n=2$, we extend the empty configuration space at $n=0$ by $\{\phi_0^i\}_{i=1}^{12}$ and the configuration space at $n=1$ by $\{\phi_1^j\}_{j=5}^{12}$. Obviously, $S_1$ trivially depends on these newly introduced spurious variables. Hence, in terms of the matrices, $a^1=c^1\equiv0$ and
\ba
b^1=\scriptsize{\left(\begin{array}{cccccccccc}2+\f{m^2}{2} & -1 & 0 & -1 & 0 & 0 & 0 & 0 & 0 & 0 \\-1 & 2+\f{m^2}{2} & -1 & 0 & 0 & 0 & 0 & 0 & 0 & 0 \\0 & -1 & 2+\f{m^2}{2} & -1 & 0 & 0 & 0 & 0 & 0 & 0 \\-1 & 0 & -1 & 2+\f{m^2}{2} & 0 & 0 & 0 & 0 & 0 & 0 \\0 & 0 & 0 & 0 & 0 & 0 & 0 & 0 & 0 & 0 \\0 & 0 & 0 & 0 & 0 & 0 & 0 & 0 & 0 & 0 \\0 & 0 & 0 & 0 & 0 & 0 & 0 & 0 & 0 & 0 \\0 & 0 & 0 & 0 & 0 & 0 & 0 & 0 & 0 & 0 \\0 & 0 & 0 & 0 & 0 & 0 & 0 & 0 & 0 & 0 \\0 & 0 & 0 & 0 & 0 & 0 & 0 & 0 & 0 & 0\end{array}\right)}.
\ea
This yields 12 pre--constraints (\ref{kincon1}) for the spurious degrees of freedom at $n=0$ 
\ba
{}^-C^0_i:={}^-p^0_i=0,\q\q i=1,\ldots,12\,\nn
\ea
%The only non-zero components of $b^1$ are given by
%\ba
%b^1_{ii}=2+\f{1}{2}m^2, \q\q b^1_{i(i+1)}=-1,\q\q i=1,\ldots,4\,,\q\q\text{where}\q\q b^1_{4(5)}:=b^1_{41}\,.\nn
%\ea
and 12 post--constraints (\ref{kincon}) at step $n=1$
\ba
{}^+C^1_i&:=&{}^+p^1_i-\left(2+\f{1}{2}m^2\right)\phi^i_1+\phi^{i-1}_1+\phi^{i+1}_1=0,\q\q i=1,\ldots,4,\label{exp-postcon1}\\
 {}^+C^1_j&:=&{}^+p^1_j=0,\q\q\q\q\q\q\q\q\q\q\q\q\q\q\,\,\,\q j=5,\ldots,12,\label{exp-postcon2}
 \ea
where, again, $\phi_1^{i+1=5}:=\phi^1_1$ and $\phi_1^{i-1=0}:=\phi_1^4$ because of periodicity. Evidently, the move $0\rightarrow1$ is totally constrained. The post--constraints (\ref{exp-postcon2}) fix the momenta of the eight artificially added spurious modes at $n=1$. The latter are unpredictable and, in fact, gauge degrees of freedom.}

\emph{In the global move $1\rightarrow2$ the square expands to the one with 12 vertices at step $n=2$, as given in figure \ref{fig_n2square}. The action associated to this global move involving eight unit squares reads 
\ba\label{exampleaction}
S_2(\{\phi_1\},\{\phi_2\})=   \f{1}{2}a^2_{ij}\phi^i_1\phi^j_1+\f{1}{2}b^2_{ij}\phi_2^i\phi_2^j+c^2_{ij}\phi^i_1\phi_2^j, %\!\!\!\!&\sum_{i=1}^4\left(3\left(1+\f{1}{4}m^2\right)(\phi_1^i)^2-\phi_1^i\phi_1^{i+1}+\left(1+\f{1}{4}m^2\right)(\phi_2^i)^2\right)+\sum_{j=5}^{12}2\left(1+\f{1}{4}m^2\right)(\phi_2^j)^2\nn\\
%&-\phi_2^1\phi^6_2-\phi_2^1\phi_2^7 - \phi_2^2\phi_2^8 - \phi_2^7\phi_2^8 - \phi_2^2\phi_2^9 - \phi_2^9\phi_2^{10} - 
% \phi_2^3\phi_2^{10} - \phi_2^3\phi_2^{11} - \phi_2^{11}\phi_2^{12}  
% -\phi_2^{4}\phi_2^5 \nn\\&- \phi_2^{4}\phi_2^{12} - \phi_2^5\phi_2^6 - 
% 2 (\phi_1^1\phi_2^6 + \phi_1^1\phi_2^7 + \phi_1^2\phi_2^8 + \phi_1^2\phi_2^9 + \phi_1^3\phi_2^{10} + \phi_1^3\phi_2^{11} + 
%    \phi_1^4\phi_2^{12} + \phi_1^4\phi_2^5).\nn%\\
\ea 
where the matrices $a^2,b^2,c^2$ are given by
\ba
a^2&=&\scriptsize{\left(
\begin{array}{cccccccccccc}
 \frac{3 m^2}{2}+6 & -1 & 0 & -1 & 0 & 0 & 0 & 0 & 0 & 0 & 0 & 0 \\
 -1 & \frac{3 m^2}{2}+6 & -1 & 0 & 0 & 0 & 0 & 0 & 0 & 0 & 0 & 0 \\
 0 & -1 & \frac{3 m^2}{2}+6 & -1 & 0 & 0 & 0 & 0 & 0 & 0 & 0 & 0 \\
 -1 & 0 & -1 & \frac{3 m^2}{2}+6 & 0 & 0 & 0 & 0 & 0 & 0 & 0 & 0 \\
 0 & 0 & 0 & 0 & 0 & 0 & 0 & 0 & 0 & 0 & 0 & 0 \\
 0 & 0 & 0 & 0 & 0 & 0 & 0 & 0 & 0 & 0 & 0 & 0 \\
 0 & 0 & 0 & 0 & 0 & 0 & 0 & 0 & 0 & 0 & 0 & 0 \\
 0 & 0 & 0 & 0 & 0 & 0 & 0 & 0 & 0 & 0 & 0 & 0 \\
 0 & 0 & 0 & 0 & 0 & 0 & 0 & 0 & 0 & 0 & 0 & 0 \\
 0 & 0 & 0 & 0 & 0 & 0 & 0 & 0 & 0 & 0 & 0 & 0 \\
 0 & 0 & 0 & 0 & 0 & 0 & 0 & 0 & 0 & 0 & 0 & 0 \\
 0 & 0 & 0 & 0 & 0 & 0 & 0 & 0 & 0 & 0 & 0 & 0 \\
\end{array}
\right)},\nn%\nn\\
\ea
\ba
b^2\!\!\!\!&=&\!\!\!\!\!\tiny{\left(
\begin{array}{cccccccccccc}
 \frac{m^2}{2}+2 & 0 & 0 & 0 & 0 & -1 & -1 & 0 & 0 & 0 & 0 & 0 \\
 0 & \frac{m^2}{2}+2 & 0 & 0 & 0 & 0 & 0 & -1 & -1 & 0 & 0 & 0 \\
 0 & 0 & \frac{m^2}{2}+2 & 0 & 0 & 0 & 0 & 0 & 0 & -1 & -1 & 0 \\
 0 & 0 & 0 & \frac{m^2}{2}+2 & -1 & 0 & 0 & 0 & 0 & 0 & 0 & -1 \\
 0 & 0 & 0 & -1 & m^2+4 & -1 & 0 & 0 & 0 & 0 & 0 & 0 \\
 -1 & 0 & 0 & 0 & -1 & m^2+4 & 0 & 0 & 0 & 0 & 0 & 0 \\
 -1 & 0 & 0 & 0 & 0 & 0 & m^2+4 & -1 & 0 & 0 & 0 & 0 \\
 0 & -1 & 0 & 0 & 0 & 0 & -1 & m^2+4 & 0 & 0 & 0 & 0 \\
 0 & -1 & 0 & 0 & 0 & 0 & 0 & 0 & m^2+4 & -1 & 0 & 0 \\
 0 & 0 & -1 & 0 & 0 & 0 & 0 & 0 & -1 & m^2+4 & 0 & 0 \\
 0 & 0 & -1 & 0 & 0 & 0 & 0 & 0 & 0 & 0 & m^2+4 & -1 \\
 0 & 0 & 0 & -1 & 0 & 0 & 0 & 0 & 0 & 0 & -1 & m^2+4 \\
\end{array}
\right)},\nn\\
c^2&=&\scriptsize{\left(
\begin{array}{cccccccccccc}
 0 & 0 & 0 & 0 & 0 & -2 & -2 & 0 & 0 & 0 & 0 & 0 \\
 0 & 0 & 0 & 0 & 0 & 0 & 0 & -2 & -2 & 0 & 0 & 0 \\
 0 & 0 & 0 & 0 & 0 & 0 & 0 & 0 & 0 & -2 & -2 & 0 \\
 0 & 0 & 0 & 0 & -2 & 0 & 0 & 0 & 0 & 0 & 0 & -2 \\
 0 & 0 & 0 & 0 & 0 & 0 & 0 & 0 & 0 & 0 & 0 & 0 \\
 0 & 0 & 0 & 0 & 0 & 0 & 0 & 0 & 0 & 0 & 0 & 0 \\
 0 & 0 & 0 & 0 & 0 & 0 & 0 & 0 & 0 & 0 & 0 & 0 \\
 0 & 0 & 0 & 0 & 0 & 0 & 0 & 0 & 0 & 0 & 0 & 0 \\
 0 & 0 & 0 & 0 & 0 & 0 & 0 & 0 & 0 & 0 & 0 & 0 \\
 0 & 0 & 0 & 0 & 0 & 0 & 0 & 0 & 0 & 0 & 0 & 0 \\
 0 & 0 & 0 & 0 & 0 & 0 & 0 & 0 & 0 & 0 & 0 & 0 \\
 0 & 0 & 0 & 0 & 0 & 0 & 0 & 0 & 0 & 0 & 0 & 0 \\
\end{array}
\right)}.\label{b2}
\ea
These matrices can also be viewed as {\it adjacency} matrices: each off-diagonal non-zero entry in the matrices corresponds to a coupling of the two field variables associated to the row/column index and thus to an edge in the square lattice of figure \ref{fig_n2square}. The diagonal components in the first two matrices correspond to self-coupling of fields at steps $n=1,2$, respectively. For the dynamics, the $c^2$ matrix is most important because it describes the adjacency or coupling between the two time steps. For instance, we see that the vertices of the four new fields $\phi_2^1,\ldots,\phi^4_2$ in the corners of the new square at $n=2$ are not connected by edges to vertices at the earlier time step $n=1$. Accordingly, $c^2_{ij}=0$ for $i,j=1,\ldots,4$ such that the new corner fields are not coupled to the fields at step $n=1$. We thus already anticipate that the values of the new corner fields cannot be predicted.}

%\emph{Now consider $1\rightarrow2$. Evidently, the only non-zero components of $c^2$ are
%\ba
%c^2_{16}=c^2_{17}=c^2_{28}=c^2_{29}=c^2_{310}=c^2_{311}=c^2_{412}=c^2_{45}=-2,\nn
%\ea
%while the only non-zero components of $a^2$ read
%\ba
%a^2_{ii}=6+\f{3}{2}m^2, \q\q a^2_{i(i+1)}=-1,\q\q i=1,\ldots,4\,,\q\q\text{where}\q\q a^2_{4(5)}:=a^2_{41}\,.\nn
%\ea
%Finally, the only non-zero components of $b^2$ are given by
%\ba
%b^2_{ii}=2+\f{1}{2}m^2,\q\q i=1,\ldots,4,\q\q b^2_{jj}=4+m^2,\q\q j=5,\ldots,12\nn\\
%b^2_{16}=b^2_{17}=b^2_{28}=b^2_{78}=b^2_{29}=b^2_{910}=b^2_{310}=b^2_{311}=b^2_{1112}=b^2_{412}=b^2_{45}=b^2_{56}=-1.\nn
%\ea
\emph{$c^2$ has an eight-dimensional left null space spanned by the basis vectors
\ba\label{L1}
(L_1)_5{}^i=\delta_{i5},\q\ldots,\q (L_1)_{12}{}^i=\delta_{i12},
\ea
which results in the following eight pre--constraints (\ref{kincon}) for the spurious modes at $n=1$
\ba\label{exp-precon}
{}^-C^1_j:={}^-p^1_j=0,\q\q j=5,\ldots,12.
\ea
Four non-trivial pre--momentum evolution equations of the Hamiltonian map $\mathfrak{H}_1$ (\ref{mom1b}) arise
\ba\label{premomsn1}
{}^-p^1_1&=&\left(6+\f{3}{2}m^2\right)\phi^1_1-\phi^{2}_1-\phi^{4}_1-2\left(\phi_2^6+\phi_2^7\right),\nn\\
{}^-p^1_2&=&\left(6+\f{3}{2}m^2\right)\phi^2_1-\phi^{1}_1-\phi^{3}_1-2\left(\phi_2^8+\phi_2^9\right),\nn\\
{}^-p^1_3&=&\left(6+\f{3}{2}m^2\right)\phi^3_1-\phi^{2}_1-\phi^{4}_1-2\left(\phi_2^{10}+\phi_2^{11}\right),\\
{}^-p^1_4&=&\left(6+\f{3}{2}m^2\right)\phi^4_1-\phi^{1}_1-\phi^{3}_1-2\left(\phi_2^5+\phi_2^{12}\right).\nn
\ea
Given the canonical data $(\phi_1,{}^-p^1)$, one can solve each of these four equations for the {\it sum} of a pair of fields residing on the nearest neighbouring vertices of the four corners of the new square in figure \ref{fig_n2square}. For example, through the first equation the sum $\phi^6_2+\phi^6_2$ can be determined given the canonical data at step $n=1$. This is the sum of the field variables directly neighbouring the upper left corner field $\phi^1_2$. The corner field itself cannot appear in the predictable variable because, as seen above, it does not couple to the fields at step $n=1$. }

\emph{If one performed a Fourier mode analysis of the scalar field on the lattice, one would find that the {\it sum} of the (real space) field variables corresponding to neighbouring vertices is written in terms of a sum over Fourier modes multiplied by a coefficient function which dominates for small mode numbers and becomes small for large mode numbers (near the Brillouin zone boundary). That is, in terms of Fourier modes, the predictable sums of (nearly) neighbouring field variables correspond to smaller Fourier modes and thereby to coarser degrees of freedom. However, we shall not carry out a Fourier analysis in detail here as this would be specific to the scalar field model and would go beyond the scope of this article which focuses rather on the general formalism. Instead, a detailed classical and quantum discussion of a scalar field on a growing lattice including a Fourier analysis will appear in \cite{biancameted} (see also \cite{Foster:2004yc}).}

%Sums of field variables correspond to coarser degrees of freedom, while (alternating) differences correspond to finer degrees of freedom. Viewing the move $1\rightarrow2$ as a refining time evolution which adds new degrees of freedom, we see that, using $\mathfrak{H}_1$ we can only predict coarser degrees of freedom, while finer degrees of freedom remain undetermined. in fourier analysis sums correspond to .....but abstain from doing so in detail here. would be a whole different topic for scalar field. will appear elsewhere \cite{}. here technical focus of formalism, less importance on specific example.}

\emph{The eight-dimensional right null space of $c^2$ is spanned by the vectors
\ba\label{R2}
(R_2)_1{}^i&=&\delta_{i12}-\delta_{i5},\q (R_2)_2{}^i=\delta_{i11}-\delta_{i10},\q (R_2)_3{}^i=\delta_{i9}-\delta_{i8},\q (R_2)_4{}^i=\delta_{i7}-\delta_{i6},\nn\\
(R_2)_5{}^i&=&\delta_{i1},\q\ldots,\q (R_2)_8{}^i=\delta_{i4}.
\ea
Using (\ref{kincon1}), this gives the following eight post--constraints at $n=2$
\ba\label{postconsn2}
{}^+C^2_1&=&{}^+p^2_{12}-{}^+p^2_5+\left(4+m^2\right)\left(\phi_2^5-\phi_2^{12}\right)+\phi_2^{11}-\phi_2^6,\nn\\
{}^+C^2_2&=&{}^+p^2_{11}-{}^+p^2_{10}+\left(4+m^2\right)\left(\phi_2^{10}-\phi_2^{11}\right)+\phi_2^{12}-\phi_2^9,\nn\\
{}^+C^2_3&=&{}^+p^2_{9}-{}^+p^2_8+\left(4+m^2\right)\left(\phi_2^8-\phi_2^{9}\right)+\phi_2^{10}-\phi_2^7,\nn\\
{}^+C^2_4&=&{}^+p^2_{7}-{}^+p^2_6+\left(4+m^2\right)\left(\phi_2^6-\phi_2^{7}\right)+\phi_2^{8}-\phi_2^5,\nn\\
{}^+C^2_5&=&{}^+p^2_4-\left(2+\f{1}{2}m^2\right)\phi_2^4+\phi_2^5+\phi_2^{12},\\
{}^+C^2_6&=&{}^+p^2_3-\left(2+\f{1}{2}m^2\right)\phi_2^3+\phi_2^{10}+\phi_2^{11},\nn\\
{}^+C^2_7&=&{}^+p^2_2-\left(2+\f{1}{2}m^2\right)\phi_2^2+\phi_2^{8}+\phi_2^{9},\nn\\
{}^+C^2_8&=&{}^+p^2_1-\left(2+\f{1}{2}m^2\right)\phi_2^1+\phi_2^{6}+\phi_2^{7},.\nn
\ea
The final four post--constraints correspond to the four unpredictable corner fields at $n=2$ in figure \ref{fig_n2square}: the momentum variable and its conjugate part are associated to the four corner vertices which are not connected to vertices at the earlier step $n=1$. On the other hand, we note that in the first four post--constraints in (\ref{postconsn2}) the momentum variables and their conjugate field variables correspond to the {\it differences} of the pairs of new field variables directly neighbouring the new corner vertices. For instance, the fourth constraint has ${}^+p^2_{7}-{}^+p^2_6$ as the momentum variable and $\phi_2^6-\phi_2^{7}$ as its conjugate (the last term $\phi_2^{8}-\phi_2^5$ commutes with the momentum part). These are the {\it differences} of the canonical data associated to the two vertices neighbouring the upper left corner vertex $1$ in figure \ref{fig_n2square} and constitute the {\it unpredictable} part of the data at $n=2$. In a Fourier mode analysis, the {\it difference} of neighbouring field variables amounts to a sum over Fourier modes multiplied by a coefficient function which dominates for large mode numbers and becomes small for small mode numbers. The unpredictable combinations appearing in the constraints thus correspond to larger Fourier modes and thereby to finer degrees of freedom. } 

\emph{The move $1\rightarrow2$ is a growing lattice time evolution which adds new but unpredictable small scale degrees of freedom at $n=2$. Using the Hamiltonian time evolution map $\mathfrak{H}_1$ and the canonical data at $n=1$ we can only predict the four coarse degrees of freedom given by the {\it sums} of field variable pairs neighbouring the corners, while their {\it differences}, representing finer degrees of freedom, remain unpredictable. }

\emph{Finally, employing (\ref{mom1b}), the following four non--trivial post--momentum evolution equations of the map $\mathfrak{H}_1$
\ba\label{postmomsn2}
{}^+p^2_{12}&=&\left(4+m^2\right)\phi_2^{12}-\phi_2^{4}-\phi_2^{11}-2\phi_1^4,\nn\\
{}^+p^2_{11}&=&\left(4+m^2\right)\phi_2^{11}-\phi_2^3-\phi_2^{12}-2\phi_1^3,\nn\\
{}^+p^2_{9}&=&\left(4+m^2\right)\phi_2^9-\phi_2^2-\phi_2^{10}-2\phi_1^2,\\
{}^+p^2_{7}&=&\left(4+m^2\right)\phi_2^7-\phi_2^1-\phi_2^8-2\phi_1^1\nn
\ea
remain. Given the canonical data $(\phi_2,{}^+p^2)$ at $n=2$, these four equations allow one to postdict the four dynamical variables $\phi_1^i$, $i=1,\ldots,4$ at step $n=1$.}

\emph{Before we discuss the constraints and evolution equations of this example model further, we firstly return to the general case and introduce a classification of the null vectors.}
\end{Example}

\section{Classification of the null vectors}\label{sec_nullvecs}

We continue by classifying the null vectors of $c^1,c^2,h^{12}$ according to eight types. Henceforth, we distinguish between five broad of the eight types by denoting the corresponding null vectors at $n=1$ by five different letters $(Y_1),(L_1),(R_1),(Z_1),(V_1)$. Sub--cases will further be distinguished by lower indices. (For the moment we drop the index $i\in1,\ldots,Q$.) Together, there are eight possibilities:
\begin{itemize}
\item[{\bf (1)}] $c^1\cdot (Y_1)=0=(Y_1)\cdot c^2$. ($(Y_1)$: both right null vector of $c^1$ and left null vector of $c^2$.) We label these null vectors by the indices $I$ or $H$ according to whether
\begin{itemize}
\item[{\bf (A)}] $(Y_1)_I\cdot h^{12}=0$,
\item[{\bf (B)}] $(Y_1)_H\cdot h^{12}\neq0$.
\end{itemize}
\item[{\bf (2)}] $(L_1)\cdot c^2=0$, but $c^1\cdot (L_1)\neq0$. ($(L_1)$: left, but not right null vector.) These null vectors are labeled by the indices $l$ or $\lambda$ according to whether
\begin{itemize}
\item[{\bf (A)}] $(L_1)_l\cdot h^{12}=0$,
\item[{\bf (B)}] $(L_1)_\lambda\cdot h^{12}\neq0$.
\end{itemize}
\item[{\bf (3)}] $c^1\cdot (R_1)=0$, but $(R_1)\cdot c^2\neq0$. ($(R_1)$: right, but not left null vector.) These right null vectors are labeled by the indices $r$ or $\rho$ according to whether
\begin{itemize}
\item[{\bf (A)}] $(R_1)_r\cdot h^{12}=0$,
\item[{\bf (B)}] $(R_1)_\rho\cdot h^{12}\neq0$.
\end{itemize}
\item[{\bf (4)}] $(Z_1)\cdot h^{12}=0$, but $c^1\cdot (Z_1)\neq0\neq (Z_1)\cdot c^2$. ($(Z_1)$: null vector of Hessian, but not left or right null vector.)
\item[{\bf (5)}] $(V_1)\cdot h^{12}\neq0$, $(V_1)\cdot c^2\neq0$ and $c^1\cdot(V_1)\neq0$. ($(V_1)$: no null vector.)
\end{itemize}

How does the three--step action (\ref{2step}) vary under transformations defined by these null vectors? Consider an arbitrary variation of the `bulk variables' $x^i_1\rightarrow x^i_1+\varepsilon W^i$ with an arbitrary vector $W^i$ and an infinitesimal order parameter $\varepsilon$. The variation of the action $S_1+S_2$ reads 
\ba
\delta S=\varepsilon\left(c^1_{ij}x^i_0W^j+h^{12}_{ij}x^i_1W^j+c^2_{ji}W^jx^i_2\right)+\frac{\varepsilon^2}{2}h^{12}_{ij}W^iW^j\underset{(\ref{eom})}{=}\frac{\varepsilon^2}{2}h^{12}_{ij}W^iW^j
\ea
and thus on-shell
\ba\label{invac}
  h^{12}\cdot W=0\q\Rightarrow\q\delta S=0\,.
\ea
That is, null vectors of the Hessian $h^{12}$ of the three--step action (\ref{2step}), $S_1+S_2$, define symmetries of this piece of action. We shall see later in section \ref{sec_classeff} that not all of the null vectors of $h^{12}$ extend to null vectors of `effective' Hessians obtained after integrating out neighbouring time steps in an evolution involving larger numbers of steps. Hence, not all null vectors of $h^{12}$ will define symmetries of `effective' actions. 

In other words, the classification of the null vectors (and thus the corresponding classification of constraints and degrees of freedom below) is {\it evolution move dependent}. However, this only occurs for systems with a temporally varying discretization and, consequently, temporally varying numbers of dynamical degrees of freedom \cite{Dittrich:2013jaa,Hoehn:2014fka,Hoehn:2014wwa,Dittrich:2013xwa}. This is not surprising in view of the fact that, at least in a spacetime context, a given evolution move corresponds to a region of (discrete) spacetime to which an action contribution can be associated. Composition of two moves to a new move is equivalent to a gluing of the corresponding spacetime regions to a new region. Every region, through its associated action, comes with its own set of propagating and gauge degrees of freedom, as well as constraints; different regions will generally contain different dynamics and degrees of freedom. In particular, {\it discretization changing evolution moves can be viewed as refining or coarse graining operations} which necessarily change the dynamics. This is the origin for the move dependence of the classification of degrees of freedom. We shall see this in detail below for the case of quadratic discrete actions. For a more profound conceptual discussion of this move or region dependence we refer the reader to \cite{Dittrich:2013jaa,Hoehn:2014fka,Hoehn:2014wwa,Dittrich:2013xwa}.

\begin{Example}
\emph{In the example of the Euclidean scalar field on an expanding square lattice, it is easy to classify the null vectors at $n=1$ according to the eight types above:
\begin{enumerate}
\item Since $c^1\equiv0$, the eight left null vectors (\ref{L1}) are also right null vectors at $n=1$. Moreover, they are also null vectors of $h^{12}=b^1+a^2$ and, hence, are of type (1)(A). 
\item We can extend the set (\ref{L1}) by the following four vectors
\ba\label{v1v4}
(R_1)_1{}^i=\delta_{i1},\q\ldots,\q (R_1)_4{}^i=\delta_{i4}
\ea
to obtain a full 12-dimensional basis. These vectors are trivially right null vectors of $c^1$, however, are neither null vectors of $c^2$ nor of $h^{12}$. Thus, these vectors are of type (3)(B).
\end{enumerate}
There are no other types of vectors at $n=1$ in this example model.}
\end{Example}

\section{Classification of the constraints}\label{sec_constraintclass}

The composition of the two moves $0\rightarrow1$ and $1\rightarrow2$ requires a momentum matching ${}^+p^1={}^-p^1$ at $n=1$, i.e.\ the two sets of pre-- and post--constraints (\ref{kincon}), respectively, must {\it both} be imposed. The pre--constraints ${}^-C^1$ arising in the move $1\rightarrow2$ now comprise conditions that must be satisfied by the canonical data of the move $0\rightarrow1$, while the post--constraints ${}^+C^1$ arising in the move $0\rightarrow1$ now constitute conditions that must be fulfilled by the canonical data of the move $1\rightarrow2$. This leads to the following general constraint characterization (for details see \cite{Dittrich:2013jaa,Hoehn:2014fka}):
\begin{itemize}
\item[(a)] A pre--constraint coincides with a post--constraint (in possibly rewritten form), $C^1:={}^+C^1={}^-C^1$. It was generally shown in \cite{Dittrich:2013jaa}---and we shall see this specifically for quadratic actions below---that such a constraint generates a gauge symmetry. 
\item[(b1)] A pre--constraint is independent of all post--constraints, but first class. Such a pre--constraint can only arise during a {\it coarse graining} or {\it lattice shrinking} time evolution: it constitutes a non-trivial {\it coarse graining condition} on the degrees of freedom which are dynamically relevant for $0\rightarrow1$ and renders those which are `finer' than the coarse graining scale set by $1\rightarrow2$ dynamically irrelevant for the latter \cite{Dittrich:2013jaa,Hoehn:2014fka,Dittrich:2013xwa}. Such a pre--constraint does {\it not generate} a symmetry, albeit being first class, but simply reduces the space of solutions by one canonical pair. The (covariant) phase space is then move dependent. 
\item[(b2)] A post--constraint is independent of all pre--constraints, but first class. This is the time-reverse of (b1) and only occurs in a {\it refining} or {\it lattice growing} time evolution. Such a post--constraint thus does not generate a symmetry but enforces that the coarser data of $0\rightarrow1$ can be consistently embedded in the finer data of $1\rightarrow2$. Equivalently, it is a non-trivial coarse graining condition on the move $1\rightarrow2$ and thereby further reduces the space of solutions or phase space associated to this move by one canonical pair.
\item[(c)] A pre--constraint (post--constraint) is independent of the post--constraints (pre--constraints), but is second class and fixes an orbit parameter of one of the latter.  
\item[(d)] A pre--constraint may be incompatible with {\it all} post--constraints (or vice versa), yielding an inconsistent dynamics.
\end{itemize}
We shall see shortly how the first four cases arise in the present context.

Before we classify the constraints arising from quadratic discrete actions further according to the above classification of the null vectors, let us generally determine when the pre-- and post--constraints of the present case are first or second class. To this end, we recall theorem 5.1 of \cite{Dittrich:2013jaa}, according to which the set of pre--constraints ${}^{-}C^1$, on the one hand, and the set of post--constraints ${}^+C^1$, on the other, each form a first class Poisson sub--algebra. In the present case this is an immediate consequence of the symmetry of the matrices $b^1,a^2$ and the sub--algebras are even abelian. However, in general, the Poisson brackets between the pre-- and post--constraints may not vanish. In our case, the brackets between ${}^{-}C^1,{}^+C^1$ of (\ref{kincon}) read\footnote{We assume momentum matching $p^1_i:={}^+p^1_i={}^-p^1_i$.} 
\ba\label{1st}
\{{}^-C^1,{}^+C^1\}=(L_1)^i(R_1)^j\{p^1_i
+a^2_{il}x^l_1,p^1_j-b^1_{jm}x^m_1\}=(L_1)^ih^{12}_{ij}(R_1)^j\,.
\ea
%which is (\ref{LHR}) in simplified form.

Consequently, pre-- and post--constraints are first class if the corresponding left or right null vector is also a null vector of the Hessian $h^{12}$. In fact, in this case the corresponding constraints are {\it abelian}. These first class constraints therefore generate symmetries of the three--step action (\ref{2step}), $S_1+S_2$. However, just like some null vectors of $h^{12}$ will fail to be null vectors of `effective' Hessians upon inclusion of additional evolution steps, we shall see later in section \ref{sec_classeff} that constraints of types (2)(A) and (3)(A), which are first class for the problem defined by $S_1+S_2$, will generally no longer be first class for a problem defined by `effective' actions for more time steps. This is the move or region dependence of the classification alluded to above.

Let us now classify the constraints and equations of motion according to the eight vector types:
\\~
\\~
{\bf (1)} Using momentum matching, $p^1_i:={}^+p^1_i={}^-p^1_i$, the pre-- and post--constraints (\ref{kincon}) corresponding to the $(Y_1)$ satisfy
\ba
{}^+C^1={}^-C^1-(Y_1)^ih^{12}_{ij}x^j_1\,.
\ea
\begin{itemize}
\item[{\bf (A)}] Denote the constraints corresponding to $(Y_1)_I$ as follows
\ba
{}^+C^1_I=(Y_1)_I{}^i\left(p^1_i-b^1_{ij}x^j_1\right)\,,\q\q\q {}^-C^1_I=(Y_1)_I{}^i\left(p^1_i
+a^2_{ij}x^j_1\right)\,.
\ea
As a result of $(Y_1)_I\cdot h^{12}=0$, these pre-- and post--constraints coincide on--shell
\ba
C^1_I:={}^+C^1_I={}^{-}C^1_I\,
\ea
and are thus an example of case (a) above. Indeed, by (\ref{1st}), the $C^1_I$ are abelian and thus {\bf first class} (this is a special case of corollary 5.1 in \cite{Dittrich:2013jaa}) and (\ref{invac}) entails that these coinciding constraints generate genuine gauge transformations of the action (\ref{2step}) (this is a special case of theorem 5.2 in \cite{Dittrich:2013jaa}). Furthermore, theorem 5.3 in \cite{Dittrich:2013jaa} implies that to each such $C^1_I$ there will be associated a genuine gauge mode such that these constraints must also be gauge generators of `effective' actions. We shall confirm this below.
\item[{\bf (B)}] In analogy, by $(Y_1)_H\cdot h^{12}\neq0$, one finds 
\ba
{}^+C^1_H={}^-C^1_H-(Y_1)_H{}^ih^{12}_{ij}x^j_1\,,
\ea
and thus ${}^+C^1_H\neq{}^-C^1_H$ such that on the constraint surface an additional (dependent) set of {\it holonomic}\footnote{Holonomic constraints are constraints that only involve configuration variables \cite{marsdenratiu,marsdenwest}. Notice that holonomic constraints can never be primary constraints because the latter are defined through the Legendre transformations which always involve the momenta.} constraints is produced
\ba\label{holcon1}
(Y_1)_H{}^ih^{12}_{ij}x^j_1=0\,.
\ea
The ${}^+C^1_H, {}^-C^1_H$ always arise in pairs, satisfy $\{{}^+C^1_H, {}^-C^1_{H'}\}=(Y_1)_H\cdot h^{12}\cdot (Y_1)_{H'}$ and are thus generally {\bf second class}. (Likewise, the dependent holonomic constraints (\ref{holcon1}) do {\it not} commute with the ${}^\pm C^1_H$ and are thus second class too.) As a consequence of (\ref{invac}), these constraints do {\it not} generate gauge transformations of the three--step action (\ref{2step}). This is an example to case (c) above.
\end{itemize}
%\\~
%\\~
{\bf (2)} Denote the corresponding pre--constraints by ${}^-C^1_l$ and ${}^-C^1_\lambda$, respectively.
\begin{itemize}
\item[{\bf (A)}] Since $(L_1)_l\cdot h^{12}=0$, the ${}^-C^1_l$ are {\bf first class} (even abelian) symmetry generators of the three--step action $S_1+S_2$. Acting with $(L_1)_l$ on (\ref{eom}), one obtains independent secondary holonomic constraints at the \underline{initial} step $n=0$ which must be satisfied \underline{on--shell},
\ba\label{holcon2}
H^0_l:=(L_1)_{l}{}^ic^1_{ji}x^j_0=0\,.
\ea
Combining these holonomic constraints with a contraction of the right equations in (\ref{mom1}) with $(L_1)_l{}^i$ yields {\it secondary} post--constraints ${}^+C^1_l$. Invoking again $(L_1)_l\cdot h^{12}=0$, one finds that (on-shell) $C^1_l:={}^+C^1_l\equiv{}^-C^1_l$. That is, at least on-shell, the ${}^-C^1_l$ are of case (a) above. However, as we shall see in section \ref{sec_redps}, the ${}^-C^1_l$ can also be viewed as coarse graining conditions on the move $0\rightarrow1$, rendering originally propagating degrees of freedom of this move into non-dynamical Lagrange multipliers of the holonomic constraints (\ref{holcon2}).
\item[{\bf (B)}] As a consequence of $(L_1)_\lambda\cdot h^{12}\neq0$ and by (\ref{invac}), the ${}^-C^1_\lambda$ do {\it not} generate symmetries of $S_1+S_2$. Because of (\ref{1st}), these pre--constraints are \underline{generally} {\bf second class} in which case they are of case (c) above. However, (\ref{1st}) may, nevertheless, vanish for a given ${}^-C^1_\lambda$, even though $(L_1)_\lambda\cdot h^{12}\neq0$. In this case, this type of constraint would be {\bf first class} and of case (b1) above.
\end{itemize}
{\bf (3)} Denote the corresponding post--constraints by ${}^+C^1_r$ and ${}^+C^1_\rho$, respectively.
\begin{itemize}
\item[{\bf (A)}] $(R_1)_r\cdot h^{12}=0$ implies that the ${}^-C^1_l$ are {\bf first class} (even abelian) symmetry generators of $S_1+S_2$. Again, projecting (\ref{eom}) with $(R_1)_r$ yields independent holonomic constraints at the \underline{final} step $n=2$ which must also be fulfilled on--shell,
\ba\label{holcon3}
H^2_r:=(R_1)_{r}{}^ic^2_{ij}x^j_2=0\,.
\ea
In analogy to type (2)(A), one finds {\it secondary} pre--constraints ${}^-C^1_r$ which coincide with the ${}^+C^1_r$ such that this type of constraint is on-shell of case (a). 
\item[{\bf (B)}] By $(R_1)_\rho\cdot h^{12}\neq0$, the transformations generated by the ${}^+C^1_\rho$ do {\it not} leave the action $S_1+S_2$ invariant. In analogy to case (2)(B), these pre--constraints are \underline{generally} {\bf second class} and of case (c) above because (\ref{1st}) does not vanish in general. However, there can still be cases when it does vanish. In this case the corresponding ${}^+C^1_\rho$ is {\bf first class} and of case (b2) above.
\end{itemize}
{\bf (4)} Since $(Z_1)$ are not null vectors of $c^1,c^2$, no canonical constraints result from the Legendre transformations (\ref{mom1}, \ref{mom1b}). However, projecting (\ref{eom}) with $(Z_1)$ yields holonomic relations between $x_0$ and $x_2$,
\ba
B^{02}_z:=(Z_1)_z{}^j\left(c^1_{ij}x^i_0+c^2_{ji}x^i_2\right)=0\,,\label{2stepcon}
\ea
and thus between initial and final steps. We shall call these {\it boundary data constraints} (see also \cite{Dittrich:2013jaa}). Notice that these are {\it secondary}. (\ref{invac}) implies that the $(Z_1)$ define symmetries of the three--step action (\ref{2step}). That is, the presence of degenerate directions of the Hessian does {\it not} necessarily imply the existence of symmetry generating canonical constraints. In section \ref{sec_classeff} we shall see that $(Z_1)$ will generally not define degenerate directions of `effective' Hessians.
\\~
\vspace*{-.2cm}
\\~
{\bf (5)} The vectors $(V_1)$ yield via (\ref{eom}) proper equations of motion, relating the three discrete time steps. No constraints arise.
\\

In summary, the Poisson bracket structure of the constraints at $n=1$ is schematically represented in table \ref{conal}. In this table we have also included the holonomic constraints of the kind (\ref{holcon2}, \ref{holcon3}), $H^1_l,H^1_r$, which may arise at $n=1$ upon including further time steps and solving the equations of motion analogous to (\ref{eom}) at $n=0$ and $n=2$. This exhausts the list of constraints at $n=1$ for the given move.\footnote{Upon further integrating out $n=-1$ and $n=3$, additional independent holonomic constraints $\tilde{H}^1_{l'},\tilde{H}^1_{r'}$ may, in principle, arise (see also figure \ref{bvpfig}). But they will be of the same shape and we shall ignore them here.} The only generally non--zero Poisson brackets of the latter holonomic constraints are
\ba
\{{}^+C^1_\rho,H^1_l\}&=&(L_2)_{l}{}^ic^2_{ji}(R_1)_{\rho}{}^j\neq0\,,\q\q\{{}^+C^1_r,H^1_l\}=(L_2)_{l}{}^ic^2_{ji}(R_1)_{r}{}^j\neq0\,,\nn\\
\{{}^-C^1_\lambda,H^1_{r}\}&=&(R_0)_{r}{}^ic^1_{ij}(L_1)_{\lambda}{}^j\neq0\,,\q\q
\{{}^-C^1_l,H^1_{r}\}\,=(R_0)_{r}{}^ic^1_{ij}(L_1)_{l}{}^j\,\neq0\,.\nn
\ea
Thus, in general, only the $C_I$ are necessarily first class constraints. Although in cases, only vectors of, say, type (3)(B) may arise at $n=1$ initially in which case their associated primary post--constraints ${}^+C^1_\rho$ are trivially first class because there are no pre--constraints. This happens only in a refining time evolution, as in the scalar field example below. However, on account of the move dependence, integrating out neighbouring steps can produce secondary pre--constraints at $n=1$ which may render the primary ${}^+C^1_\rho$ second class.  
%We will see an example of this scenario in section \ref{sec_examp} and discuss the consequences for the classification of degrees of freedom in the general case in section \ref{sec_minstep}.

\begin{table*}[h]
\centering \caption{\small Schematic summary of the Poisson bracket structure of the constraints at $n=1$. First
terms in the Poisson bracket are labeled by rows, second terms are labeled
by columns. An $X$ means that the corresponding constraints generally do not Poisson commute with each other. \label{conal}}
\begin{tabular}{|c|| c|c|c|c|c|c|c|c|c|} \hline & $C^1_I$
& ${}^+C^1_H$ & ${}^-C^1_H$ &
${}^-C^1_l$ & ${}^-C^1_\lambda$ &
${}^+C^1_r$ & ${}^+C^1_\rho$ &
$H^1_l$ & $H^1_r$  \\ \hline\hline 
$C^1_I$ & ${\scriptstyle 0}$ & ${\scriptstyle 0}$ &
${\scriptstyle 0}$ & ${\scriptstyle 0}$ &
${\scriptstyle 0}$ & ${\scriptstyle 0}$ & ${\scriptstyle 0}$ &
${\scriptstyle 0}$ & ${\scriptstyle 0}$ \\ \hline
${}^+C^1_H$ & ${\scriptstyle 0}$ & ${\scriptstyle 0}$ &
${\scriptstyle X}$ & ${\scriptstyle 0}$ &
${\scriptstyle X}$ & ${\scriptstyle 0}$ & ${\scriptstyle 0}$ &
${\scriptstyle 0}$ & ${\scriptstyle 0}$ \\ \hline
${}^-C^1_H$ & ${\scriptstyle 0}$ & ${\scriptstyle X}$ &
${\scriptstyle 0}$ & ${\scriptstyle 0}$ &
${\scriptstyle 0}$ & ${\scriptstyle 0}$ & ${\scriptstyle X}$ &
${\scriptstyle 0}$ & ${\scriptstyle 0}$ \\ \hline
${}^-C^1_l$ & ${\scriptstyle 0}$ & ${\scriptstyle 0}$ &
${\scriptstyle 0}$ & ${\scriptstyle 0}$ &
${\scriptstyle 0}$ & ${\scriptstyle 0}$ & ${\scriptstyle 0}$ &
${\scriptstyle 0}$ & ${\scriptstyle X}$ \\ \hline
${}^-C^1_\lambda$ & ${\scriptstyle 0}$ & ${\scriptstyle X}$ &
${\scriptstyle 0}$ & ${\scriptstyle 0}$ &
${\scriptstyle 0}$ & ${\scriptstyle 0}$ & ${\scriptstyle X}$ &
${\scriptstyle 0}$ & ${\scriptstyle X}$ \\ \hline
${}^+C^1_r$ & ${\scriptstyle 0}$ & ${\scriptstyle 0}$ &
${\scriptstyle 0}$ & ${\scriptstyle 0}$ &
${\scriptstyle 0}$ & ${\scriptstyle 0}$ & ${\scriptstyle 0}$ &
${\scriptstyle X}$ & ${\scriptstyle 0}$ \\ \hline
${}^+C^1_\rho$ & ${\scriptstyle 0}$ & ${\scriptstyle 0}$ &
${\scriptstyle X}$ & ${\scriptstyle 0}$ &
${\scriptstyle X}$ & ${\scriptstyle 0}$ & ${\scriptstyle 0}$ &
${\scriptstyle X}$ & ${\scriptstyle 0}$ \\ \hline
$H^1_l$ & ${\scriptstyle 0}$ & ${\scriptstyle 0}$ &
${\scriptstyle 0}$ & ${\scriptstyle 0}$ &
${\scriptstyle 0}$ & ${\scriptstyle X}$ & ${\scriptstyle X}$ &
${\scriptstyle 0}$ & ${\scriptstyle 0}$ \\ \hline
$H^1_r$ & ${\scriptstyle 0}$ & ${\scriptstyle 0}$ &
${\scriptstyle 0}$ & ${\scriptstyle X}$ &
${\scriptstyle X}$ & ${\scriptstyle 0}$ & ${\scriptstyle 0}$ &
${\scriptstyle 0}$ & ${\scriptstyle 0}$ \\ \hline
\end{tabular}
\end{table*}

\begin{Example}
\emph{The eight post--constraints (\ref{exp-postcon2}) and the eight pre--constraints (\ref{exp-precon}) for the spurious modes at $n=1$ coincide upon momentum matching ${}^+p^1={}^-p^1$, and are of type (1)(A), i.e., are gauge generators. The four remaining post--constraints (\ref{exp-postcon1}) are associated to the vectors (\ref{v1v4}) and, hence, are of type (3)(B). Since there do not exist any further pre--constraints at $n=1$, \emph{all} constraints are first class at $n=1$. In particular, the post--constraints (\ref{exp-postcon1}) are of the general case (b2) and thus coarse graining conditions for the move $1\rightarrow2$. They ensure that the `coarser' field data of step $n=1$ can be consistently represented and mapped to step $n=2$.}
\end{Example}

\section{Classification of the degrees of freedom}\label{sec_classdof}

The next task is to classify the degrees of freedom appearing in the three--step action (\ref{2step}) into gauge and propagating modes for the evolution $0\rightarrow1\rightarrow2$, according to \cite{Dittrich:2013jaa}. Subsequently, we shall worry about how these degrees of freedom behave upon inclusion of additional time steps.

In order to make the different types of degrees of freedom explicit, it is useful to introduce a linear canonical transformation on the phase space $T^*\cq_n$ which takes the classification of the null vectors of section \ref{sec_nullvecs} suitably into account and produces canonical variables according to the eight types of vectors.

Namely, at step $n$ introduce an invertible transformation matrix $(T_n)_\Gamma{}^i$, where $i=1,\ldots,Q$ and $\Gamma$ enumerates a suitable basis of $Q$ vectors of the eight types of section \ref{sec_nullvecs}. The choice of this basis is, of course, non--unique and not all eight types of vectors are, in principle, necessary. For instance, when adding to a vector $(L_n)_l$ of type (2)(A) a vector $(L_n)_\lambda$ of type (2)(B) one obtains another vector $(L_n)_{\lambda'}$ of type (2)(B) such that one could disregard vectors of type (2)(A) in the basis from the start. By similar linear combinations one could equally well disregard vectors of types (3)(A) and (4) and so on. 

\begin{figure}[hbt!]
\begin{center}
\psfrag{1}{$1$}
\psfrag{2}{$2$}
\psfrag{3}{ $3$}
\psfrag{4}{ $4$}
\psfrag{5}{$5$}
\psfrag{l1}{$L_\lambda$}
\psfrag{l2}{$L_l$}
\psfrag{r1}{$R_\rho$}
\psfrag{r2}{$R_r$}
\psfrag{v}{$V_\gamma$}
\psfrag{z}{$Z_z$}
\psfrag{y1}{$Y_I$}
\psfrag{yh}{$Y_H$}
\psfrag{c1}{$c^1$}
\psfrag{c2}{$c^2$}
\psfrag{h}{$h^{12}$}
\includegraphics[scale=.6]{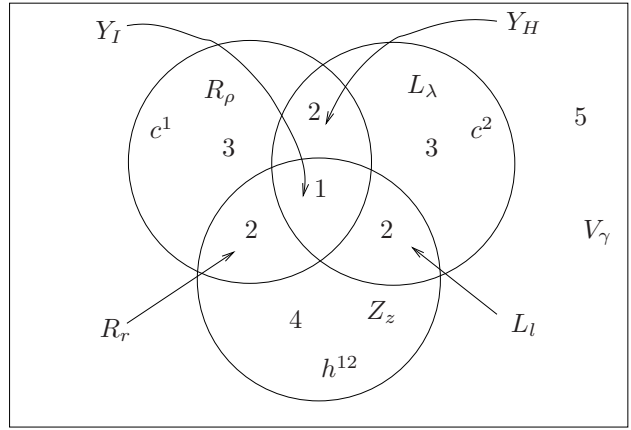}
\caption{\small Schematic characterization of the vector subspaces at $n=1$ into eight sets corresponding to the vector classification in section \ref{sec_nullvecs}. The numbers indicate the step number of the procedure to choose a basis $(T_1)_\Gamma$.}\label{fig_sets}
\end{center}
\end{figure}

Nevertheless, we would like to make a complete and independent set of different types of constraints and degrees of freedom explicit and, additionally, choose the basis $(T_n)_\Gamma$ so as to separate the first class from the second class constraints (according to table \ref{conal}). To this end, it is necessary to isolate a maximal set of independent $(Y_n)_I,(L_n)_l,(R_n)_r, (Y_n)_H$ and include them in the basis $(T_n)_\Gamma$. We therefore choose the column vectors of the transformation matrix $(T_n)_\Gamma{}^i$ according to the following (still non--unique) procedure (see also figure \ref{fig_sets} for an illustration):
\begin{itemize}
\item[0.] Choose a maximal number of linearly independent null vectors of the Hessian $h^{12},c^1,c^2$.
\item[1.] From this set of null vectors construct a maximal number of linearly independent vectors $(Y_n)_I$ of type (1)(A).
\item[2.] Of the remaining null vectors of $c^1,c^2$ choose a maximal number of linearly independent vectors $(Y_n)_H$, $(L_n)_l$ and $(R_n)_r$ of types (1)(B), (2)(A) and (3)(A).
\item[3.] From the rest of the null vectors of $c^1,c^2$ choose a maximal number of independent vectors $(L_n)_\lambda,(R_n)_\rho$ of type (2)(B) or (3)(B).
\item[4.] Of the remaining null vectors of $h^{12}$ choose a maximally independent set of vectors $(Z_n)_z$ of type (4) and enumerate them by index $z$.
\item[5.] Among the remaining vectors of type (5), i.e.\ $(V_n)_\gamma$, choose a maximally independent set and enumerate them by index $\gamma$.
\end{itemize}
Accordingly, $\Gamma$ runs over the indices $I,H,l,\lambda,r,\rho,z$ and $\gamma$ which enumerate the $Q$ basis vectors such that $(T_n)_I{}^i=(Y_n)_I{}^i$,...,$(T_n)_z{}^i=(Z_n)_z{}^i$ and $(T_n)_\gamma{}^i=(V_n)_\gamma{}^i$. 

This leads to the linear transformation\footnote{To keep the notation as simple as possible, we use the same indices for the various types of vectors at the different $n$, despite the fact that, e.g.\ $H$ at $n=0$ may run over less values than $H$ at $n=1$. It should be clear from the $n$ label at the vectors to which set each index refers.}
%\ba\label{lindecomp}
%x^i_n&=&(Y_n)^i_{I}x^I_n+(Y_n)^i_{H}x^H_n+(L_n)^i_{l}x^l_n+(L_n)^i_\lambda x^\lambda_n+(R_n)^i_rx^r_n+(R_n)^i_\rho x^\rho_n\nn\\
%&&\q\q\q\q\q\q\q\q\q\q+(Z_n)^i_zx^z_n+(V_n)^i_\gamma x^\gamma_n\,,\\
%p^n_i&=&(T_n^{-1})^I_ip^n_I+(T_n^{-1})^H_ip^n_H+(T_n^{-1})^l_ip^n_l+(T_n^{-1})^\lambda_i p^n_\lambda+(T_n^{-1})^r_ip^n_r\nn\\
%&&\q\q\q\q\q\q\q\q\q\q+(T_n^{-1})^\rho_ip^n_\rho+(T_n^{-1})^z_ip^n_z+(T_n^{-1})^\gamma_ip^n_\gamma\,.\nn
%\ea
%and the transformation
\ba\label{lindecomp}
x^\Gamma_n=((T_n^{-1})^T)^\Gamma{}_ix^i_n\,,\q\q\q\q p^n_\Gamma=(T_n)_\Gamma{}^i p^n_i\,,
\ea
where the superscript $T$ stands for transposition. This transformation is canonical because 
\ba
\{x^\Gamma_n,p^n_{\Gamma'}\}=((T_n^{-1})^T)^\Gamma{}_i (T_n)_{\Gamma'}{}^j\{x^i_n,p^n_j\}= (T_n^{-1})_i{}^\Gamma(T_n)_{\Gamma'}{}^i=  \delta^\Gamma_{\Gamma'}.\nn
\ea
Indeed, it can be easily checked that
\ba
A=\left(\begin{array}{cc}(T^{-1})^T & 0 \\0 & T\end{array}\right)\nn
\ea
acting on (tangent spaces of) $T^*\cq_n$ is an element of the symplectic group $\rm{Sp}(2Q,\mathbb{R})$, as necessary for a linear canonical transformation (e.g., see \cite{marsdenratiu}).

In fact, for a simple characterization of the degrees of freedom, it is useful to perform yet another canonical transformation on $T^*\cq_n$ {\it prior to} momentum matching, i.e.\ imposing the equations of motion. It proceeds differently for pre-- and post--momenta (we shall discuss momentum matching below in section \ref{sec_redps})
\ba
x^\Gamma_n&\rightarrow& x^\Gamma_n \q\q\q\q {}^-p^n_\Gamma\,\rightarrow\,{}^-\pi^n_\Gamma:={}^-p^n_\Gamma+(T_n)_\Gamma{}^i a^{n+1}_{ij}x^j_n\,,\nn\\
x^\Gamma_n&\rightarrow& x^\Gamma_n \q\q\q\q {}^+p^n_\Gamma\,\rightarrow\,{}^+\pi^n_\Gamma:={}^+p^n_\Gamma-(T_n)_\Gamma{}^i b^{n}_{ij}x^j_n\,.\label{newcantrans}
\ea
The reason for these two transformations will become clear momentarily. It is straightforward to check that both these transformations are canonical, i.e.\
\ba
\{x^\Gamma_n,{}^-\pi^n_{\Gamma'}\}&=&\delta^\Gamma_{\Gamma'}\,,\q\q\q\q\{x^\Gamma_n,{}^+\pi^n_{\Gamma'}\}=\delta^\Gamma_{\Gamma'}\,,\nn\\
\{x^\Gamma_n,x^{\Gamma'}_n\}&=&0\,,\q\q\q\q\{{}^\pm\pi^n_\Gamma,{}^\pm\pi^n_{\Gamma'}\}=0\,.\nn
\ea
The pre-- and post--constraints (\ref{kincon}) at $n$ now take a particularly simple form: they are trivialized,
\ba
{}^-C^n_L={}^-\pi^n_L\,,\q L=I,H,l,\lambda,\q\q\q{}^+C^n_R={}^+\pi^n_R\,,\q R=I,H,r,\rho,\label{newpcons}
\ea
where we have combined all left null vector indices into a new index $L$ and all right null vector indices into a new index $R$. Hence,
\ba
\{x^\Gamma_n,{}^-C^n_L\}&=&\delta^\Gamma_L\,,\q\q\q\q\{{}^-\pi^n_\Gamma,{}^-C^n_L\}=0\,,\nn\\\label{xcommute}
\{x^\Gamma_n,{}^+C^n_R\}&=&\delta^\Gamma_R\,,\q\q\q\q\{{}^+\pi^n_\Gamma,{}^+C^n_R\}=0\,.
\ea
The new canonical pairs can therefore also be classified according to the eight types of vectors. %In particular, each type of $x^\Gamma_n$ is chosen such that it Poisson commutes with all constraints except those which are of the same type: e.g., by (\ref{xcommute}), $x^I_n$ commutes with all constraints except $C^n_I$, etc. Likewise, all pre-- and post--momenta commute with all pre-- and post--constraints, respectively. (Since we do not yet consider momentum matching, i.e.\ the equations of motion (\ref{eom}), we need not worry about the holonomic constraints $H^n_l,H^n_r$ at this stage which only arise on--shell.)

Let us now determine a complete set of propagating degrees of freedom. By propagating degrees of freedom we refer to canonical data at a given time step that, using the Hamiltonian time evolution map $\mathfrak{H}_n$, can be uniquely pre- or postdicted (given some initial or final data). This notion of propagation requires two time steps---in contrast to the continuum---and is move or region dependent. We recall the following characterization of degrees of freedom  \cite{Dittrich:2013jaa}:
\begin{itemize}
\item A {\it pre--observable} at step $n$ is a phase space function which Poisson commutes with all pre--constraints at $n$. This is a propagating canonical datum whose value can be uniquely postdicted, using $\mathfrak{H}_n$, given sufficient final data at step $n+1$.
\item A {\it post--observable} at step $n$ is a phase space function which Poisson commutes with all post--constraints at $n$. This is a propagating canonical datum whose value can be uniquely predicted, using $\mathfrak{H}_{n-1}$, given sufficient initial data at step $n-1$.
\item An {\it a priori free} parameter at $n$ is a degree of freedom which, using $\mathfrak{H}_{n-1}$, cannot be predicted, given initial data at step $n-1$. It thus does not correspond to a propagating datum for $n-1\rightarrow n$.
\item An {\it a posteriori free} parameter at $n$ is a degree of freedom which, using $\mathfrak{H}_{n}$, cannot be postdicted, given final data at step $n+1$. It thus does not correspond to a propagating datum for $n\rightarrow n+1$.
\end{itemize}
In particular, a gauge parameter is both unpre- and unpostdictable and thus both {\it a priori} and {\it a posteriori free}.

What are the pre-- and post--observables associated to the evolution move $0\rightarrow1$? The corresponding Hamiltonian time evolution map $\mathfrak{H}_0$ is given by (\ref{mom1}). %Projecting the equation for the pre--momenta ${}^-p^0$ in (\ref{mom1}) with the left null vectors at $n=0$ and the equation for the post--momenta ${}^+p^1$ in (\ref{mom1}) with the right null vectors at $n=1$ yields the pre-- and post--constraints (\ref{newpcons}). On the other hand, 
Projecting the ${}^-p^0$ in (\ref{mom1}) with $(T_0)_{\Gamma\neq L}$ and the ${}^+p^1$ in (\ref{mom1}) with $(T_1)_{\Gamma\neq R}$ gives the proper Hamiltonian time evolution equations of $\mathfrak{H}_0$ in the form
\ba
{}^-\pi^0_A&=&-(T_0)_A{}^jc^1_{ji}(T_1^T)^i{}_Bx^B_1,\q\q A=r,\rho,\gamma,z,\nn\\
{}^+\pi^1_B&=&\,\,\,\,\,(T_1)_B{}^jc^1_{ij}(T_0^T)^i{}_Ax^A_0,\q\q B=l,\lambda,\gamma,z\,.\label{h0}
\ea
Note that $c^1_{AB}:=(T_0)_A{}^jc^1_{ji}(T_1^T)^i{}_B$ is a square matrix. This can be seen as follows: denote the number of linearly independent $(Y_n)_I, (Y_n)_H,(L_n)_l,\ldots,(V_n)_\gamma$ by $N^n_{I},N^n_{H},N^n_{l},\ldots,N^n_{\gamma}$. $c^1$ possesses as many left as right null vectors and, according to our prescription, a maximal number of linearly independent null vectors of $c^1$ is contained in $(T_0)_{\Gamma'},(T_1)_\Gamma$.  Hence, 
\ba\label{leqr}
N^0_{I}+N^0_{H}+N^0_{l}+N^0_{\lambda}=N^1_{I}+N^1_{H}+N^1_{r}+N^1_{\rho}\,.\nn
\ea
Thanks to $Q=N^n_{I}+N^n_{H}+N^n_{l}+N^n_{\lambda}+N^n_{r}+N^n_{\rho}+N^n_{z}+N^n_{\gamma}=const$, one finds
\ba
N^0_{r}+N^0_{\rho}+N^0_{z}+N^0_{\gamma}= N^1_{l}+N^1_{\lambda}+N^1_{z}+N^1_{\gamma}\,,\nn
\ea
such that $A$ and $B$ run over equally many indices and $c^1_{AB}$ is a square matrix. Since the $(T_0)_A,(T_1)_B$ are not null vectors of $c^1$, $c^1_{AB}$ is generally invertible. 

That is, given the initial data $(x^A_0,{}^-\pi^0_A)$, $A=r,\rho,\gamma,z$, and using $\mathfrak{H}_0$ in the form (\ref{h0}), one can uniquely determine $(x^B_1,{}^+\pi^1_B)$, $B=l,\lambda,\gamma,z$, and vice versa. Furthermore, by (\ref{xcommute}), $(x^A_0,{}^-\pi^0_A)$ are a maximal set of independent canonical data that commute with all pre--constraints at $n=0$. They are therefore pre--observables. Similarly, $(x^B_1,{}^+\pi^1_B)$ are a maximally independent set of data that commute with all post--constraints at $n=1$ and thus post--observables. Hence, a complete set of propagating pre-- and post--observables for the move $0\rightarrow1$ reads
\ba
(x^A_0,{}^-\pi^0_A)\,,\q\q A=r,\rho,\gamma,z\q\q\overset{\tiny{\mathfrak{H}_0}}{\longrightarrow} \q\q(x^B_1,{}^+\pi^1_B)\,,\q\q B=l,\lambda,\gamma,z\,.\label{propobs1}
\ea

On the other hand, clearly, the $x^I_1,x^H_1,x^r_1,x^\rho_1$ are {\it a priori} free variables that cannot be predicted via $\mathfrak{H}_0$ by the initial data $x_0,p^0$ at $n=0$. Their conjugate momenta are simply the post--constraints ${}^+C^1_R$ in (\ref{newpcons}). Likewise, the $x^I_0,x^H_0,x^l_0,x^\lambda_0$ are {\it a posteriori} free variables of the time evolution map $\mathfrak{H}_0$ which cannot be postdicted by the canonical data $x_1,p^1$ at $n=1$. Their conjugate momenta are just the pre--constraints ${}^-C^0_L$ as given in (\ref{newpcons}).

In complete analogy, one finds that the set $(x^A_1,{}^-\pi^1_A)$ is a complete set of pre--observables for the move $1\rightarrow2$ which under $\mathfrak{H}_1$ propagates into the post--observables $(x^B_2,{}^+\pi^2_B)$.

\begin{Example}
\emph{In our scalar field example the move $0\rightarrow1$ is totally constrained, such that all 12 $\{\phi_1\}$ are \emph{a priori} free, all 12 $\{\phi_0\}$ are \emph{a posteriori} free and no observables propagate between $n=0$ (i.e., `nothing') and $n=1$. We therefore focus on $1\rightarrow2$.}

\emph{We can choose $(T_1)_\Gamma{}^i=\delta^i_\Gamma$. This yields (\ref{newcantrans}) for the pre--momenta in the form
\ba
{}^-\pi^1_i&:=&{}^-p^1_i+\left(6+\f{3}{2}m^2\right)\phi^i_1-\phi^{i-1}_1-\phi^{i+1}_1,\q\q i=1,\ldots,4,\label{ex1}\\
 {}^-\pi^1_j&:=&{}^-p^1_j=0,\q\q\q\q\q\q\q\q\q\q\q\q\q\,\,\,\, j=5,\ldots,12.\label{ex1b}
 \ea
At step $n=2$, we use (\ref{R2}) to choose
\ba\label{T2}
(T_2)_R{}^i&=&(R_2)_R{}^i,\q\q R=1,\ldots,8,\q\q\text{and}\nn\\
 (T_2)_9{}^i&=&\delta_{i9},\q (T_2)_{10}{}^i=\delta_{i7},\q (T_2)_{11}{}^i=\delta_{i11},\q(T_2)_{12}{}^i=\delta_{i12}.
\ea
This gives (\ref{newcantrans}) for the post--momenta at $n=2$, 
\ba
{}^+\pi^2_R&\equiv&{}^+C^2_R=0,\q\q R=1,\ldots,8,\\
{}^+\pi^2_9&=&{}^+p^2_{9}-\left(4+m^2\right)\phi_2^9+\phi_2^2+\phi_2^{10},\nn\\
{}^+\pi^2_{10}&=&{}^+p^2_{7}-\left(4+m^2\right)\phi_2^7+\phi_2^1+\phi_2^8,\\
{}^+\pi^2_{11}&=&{}^+p^2_{11}-\left(4+m^2\right)\phi_2^{11}+\phi_2^3+\phi_2^{12},\nn\\
{}^+\pi^2_{12}&=&{}^+p^2_{12}-\left(4+m^2\right)\phi_2^{12}+\phi_2^{4}+\phi_2^{11},\nn
\ea
where the ${}^+C^2_R$ are shown in (\ref{postconsn2}), such that (\ref{h0}) takes the simple form
\ba\label{h1}
{}^-\pi^1_1&=&2\Phi_2^{12},\q\,\,\,\,{}^-\pi^1_2=2\Phi_2^{11},\q\,\,\,{}^-\pi^1_3=2\Phi_2^{10},\q\,\,\,\,{}^-\pi^1_4=2\Phi_2^{9},\nn\\
{}^+\pi^2_9&=&-2\phi_1^4,\q{}^+\pi^2_{10}=-2\phi_1^3,\q{}^+\pi^2_{11}=-2\phi_1^2,\q{}^+\pi^2_{12}=-2\phi_1^1,
\ea
where $\Phi_2^\Gamma=((T^{-1}_2)^T)^\Gamma{}_i\phi_2^i$, following (\ref{lindecomp}). Note that (\ref{h1}) defines a canonical transformation.}

\emph{The four canonical pairs $(\phi_1^i,{}^-\pi^1_i)$, $i=1,\ldots,4$, are a complete set of pre--observables which Poisson-commute with all eight pre--constraints (\ref{exp-precon}) at $n=1$ and propagate under $\mathfrak{H}_1$ to $n=2$ to uniquely determine the four canonical pairs $(\Phi_2^B,{}^+\pi^2_B)$, $B=9,\ldots,12$. Likewise, the latter are a complete set of post--observables that Poisson-commute with all eight post--constraints ${}^+C^2_R$, $R=1,\ldots,8$, at $n=2$. Finally, the spurious $\{\phi^j_1\}_{j=5}^{12}$ are both \emph{a priori} and \emph{a posteriori} free (gauge) parameters, while the $\{\Phi_2^R\}_{R=1}^8$ are \emph{a priori} free parameters. In light of our earlier discussion, the $(\Phi_2^B,{}^+\pi^2_B)$ represent predictable `large scale' degrees of freedom of the growing lattice, while the $\{\Phi_2^R\}_{R=1}^8$ are unpredictable `smaller scale' degrees of freedom.    }
\end{Example}

\subsection{The reduced phase space}\label{sec_redps}

Next, let us ask the questions: `what are the gauge modes at $n=1$ and what are the observables that propagate from $0$ {\it through} $1$ to $2$?' We therefore consider the dynamics here as an initial value problem. As explained in \cite{Dittrich:2013jaa}, in order to answer these questions, we need to consider the matching of the symplectic structures and the reduced phase space at $n=1$. This will require some details in the present case.

For notational simplicity, let us define
\ba
h^{12}_{\Gamma\Gamma'}:=(T_1)_\Gamma{}^i h^{12}_{ij}(T_1^T)^j{}_{\Gamma'}\,,\q\q\q c^n_{\Gamma\Gamma'}:=(T_{n-1})_\Gamma{}^i c^n_{ij}(T_n^T)^j{}_{\Gamma'}\,.\nn
\ea
One easily checks that momentum matching, ${}^-p^1={}^+p^1$, implies the following relation for the new canonical variables as given in (\ref{newcantrans}):
\ba
{}^-\pi^1_\Gamma={}^+\pi^1_\Gamma+h^{12}_{\Gamma\alpha}x^\alpha_1\,,\q\q\q\q \alpha=H,\lambda,\rho,\gamma\,,\label{prepost}
\ea
where we collect the indices of all non-null vectors of $h^{12}$ in the common index $\alpha$. Let us now characterize the different canonical pairs according to the eight types:
\\~
\\~
{\bf (1)} A necessary condition for gauge modes is that they are both unpre-- and unpostdictable such that they must be both {\it a priori} and {\it a posteriori} free. The only variables at $n=1$ fulfilling this condition {\it before} momentum matching are $x^I_1,x^H_1$. 
\begin{itemize}
\item[{\bf (A)}] Theorem 5.3 in \cite{Dittrich:2013jaa} implies that the conjugate variable to a constraint which is both a pre-- and post--constraint is a gauge mode which will never appear in any equation of motion (this is case (a) of section \ref{sec_constraintclass}). The $x^I_1$ are thus genuine gauge modes.
\item[{\bf (B)}] Consider $x^H_1$. Note that on--shell we now have the holonomic constraints (\ref{holcon1})
\ba
h^{12}_{H\alpha} x^\alpha_1=0\,.\label{Hcon}
\ea
(This also follows from (\ref{prepost}) and noting that ${}^-\pi^1_H={}^+\pi^1_H=0$ are both constraints.) The square matrix $h^{12}_{HH'}$ is generally invertible (otherwise at least a pair of ${}^+C^1_H,{}^-C^1_H$ commute with each other, see the discussion below (\ref{holcon1})). Assuming invertibility of $h^{12}_{HH'}$ and denoting the inverse by $h_{12}^{HH'}$, one can solve (\ref{Hcon}) for $x^H_1$, 
\ba
x^H_1=-h_{12}^{HH'}h^{12}_{H'\tilde{\alpha}}x^{\tilde{\alpha}}_1\,,\q\q\q\tilde{\alpha}=\lambda,\rho,\gamma\,.\label{xH}
\ea
Hence, the $x^H_1$ are neither propagating degrees of freedom, nor free gauge modes. These modes are {\it a priori} free parameters of the ${}^+C^1_H$ at $n=1$, however, get fixed by the pre--constraints ${}^-C^1_H$ which render the ${}^+C^1_H$ second class. Nevertheless, these modes do not propagate because they are {\it a posteriori} free variables of the map $\mathfrak{H}_1$.\footnote{The $x^H_1$ are therefore an example of the special situation discussed for case (c) in section V.D.2 in \cite{Dittrich:2013jaa} in which a variable that is both {\it a priori} and {\it a posteriori} free gets fixed, yet does not propagate.}
\end{itemize}
Therefore, only the $x^I_1$ are genuine gauge modes that always remain free. %We shall see shortly that on--shell also the $x^l_1,x^r_1$ are {\it a priori} and {\it a posteriori} free parameters corresponding to symmetries of $S_1+S_2$. However, in section \ref{sec_classeff} we will explain that the latter---in contrast to $x^I_1$---do not in general correspond to symmetries of effective actions (or Hamilton's principle functions) involving larger numbers of steps and are thus not genuine gauge modes. %In fact, they turn out to become Lagrange multipliers of the holonomic constraints.
\\~
\\~
{\bf (2)} Both $x^l_1,x^\lambda_1$ are pre--observables that propagated from $n=0$ via $\mathfrak{H}_0$ to $n=1$ (see (\ref{propobs1})). However, they are {\it a posteriori} free parameters for the time evolution map $\mathfrak{H}_1$ and thus will not continue to propagate to $n=2$.
\begin{itemize}
\item[{\bf (A)}] Matching the symplectic structures at $n=1$ leads to non--trivial conditions. In fact, the pre--constraints ${}^-C^1_l$ are necessarily first class at $n=1$ (see table \ref{conal} and note that we are only considering the evolution $0\rightarrow1\rightarrow2$ and so holonomic constraints $H^1_l,H^1_r$ do not arise at $n=1$). All $N^1_{l}$ pre--constraints ${}^-C^1_l$, by imposing momentum matching (\ref{eom}), propagate back to $n=0$ and arise there in the form of the $N^1_{l}$ secondary holonomic constraints (\ref{holcon2}), which now read
\ba
H^0_l=x^A_0c^1_{Al}=0\,,\q\q A=r,\rho,\gamma,z\,.\label{hcon2}
\ea
Note that (\ref{h0}, \ref{prepost}) then implies ${}^+C^1_l:={}^+\pi^1_l\equiv{}^-C^1_l={}^-\pi^1_l=0$. %, such that on solutions to the equations of motion at $n=1$ the $x^l_1$ no longer are propagating degrees of freedom---even for $0\rightarrow1$. 
%In fact, on--shell the $x^l_1$ are now both {\it a priori} and {\it a posteriori} free in agreement with the fact that 
The type (2)(A) constraints ${}^-C^1_l$ are symmetry generators of the three--step action $S_1+S_2$ on--shell. Correspondingly, the $N^1_{l}$ holonomic constraints (\ref{hcon2}) commute with all pre--constraints at $n=0$ (see table \ref{conal}) such that $2N^1_{l}$ propagating phase space observables among the $(x^A_0,{}^-\pi^0_A)$ are eliminated for the `effective' move $0\rightarrow2$. However, the $x^l_1$ are not genuine gauge degrees of freedom: the $x^l_1$ are \underline{not} {\it a priori free}: using (\ref{h0}) the $x^l_1$ can still be predicted in a canonical initial value problem. By contrast, in a configuration boundary value problem, the $x^l_1$ cannot be determined by the configuration data at $n=0,2$ because they correspond to null vectors of the Hessian. We shall see in section \ref{sec_unique} that the {\it a posteriori free} $x^l_1$ conjugate to ${}^-C^1_l$ and originally propagating in the move $0\rightarrow1$ becomes on-shell a Lagrange multiplier of (\ref{hcon2}) in the effective action associated to the `effective' move $0\rightarrow2$. The ${}^-C^1_l$ can thereby also be interpreted as {\it coarse graining} or {\it lattice shrinking} conditions. 
\item[{\bf (B)}] A pre--constraint ${}^-C^1_\lambda={}^-\pi^1_\lambda=0$ either remains a first class {\it coarse graining} or {\it lattice shrinking} condition which restricts the space of solutions (case (b1) of section \ref{sec_constraintclass}) or it becomes second class and fixes {\it a priori} free variables (case (c) of section \ref{sec_constraintclass}). In particular, in the latter case, ${}^-C^1_\lambda$ fixes the flows of a post--constraint ${}^+C^1_\rho$ (see table \ref{conal}) and thereby the conjugate $x^\rho_1$. Namely, via (\ref{prepost}), ${}^-C^1_\lambda$ translates into
\ba
{}^+\pi^1_\lambda+h^{12}_{\lambda\alpha}x^\alpha_1=0\,,\q\q \alpha=H,\lambda,\rho,\gamma\,.\label{lapcon}
\ea
Recall that ${}^+\pi^1_\lambda,x^\lambda_1,x^\gamma_1$ are among the post--observables (\ref{propobs1}) that can be predicted by the data at $n=0$, while $x^H_1$ can (generally) be determined via (\ref{xH}). Hence, (\ref{lapcon}) constitute $N^1_{\lambda}$ equations for determining the $N^1_{\rho}$ unknown---under $1\rightarrow2$ propagating---$x^\rho_1$ (or combinations thereof) as functions of propagating data of the move $0\rightarrow1$. Notice that $\{{}^-C^1_\lambda,{}^+C^1_\rho\}=h^{12}_{\lambda\rho}$. Let $\rm{Rank}(h^{12}_{\lambda\rho})=m^1_{\lambda\rho}\leq N^1_\lambda$. Accordingly, for each of the $m^1_{\lambda\rho}$ pairs of ${}^+C^1_\rho,{}^-C^1_\lambda$ that does not commute, one {\it a priori} free $x^\rho_1$ becomes fixed (or predicted) via (\ref{lapcon}) (while all {\it a posteriori} free $x^\lambda_1$ are determined via (\ref{h0})). The corresponding $(x^\lambda_1,{}^+\pi^1_\lambda)$ are a canonical pair of post--observables that propagate from $n=0$ to $n=1$, but not further to $n=2$. Nevertheless, the fixing of $x^\rho_1,x^\lambda_1$ at $n=1$ transfers the propagating data to a new pair of pre--observables $(x^\rho_1,{}^-\pi^1_\rho)$ that continue to propagate to $n=2$ (see also the discussion for type (3)(B) below). The remaining $(N_\lambda^1-m^1_{\lambda\rho})$ pre--constraints ${}^-C^1_\lambda$ remain first class and of case (b1).
\end{itemize}
{\bf (3)} Both $x^r_1,x^\rho_1$ are {\it a priori} free data for the map $\mathfrak{H}_0$ which, however, propagate under $1\rightarrow2$ via the map $\mathfrak{H}_1$. 
\begin{itemize}
\item[{\bf (A)}] This case is essentially the time reverse of (2)(A) above. Namely, the $N^1_{r}$ constraints ${}^+C^1_r$ are first class and the $N^1_{r}$ secondary holonomic constraints $H^2_r$ at $n=2$ now read
\ba
H^2_r=c^2_{rB}x^B_2=0\,,\q\q B=l,\lambda,\gamma,z\,,\label{hcon3}
\ea
and imply the secondary ${}^-C^1_r:={}^-\pi^1_r\equiv{}^+C^1_r={}^+\pi^1_r=0$. %Consequently, on--shell $x^r_1$ are both {\it a priori} and {\it a posteriori} free and thus are no longer propagating degrees of freedom even for $1\rightarrow2$. Indeed, o
On--shell type (3)(A) constraints ${}^+C^1_r$ are symmetry generators of the three--step action $S_1+S_2$. However, the conjugate $x^r_1$ are not gauge modes: they are \underline{not} {\it a posteriori free} and can thus be postdicted, using $\mathfrak{H}_1$. The $x^r_1$ are only undeterminable when considered in a configuration boundary value problem. In section \ref{sec_unique} we shall see that the $x^r_1$ become Lagrange multipliers of the holonomic constraints (\ref{hcon3}) in the effective action of the move $0\rightarrow2$. The ${}^+C^1_r$ can be interpreted as {\it refining} or {\it lattice growing} conditions.
\item[{\bf (B)}] The $x^\rho_1,{}^-\pi^1_\rho$ are canonical pre--observable pairs that propagate under $\mathfrak{H}_1$ from $n=1$ to $n=2$. The question is: `how many of these can be predicted by data at $n=0$?' The answer was given above: if there are $m^1_{\lambda\rho}\leq N^1_{\lambda}$ non--commuting pairs ${}^+C^1_\rho,{}^-C^1_\lambda$, $m^1_{\lambda\rho}$ of the $x^\rho_1$ can be determined via (\ref{lapcon}) as functions of initial data at $n=0$. Noting that ${}^+\pi^1_\rho=0$, this can be employed in
\ba
{}^-\pi^1_\rho=h^{12}_{\rho\alpha}x^\alpha_1\,,\q\q\q\q \alpha=H,\lambda,\rho,\gamma\,, \label{rhoobs}
\ea
in order to determine $m^1_{\lambda\rho}$ propagating momentum observables. Assume that (\ref{Hcon}) can be solved for all of the $x^H_1$. If $m^1_{\lambda\rho}=N^1_{\rho}$, we are done and (\ref{rhoobs}) already determines all ${}^-\pi^1_\rho$ as functions of variables that can be predicted by the canonical data at $n=0$. If, on the other hand, $m^1_{\lambda\rho}<N^1_{\rho}$, not all ${}^-\pi^1_\rho$ are uniquely determined. However, in this case we can cheat a little bit. Using (\ref{xH}), (\ref{rhoobs}) becomes
\ba
{}^-\pi^1_\rho=\tilde{h}^{12}_{\rho\tilde{\alpha}}x^{\tilde{\alpha}}_1\,,\q\q\tilde{\alpha}=\lambda,\rho,\gamma\,,\nn
\ea
where we have defined the new (`effective') Hessian on solutions to (\ref{Hcon})
\ba
\tilde{h}^{12}_{\rho\tilde{\alpha}}=h^{12}_{\rho\tilde{\alpha}}-h^{12}_{\rho H}h^{HH'}_{12}h^{12}_{H'\tilde{\alpha}}\,.\nn
\ea
Now redefine ${}^-\tilde{\pi}^1_\rho:={}^-\pi^1_\rho-\tilde{h}^{12}_{\rho{\rho}'}x^{\rho'}_1-\tilde{h}^{12}_{\rho\gamma}x^{\gamma}_1$ such that the new ${}^-\tilde{\pi}^1_{{\rho}}$ are combinations of data that propagate under $\mathfrak{H}_1$ from $n=1$ to $n=2$. It is clear that ${}^-\pi^1_\rho\rightarrow{}^-\tilde{\pi}^1_\rho$ defines a canonical transformation if we perform a similar transformation ${}^-\pi^1_\gamma\rightarrow{}^-\tilde{\pi}^1_\gamma$ below for type (5). (\ref{rhoobs}) then reads
\ba
{}^-\tilde{\pi}^1_\rho:={}^-\pi^1_\rho-\tilde{h}^{12}_{\rho{\rho}'}x^{\rho'}_1-\tilde{h}^{12}_{\rho\gamma}x^{\gamma}_1=\tilde{h}^{12}_{\rho\lambda}x^{\lambda}_1\,, \label{rhoobs2}
\ea
such that the new ${}^-\tilde{\pi}^1_{{\rho}}$ can be purely written in terms of data that can be predicted by the initial data at $n=0$ and still Poisson commute with all pre--constraints at $n=1$ because ${}^-\pi^1_\rho$ and $x^\rho_1,x^\gamma_1$ commuted with all pre--constraints (see (\ref{xcommute})). Label by $\tilde{\rho}$ those $m^1_{\lambda\rho}$ of the $x^\rho_1$ that can be determined via (\ref{lapcon}). The $m^1_{\lambda\rho}$ canonical pre--observable pairs that can be written purely in terms of canonical data of step $n=0$ and which continue to propagate to $n=2$ are therefore the $(x^{\tilde{\rho}}_1,{}^-\tilde{\pi}^1_{\tilde{\rho}})$.
\end{itemize}
{\bf (4)} The $(x^z_1,{}^+\pi^1_z)$ are canonical observable pairs that propagated under $\mathfrak{H}_0$ from $n=0$ to $n=1$ (see (\ref{propobs1})). Notice that (\ref{prepost}) implies ${}^+\pi^1_z={}^-\pi^1_z$ (recall $(Z_1)_z\cdot h^{12}=0$) and so the $(x^z_1,{}^+\pi^1_z)$ also continue to propagate via $\mathfrak{H}_1$ to $n=2$. This is embodied in the secondary boundary data constraints $B_z^{02}$ in (\ref{2stepcon}) which are equivalent to ${}^+\pi^1_z={}^-\pi^1_z$: given initial configuration data at $n=0$ one can, using (\ref{2stepcon}), predict $N^1_z$ configuration data at $n=2$. Indeed, $(x^z_1,{}^+\pi^1_z)$ Poisson commute with all constraints at $n=1$. 

However, viewing the situation as a boundary value problem, i.e., given the configuration data at steps $n=0,2$ and attempting to determine the data at $n=1$, one will not be able to determine the $x^z_1$ because of the $B^{02}_z$ which render the boundary data interdependent. Indeed, we shall see in section \ref{sec_unique} that the $x^z_1$ become the Lagrange multipliers of the $B_z^{02}$ in the effective action of the move $0\rightarrow2$. In a boundary value problem, these Lagrange multipliers cannot be determined.

{The fact that the $(x^z_1,{}^+\pi^1_z)$ are propagating at all for an initial value problem may seem surprising at first sight in light of the fact that $(Z_1)_z$ are null vectors of the Hessian and thus define symmetries of the three--step action $S_1+S_2$, see (\ref{invac}). It should be noted, however, that the conclusion that $(Z_1)_z$ are symmetries of the action $S_1+S_2$ involves {only} the {\it configuration} data of the three steps $n=0,1,2$. On the other hand, the determination of $x^z_1$ in (\ref{h0}) via an initial value problem involves the {\it canonical} data at $n=0$, i.e.\ more information than just the configuration data. In fact, if one included earlier steps $n=-1,-2$ into the evolution, by momentum matching, the momenta ${}^-\pi^0_A$ in (\ref{h0}) which determine the $x^z_1$ would contain information about configuration data at $n<0$. From this we can already infer that for moves or spacetime regions involving larger numbers of evolution steps, the $(Z_1)_z$, in fact, will generally no longer be null vectors of `effective' Hessians and thus no longer define symmetries of `effective' actions. We will confirm this in section \ref{sec_classeff}.}%That is, the $(x^z_0,{}^-\pi^0_z)$ are observables that propagate directly through to $(x^z_2,{}^+\pi^2_z)$.
\\~
\\~
{\bf (5)} The canonical pairs $(x^\gamma_1,{}^+\pi^1_\gamma)$ propagate under $\mathfrak{H}_0$ to $n=1$ (see (\ref{propobs1})). Likewise, $(x^\gamma_1,{}^-\pi^1_\gamma)$ propagate under $\mathfrak{H}_1$ to $n=2$. ($x^\gamma_1$ commutes with all constraints at $n=1$.) In the present case, the pre-- and post--observable momenta at $n=1$ are related by 
\ba
{}^-\pi^1_\gamma={}^+\pi^1_\gamma+h^{12}_{\gamma\alpha}x^\alpha_1\,,\q\q \alpha=H,\lambda,\rho,\gamma\,.\nn
\ea
Since in the general case not all of the $x^\rho_1$ on the right hand side can be determined via (\ref{lapcon}), we can, again, cheat a little bit as in (\ref{rhoobs2}). Redefine ${}^-\tilde{\pi}^1_\gamma:={}^-\pi^1_\gamma-\tilde{h}^{12}_{\gamma\rho}x^\rho_1-\tilde{h}^{12}_{\gamma\gamma'}x^{\gamma'}_1$, such that the new ${}^-\tilde{\pi}^1_\gamma$ are combinations of data that propagate under $\mathfrak{H}_1$ from $n=1$ to $n=2$. Notice that ${}^-\pi^1_\gamma\rightarrow{}^-\tilde{\pi}^1_\gamma$ in combination with ${}^-\pi^1_\rho\rightarrow{}^-\tilde{\pi}^1_\rho$ for type (3)(B) above defines a canonical transformation. Moreover, the new ${}^-\tilde{\pi}^1_\gamma$---just like the old ${}^-\pi^1_\gamma$---commute with all pre--constraints at $n=1$ and we now have
\ba
{}^-\tilde{\pi}^1_\gamma={}^+\pi^1_\gamma+\tilde{h}^{12}_{\gamma\lambda}x^{\lambda}_1\,,\label{gamobs}
\ea
so that all ${}^-\tilde{\pi}^1_\gamma$ can be determined entirely by data that propagated from $n=0$ to $n=1$. Consequently, the canonical post--observable pairs $(x^\gamma_1,{}^+\pi^1_\gamma)$ propagate under $\mathfrak{H}_0$ to $n=1$. At $n=1$, these data transfer via (\ref{gamobs}) to a new set of canonical pre--observable pairs $(x^\gamma_1,{}^-\tilde{\pi}^1_\gamma)$ that continue to propagate under $\mathfrak{H}_1$ to $n=2$ and no propagating data of type (5) are lost.
\\

Combining all of the above, there are $N_{0\rightarrow1\rightarrow2}=2N^1_{\gamma}+2N^1_{z}+2m^1_{\lambda\rho}$ phase space observables that propagate from $n=0$ {\it through} $n=1$ to $n=2$. This number coincides with the reduced phase space dimension at $n=1$, 
\ba
N_{0\rightarrow1\rightarrow2}\hspace{-.2cm}&=&2Q-2\#(\text{1st class constraints at $n=1$})-\#(\text{2nd class constraints at $n=1$})\nn\\
&=&2Q-2\left(N^1_{I}+N^1_{l}+N^1_{r}+(N^1_{\lambda}+N^1_{\rho}-2m^1_{\lambda\rho})\right)-\left(2N^1_{H}+2m^1_{\lambda\rho}\right)\nn\\
&=&2N^1_{\gamma}+2N^1_{z}+2m^1_{\lambda\rho}\,,\nn
\ea
thus confirming the general discussion of \cite{Dittrich:2013jaa}. The reduced phase space at a given step $n$ is dependent on the evolution move $i\rightarrow n$ leading from an initial step $i$ to $n$ and the evolution move $n\rightarrow f$ leading from $n$ to the final step $f$. It corresponds to the canonical data propagating from $i$ {\it via} $n$ to $f$. Hence, an empty reduced phase space at $n$ does {\it not} imply that there are no propagating degrees of freedom. It only implies that nothing propagates {\it through} $n$. But there may still be propagation from $i$ to $n$ or from $n$ to $f$ separately.

\begin{Example}
\emph{We return to the example of the scalar field on the expanding square lattice. Clearly, the type (1)(A) spurious degrees of freedom $\{\phi_1^j\}_{j=5}^{12}$, which were artificially introduced during the phase space extension, are non-propagating gauge modes. Since the move $0\rightarrow1$ is totally constrained, $N_{0\rightarrow1}=0$, where $N_{0\rightarrow1}$ denotes the number of propagating observables of the move $0\rightarrow1$, and it is also clear that the reduced phase space at $n=1$ of the evolution $0\rightarrow1\rightarrow2$ is empty. Hence, $N_{0\rightarrow1\rightarrow2}=0$. On the other hand, the \emph{a priori} free type (3)(B) degrees of freedom $\{\phi_1^i\}_{i=1}^{4}$ propagate under $\mathfrak{H}_1$ from $1$ to $2$ as described in (\ref{h1}), such that $N_{1\rightarrow2}=8$. }
\end{Example}

\section{`Effective actions' and uniqueness of the symplectic structure}\label{sec_unique}

%Before we move on and examine the behaviour of the degrees of freedom just classified under the inclusion of additional time steps, let us check under which circumstances the symplectic structures at an initial step $n_i$ and final step $n_f$ are unique in the following sense:\footnote{Two time steps are needed in order to define the symplectic structure via discrete Legendre transforms \cite{Dittrich:2013jaa}.} on solutions to intermediate time steps the momenta and constraints at $n_i$ and $n_f$ do {\it not} depend on the choice of evolution moves leading from $n_i$ to $n_f$. To this end, 

For later purpose we need the form of the `effective' action, i.e.\ Hamilton's principal function evaluated on solutions to intermediate steps. This will clarify the role of the $x^l_1,x^r_1,x^z_1$ further. As an aside we shall show that the momenta and constraints at $n=0,2$ on solutions to the equations of motion at $n=1$ do {\it not} depend on whether one
\begin{itemize}
\item[(i)] evolves canonically by two evolution moves $0\rightarrow1\rightarrow2$ from $n=0$ via $n=1$ to $n=2$, or 
\item[(ii)] integrates out the $x^i_1$ at $n=1$ first at the Lagrangian level and then considers the corresponding `effective' action as the action of a single (`effective') evolution move $0\rightarrow2$. 
\end{itemize}
(This is a special case of theorem 3.2 in \cite{Dittrich:2013jaa} which holds for arbitrary variational discrete systems.) %From this it subsequently follows that this holds for any evolution move and arbitrarily many discrete steps, such that the construction is unique. 
We begin by considering $0\rightarrow1\rightarrow2$ on--shell, i.e.\ after momentum matching at $n=1$.
\\~

(i) We solve (\ref{eom}) as a boundary value problem in the general case. Note that $h^{12}_{\alpha\alpha'}=(T_1)_\alpha{}^i h^{12}_{ij}(T_1^T)^j{}_{\alpha'}$, $\alpha=H,\lambda,\rho,\gamma$, is an invertible matrix. Denote its inverse by $h_{12}^{\alpha\alpha'}:=\left((h^{12})^{-1}\right)^{\alpha\alpha'}$. The equations of motion at $n=1$ (\ref{eom}) are solved as follows:
\ba\label{xalpha}
x^\alpha_1=-h_{12}^{\alpha\alpha'}(T_1)_{\alpha'}{}^i\left(c^1_{ji}x^j_0+c^2_{ij}x^j_2\right)\,.
\ea
Now substituting (\ref{lindecomp}) with $x^\alpha_1$ given by (\ref{xalpha}) into the defining equations for ${}^-p^0_i$ and ${}^+p^2_i$ in (\ref{mom1}, \ref{mom1b}) yields (the tilde signifies that momentum matching at $n=1$ has taken place)
\ba\label{mom2}
{}^-\tilde{p}^0_i&=&-\tilde{a}^{02}_{ij}x^j_0-\tilde{c}^{02}_{ij}x^j_2-c^1_{ij}\left((L_1)_l{}^jx^l_1+(Z_1)_z{}^jx^z_1\right)\,,\nn\\ {}^+\tilde{p}^2_i&=&\,\,\,\,\tilde{b}^{02}_{ij}x^j_2+\tilde{c}^{02}_{ji}x^j_0+c^2_{ji}\left((R_1)_r{}^jx^r_1+(Z_1)_z{}^jx^z_1\right)\,.
\ea 
%where 
%\ba\label{mom3}
%{}^-\tilde{p}^0_i=-\tilde{a}^{02}_{ij}x^j_0-\tilde{c}^{02}_{ij}x^j_2\,,\q\q\q
%{}^+\tilde{p}^2_i=\tilde{b}^{02}_{ij}x^j_2+\tilde{c}^{02}_{ji}x^j_0\,.
%\ea
%The reason for distinguishing between $\tilde{p}$ and $\tilde{p}'$ will become clear shortly. Observe that the differences between $\tilde{p}$ and $\tilde{p}'$ 
The terms on the right hand side involving step $n=1$ depend solely on the $x^l_1,x^r_1,x^z_1$ which, given boundary data at $n=0,2$, cannot be determined by (\ref{eom}). The `effective' coefficient matrices read
\ba\label{effcoeff}
\tilde{a}^{02}_{ij}&=&a^1_{ij}-h^{nm}_{12} c^1_{in}c^1_{jm}\,,\nn\\
\tilde{b}^{02}_{ij}&=&b^2_{ij}-h^{nm}_{12} c^2_{ni}c^2_{mj}\,,\nn\\
\tilde{c}^{02}_{ij}&=&-h^{nm}_{12} c^1_{in}c^2_{mj}\,,
\ea
and
\ba\label{hinv}
h^{nm}_{12}:=(T_1^T)^n{}_\alpha h_{12}^{\alpha\alpha'}(T_1)_{\alpha'}{}^m\,.
\ea
Note the ordering of the indices in the $c^n$ which preserves the structure of the coefficient matrices. In particular, $\tilde{a}^{02}_{ij}=\tilde{a}^{02}_{ji}$ and $\tilde{b}^{02}_{ij}=\tilde{b}^{02}_{ji}$ and, furthermore, if $(L_0)^ic^1_{ij}=0$ then $(L_0)^i\tilde{c}^{02}_{ij}=0$ and, likewise, if $c^2_{ij}(R_2)^j=0$, then $\tilde{c}^{02}_{ij}(R_2)^j=0$.
\\~

(ii) Now combine the two evolution moves $0\rightarrow1$ and $1\rightarrow2$ into a single move $0\rightarrow2$. To this end, we must firstly integrate out the internal variables $x_1$ in $S(x_0,x_1,x_2)$ as given in (\ref{2step}). The equations of motion for the $x_1$ are, of course, (\ref{eom}) with solution (\ref{xalpha}). Substituting this into (\ref{2step}), one obtains the `effective action',
\ba\label{effact}
\tilde{S}_{02}(x_0,x_2)\!\!\!&=&\!\!\!\frac{1}{2}\tilde{a}^{02}_{ij}x^i_0x^j_0+\frac{1}{2}\tilde{b}^{02}_{ij}x^i_2x^j_2+\tilde{c}^{02}_{ij}x^i_0x^j_2+c^1_{ij}x^i_0\left((L_1)_l{}^jx^l_1+(Z_1)_z{}^jx^z_1\right)\nn\\
&&+c^2_{ji}\left((R_1)_r{}^jx^r_1+(Z_1)_z{}^jx^z_1\right)x^i_2\\
&\underset{(\ref{holcon2},\ref{holcon3},\ref{2stepcon})}{=}&\!\!\!\frac{1}{2}\tilde{a}^{02}_{ij}x^i_0x^j_0+\frac{1}{2}\tilde{b}^{02}_{ij}x^i_2x^j_2+\tilde{c}^{02}_{ij}x^i_0x^j_2+x^l_1\,H^0_l+x^r_1\,H^2_r+x^z_1\,B^{02}_z\,,\nn
\ea
with the effective coefficients again given by (\ref{effcoeff}).\\~

Thus, the pre-- and post--momenta conjugate to $x^i_0$ and $x^i_2$ arising from this effective action directly coincide with the ${}^-\tilde{p}^0_i,{}^+\tilde{p}^2_j$ in (\ref{mom2}) of (i). %However, coincidence of (\ref{mom2}) and (\ref{mom3}) can be obtained by choosing the `Lagrange multipliers' $x^l_1,x^r_1,x^z_1$ suitably. 
In both cases the holonomic and boundary data constraints $H^0_l,H^2_r,B^{02}_z$ must still be solved. This would yield a different effective action which only depends on the remaining independent variables at steps $n=0,2$. However, we shall not carry this out explicitly as the final expression depends upon one's choice of which variables the holonomic and boundary data constraints should be solved for.

We also see that the type (2)(A), (3)(A) and (4) variables $x^l_1,x^r_1,x^z_1$ turn out to be the Lagrange multipliers of the type (2)(A), (3)(A) and (4) holonomic constraints $H^0_l,H^2_r,B^{02}_z$ in the effective action (\ref{effact}) of the effective move $0\rightarrow2$.

\begin{Example}
\emph{In the example of the scalar field on the expanding square one finds for the `effective' coefficient matrices (\ref{effcoeff}) $\tilde{a}^{02}_{ij}=\tilde{c}^{02}_{ij}\equiv0$, while $\tilde{b}^{02}_{ij}$ is a non-vanishing matrix whose explicit shape is not particularly illuminating such that we shall not reproduce it here. The effective action (\ref{effact}) (or Hamilton's principal function) for $0\rightarrow2$ is therefore a boundary term, depending only on data from $n=2$. }
\end{Example}

\section{Varying numbers of constraints}\label{sec_varcon}

The `effective' pre-- and post--constraints of the `effective' move $0\rightarrow2$ are determined by the degenerate directions of the `effective' Lagrangian two--form $\tilde{c}^{02}$. Denote the left and right null vectors of $\tilde{c}^{02}$ by $(\tilde{L}_0)^i$ and $(\tilde{R}_2)^i$ (for the moment, we suppress additional indices labeling the different types of null vectors), respectively, and contract (\ref{mom2}) with these null vectors in order to obtain the (`effective') pre-- and post--constraints,
\ba\label{effcons}
{}^-\tilde{C}^0&=&(\tilde{L}_0)^i\left(\tilde{p}^0_i+\tilde{a}^{02}_{ij}x^j_0+c^1_{ij}\left((L_1)_l{}^jx^l_1+(Z_1)_z{}^jx^z_1\right)\right)\,,\nn\\ {}^+\tilde{C}^2&=&(\tilde{R}_2)^i\left(\tilde{p}^2_i-\tilde{b}^{02}_{ij}x^j_2-c^2_{ji}\left((R_1)_r{}^jx^r_1+(Z_1)_z{}^jx^z_1\right)\right)\,.
\ea
The Lagrange multipliers $x^l_1,x^r_1,x^z_1$ are no longer dynamical for this problem and can be chosen arbitrarily. The expressions in (\ref{effcons}) are thus constraints. These have to be supplemented by the holonomic and boundary data constraints $H^0_l,H^2_r,B^{02}_z$ (if present). 

Note that solving the holonomic and boundary data constraints will generally change the structure of the `effective' constraints; in particular, after solving the boundary data constraints, some of the effective constraints in (\ref{effcons}) may no longer involve data from one time step only on account of the relations among data from $n=0$ and $n=2$. %The surviving set of pre-- and post--constraints would be the same as if one solved the holonomic and boundary data constraints in the action (\ref{effact}) and subsequently derived the pre-- and post--constraints corresponding to the ensuing effective action. 
Again, we shall not explicitly solve $H^0_l,H^2_r,B^{02}_z$ here because the solutions depend on one's choice of independent variables.

For those $(\tilde{L}_0)$ which are also left null vectors of $c^1$, i.e.\ which satisfy $(\tilde{L}_0)=(L_0)$, the ${}^-\tilde{C}^0$ in (\ref{effcons}) coincide with the ${}^-C^0$ in (\ref{kincon1}) since $(L_0)^i\tilde{a}^{02}_{ij}=(L_0)^ia^1_{ij}$ and $(L_0)^ic^1_{ij}=0$. Similarly, for those $(\tilde{R}_2)$ with $(\tilde{R}_2)=(R_2)$ one finds ${}^+\tilde{C}^2={}^+C^2$ as given in (\ref{kincon1}). (This conclusion is also not affected by solving the holonomic constraints since, by the analogue of table \ref{conal} for steps $n=0,2$, they commute with the previous pre--constraints at $n=0$ and previous post--constraints at $n=2$.) That is, the primary pre-- and post--constraints at $n=0,2$, respectively, have been preserved and are contained in the set of `effective' constraints (constraint preservation has been generally proven in \cite{Dittrich:2013jaa}). 
%\footnote{Note that for the left and right null vectors satisfying $(\tilde{L}_0)=(L_0)$ and $(\tilde{R}_2)=(R_2)$ one would have obtained the same `effective constraints' (\ref{effcons}) from using (\ref{mom2}) instead of (\ref{mom3}): the differences are projected out by the $(L_0)$ and $(R_2)$. These constraints do not depend on whether evolution (i) or (ii) has been chosen.}% but, in contrast to (\ref{kincon1}), require the equations of motion. 

However, in general, the set of `effective' constraints at $n=0,2$ will be {\it larger} than the original set at the same steps prior to momentum matching at $n=1$. Indeed, equations of motion at other steps may act as secondary constraints that can increase (but not decrease) the total number of constraints at a fixed $n$ \cite{Dittrich:2013jaa}. More precisely, if one extends the evolution, say, from $n_i$ via $n_f$ to $n_f'>n_f$, it is possible that the number of constraints at both $n_i$ and $n_f'$ is {\it larger} than at both $n_i$ and $n_f$ when only evolving between $n_i$ and $n_f$. For quadratic discrete actions it is not difficult to see how this arises. 

For instance, consider steps $n=0,2$ after integrating out internal variables at $n=1$, as performed in section \ref{sec_unique}. The number of pre--constraints at $n=0$ and of post--constraints at $n=2$ is now determined by the rank of $\tilde{c}^{02}=- c^1_{in}h^{nm}_{12}c^2_{mj}$ as given in (\ref{effcoeff}). We have seen above that left null vectors of $c^1$ and right null vectors of $c^2$ are also left and right null vectors of $\tilde{c}^{02}$, respectively. However, the number of left null vectors of $c^1$ need not coincide with the number of right null vectors of $c^2$ (one or both numbers could even be zero). Denote by $D_1,D_2,D_h,D_{02}$ the number of degenerate directions of $c^1,c^2,h_{12},\tilde{c}^{02}$, respectively. Obviously, $D_{02}\geq\max\{D_1,D_2,D_{h}\}$. This immediately implies that the numbers of pre-- and post--constraints at $n=0$ and $n=2$, respectively, can {\it only increase or remain the same} after integrating out the variables at the intermediate step $n=1$. But clearly, the number of left and right null vectors of $\tilde{c}^{02}$ coincide. 

We emphasize that, as a consequence of the rank of $c^n$ generally varying with $n$, the total number of both pre-- and post--constraints at some step $n$ need not coincide with the total number of constraints at some other step $n'$---even if equations of motion are imposed. 

The number of constraints at a given time step is thus highly move or spacetime region dependent. This is not surprising in a spacetime context since each region comes with its own action contribution and thus dynamics. It goes well in hand with the interpretation of discretization changing dynamics as refinement or coarse graining operations. These necessarily change the dynamics and lead to `constraint propagation'. We shall not further elaborate on this, as a detailed discussion of this topic can be found in \cite{Dittrich:2013jaa,Dittrich:2013xwa,Hoehn:2014fka,Hoehn:2014wwa}.

%Theorem \ref{thm_conalgeneral} also holds for `effective' constraints: indeed, since both matrices $\tilde{a}^{02},\tilde{b}^{02}$ are symmetric, the pre--constraints ${}^-\tilde{C}^0$ Poisson commute among themselves and, likewise, the post--constraints ${}^+\tilde{C}^2$ form an abelian Poisson (sub--)algebra. 

%As pointed out at the end of section \ref{sec_unique}, in the absence of types (2)(a), (3)(a) and (4), the present analysis directly extends to an arbitrary number of evolution steps because the effective action $\tilde{S}_{02}$ is in shape identical to $S_n$ as given in (\ref{Sk}). 

\begin{Example}
\emph{For the scalar field on the expanding square lattice it is easy to see how the equations of motion at $n=1$ (i.e., momentum matching ${}^+p^1={}^-p^1$) act as secondary constraints that restrict the dynamics of $1\rightarrow2$ and thereby of the `effective' evolution move $0\rightarrow2$. Since $\tilde{c}^{02}\equiv0$, $0\rightarrow2$ is totally constrained with 12 pre--constraints at $n=0$ and 12 post--constraints at $n=2$, although the move $1\rightarrow2$ was only subject to 8 pre--constraints (\ref{exp-precon}) at $n=1$ and 8 post--constraints (\ref{postconsn2}) at $n=2$. The four new post--constraints at $n=2$ arise because the matching of the symplectic structures at $n=1$ means the Hamiltonian time evolution map $\mathfrak{H}_1:T^*\cq_1\rightarrow T^*\cq_2$ defined by (\ref{premomsn1}, \ref{postmomsn2}) has to be further restricted by the four type (3)(B) post--constraints (\ref{exp-postcon1}) at $n=1$. The latter `propagate' under $\mathfrak{H}_1$ to $n=2$ because $\mathfrak{H}_1$ preserves the rank of the symplectic form restricted to the constraint surface (see theorem 6.1 in \cite{Dittrich:2011ke}). In fact, the type (3)(B) post--constraints (\ref{exp-postcon1}) are refining post--constraints of case (a) in section \ref{sec_constraintclass}. They propagate under $\mathfrak{H}_1$ as expansion consistency conditions to $n=2$ to ensure that the larger lattice at $n=2$ does not carry more dynamical information than the smaller lattices of the move $0\rightarrow1$ can support. Accordingly, nothing propagates from `nothing' at $n=0$ to $n=2$, i.e.\ $N_{0\rightarrow2}=N_{0\rightarrow1\rightarrow2}=0$, despite $N_{1\rightarrow2}=8$.} %This model serves as a specific example to the qualitative examples 5.1 and 5.2 in the companion paper \cite{Dittrich:2013jaa}.}
\end{Example}

\section{Classification and effective actions}\label{sec_classeff}

%\subsection{Boundary value problem}\label{sec_bvp}

%Given suitable boundary data at $k=0,2$, satisfying the two types of holonomic constraints (\ref{decompeom3}, \ref{decompeom5}) and equation (\ref{decompeom7}), one can always uniquely solve for the remaining $x^H_1,x^\lambda_1,x^\rho_1,x^\gamma_1$ via (\ref{decompeom2}, \ref{decompeom4}, \ref{decompeom6}, \ref{decompeom8}) ($h^{12}_{\alpha\alpha'}$ is invertible). Note, however, that these are not all independent. The holonomic constraints (\ref{decompeom2}) can be solved ($(Y_1)^i_Hh^{12}_{ij}(Y_1)^j_{H'}$ is an invertible matrix) for the $x^H_1$ as combinations of the $x^\lambda_1,x^\rho_1,x^\gamma_1$. The latter are, therefore, the independent gauge invariant observables for this BVP defined by $k=0,1,2$ only.% Indeed, the $x^\lambda_1,x^\rho_1,x^\gamma_1$ Poisson commute with the first class constraints $C^1_I$.

%\emph{one can establish that these indeed commute with the first class constraints. \marginpar{\bf!!!}not difficult to show, but maybe do later....otherwise also have to worry about observable momenta etc...}

Thus far, we have classified the null vectors, constraints and degrees of freedom for the three--step action $S_1+S_2$ (\ref{2step}), describing the evolution $0\rightarrow1\rightarrow2$. Let us now investigate the move dependence of this classification, i.e.\ how it is affected by inclusion of additional time steps and action contributions and, in particular, by solving the equations of motion at neighbouring time steps.

\begin{figure}[hbt!]
\begin{center}
\psfrag{0}{ $0$}
\psfrag{1}{$1$}
\psfrag{2}{$2$}
\psfrag{3}{ $3$}
\psfrag{4}{ $4$}
\psfrag{h4r}{$H^4_r$}
\psfrag{h2l}{$H^2_l$}
\psfrag{h2r}{$H^2_r$}
\psfrag{h0l}{$H^0_l$}
\psfrag{k}{$n$}
\psfrag{ht4r}{$\tilde{H}^4_r$}
\psfrag{ht0l}{$\tilde{H}^0_l$}
\psfrag{x1}{$x_1(x_2,x_0)$}
\psfrag{x3}{$x_3(x_4,x_2)$}
\psfrag{x2}{$x_2(x_4,x_0)$}

\subfigure[]{\label{bvpfig1}\includegraphics[scale=.6]{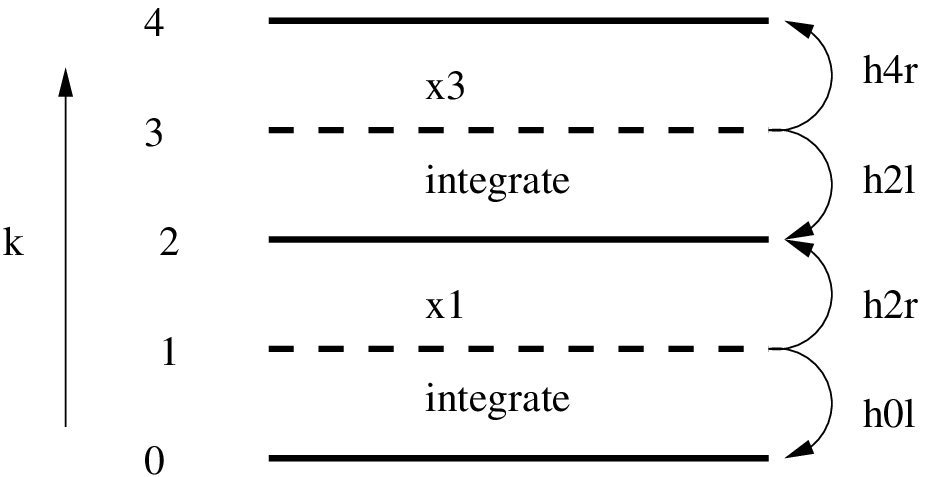}}\hspace*{1.4cm}
\subfigure[]{\label{bvpfig2}\includegraphics[scale=.6]{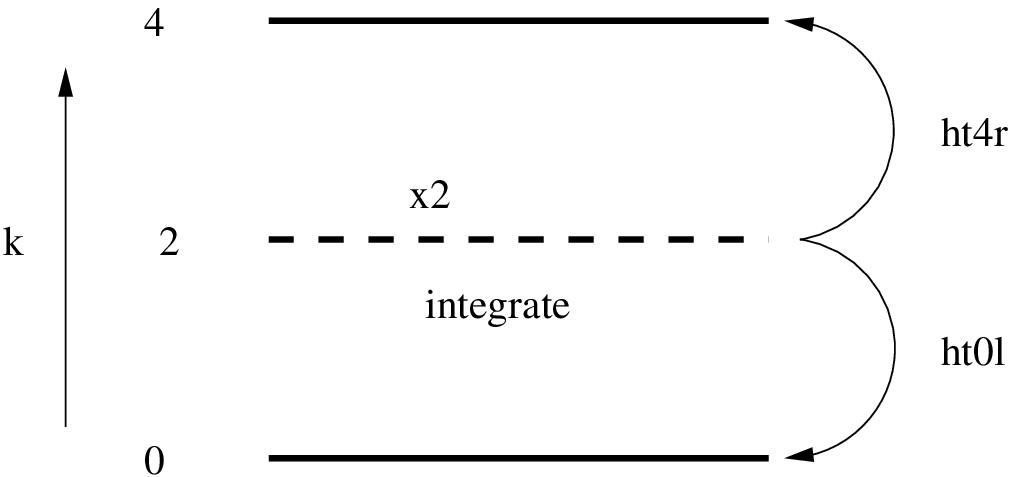}}
\caption{\small A boundary value problem involving five discrete steps. (a) Begin by solving the equations of motion at $n=1,3$ for $x_1(x_2,x_0)$ and $x_3(x_4,x_2)$. These equations of motion produce the holonomic constraints $H^0_l,H^2_r,H^2_l$ and $H^4_r$ at steps $n=0,2,4$, respectively. (b) Continue by solving for $x_2(x_4,x_0)$. The equations of motion at $n=2$ now involve effective actions and potentially produce new holonomic constraints at $n=0,4$. Notice that $H^0_l=(L_1)_{l}{}^ic^1_{ji}x^j_0$ and $\tilde{H}^0_{l'}=(\tilde{L}_2)_{l'}{}^i\tilde{c}^{02}_{ij}x^j_0$ are, in general, inequivalent. Analogously, $H^4_r$ and $\tilde{H}^4_{r'}$ are generally different.}\label{bvpfig}
\end{center}
\end{figure}

To this end, include steps $n=0,1,2$ in a larger boundary value problem also involving $n=3,4$ (see figure \ref{bvpfig}).
\noindent
Assume the first step in figure \ref{bvpfig1} has been carried out, that is, we now have $x_1(x_2,x_0)$ and the effective action (\ref{effact}). Consider step $n=2$ and the effective three--step action $\tilde{S}_{02}(x_0,x_2)+S_3(x_2,x_3)$, describing the evolution $0\rightarrow2\rightarrow3$. For the move $2\rightarrow3$ we still have the Lagrangian two--form $c^3$ such that the left null vectors at step $n=2$ are the same as before solving the equations of motion at $n=1$. However, we now also have the effective Lagrangian two--form $\tilde{c}^{02}=- c^1_{in}h^{nm}_{12}c^2_{mj}$ for the move $0\rightarrow2$. As just discussed in section \ref{sec_varcon}, all right null vectors of $c^2$ are still right null vectors of $\tilde{c}^{02}$, but there may now be additional right null vectors as a result of some of the equations of motion at $n=1$ acting as secondary constraint at $n=0,2$. 

Furthermore, using (\ref{effcoeff}), the effective Hessian at $n=2$ corresponding to the effective action $\tilde{S}_{02}(x_0,x_2)+S_3(x_2,x_3)$ is given by 
\ba\label{effhess}
\tilde{h}^{(02)3}_{ij}:=\tilde{b}^{02}_{ij}+a^3_{ij}=h^{23}_{ij}-h^{nm}_{12} c^2_{ni}c^2_{mj}\,,
\ea
such that generally
\ba\label{effhessdeg}
(L_2)_l{}^i\tilde{h}^{(02)3}_{ij}&=&-(L_2)_l{}^ih^{nm}_{12} c^2_{ni}c^2_{mj}\neq0\,,\nn\\
(Z_2)_z{}^i\tilde{h}^{(02)3}_{ij}&=&-(Z_2)_z{}^ih^{nm}_{12} c^2_{ni}c^2_{mj}\neq0\,,\\
(Y_2)_I{}^i\tilde{h}^{(02)3}_{ij}&=&0\,.\nn
\ea
That is, the $(L_2)_l,(Z_2)_z$ can fail to be degenerate directions of the effective Hessian $\tilde{h}^{(02)3}$ and, thus, by the analogue of (\ref{invac}), may also no longer define symmetries of the effective action $\tilde{S}_{02}(x_0,x_2)+S_3(x_2,x_3)$. 

As an example, consider the pre--constraints ${}^-C^2_l$ at step $n=2$. Before integrating out the variables at $n=1$, these constraints were first class and $(L_2)_l$ symmetries of the action $S_2+S_3$. Accordingly, the $x^l_2$ are initially {\it a posteriori} free parameters and do not appear in the equations of motion at $n=2$ arising from $S_2+S_3$ (just like the $x^l_1$ do {\it not} appear in (\ref{eom}) at $n=1$ because they are associated to null vectors of $h^{12}$). However, the $x^l_2$ {\it do} feature in the equations of motion (\ref{eom}) at $n=1$ arising from the action pieces $S_1+S_2$ because these type (2)(A) variables are not associated to right null vectors at $n=2$. Indeed, the equations of motion (\ref{eom}) at $n=1$ produce the additional holonomic constraint $H^2_r$ (\ref{holcon2}) at $n=2$ which can render ${}^-C^2_l$ (or, equivalently, ${}^-\tilde{C}^2_l$ in (\ref{effcons})) {\bf second class} (see table \ref{conal}). In addition, the $x^l_2$ also generally feature in the equations of motion at $n=2$ arising from the {\it effective} action $\tilde{S}_{02}(x_0,x_2)+S_3(x_2,x_3)$,
\ba
\tilde{h}^{(02)3}_{ij}x^j_2=-\tilde{c}^{02}_{ji}x^j_0-c^3_{ij}x^j_3\,,\nn
\ea
because by (\ref{effhessdeg}) the $(L_2)_l$ generally are not degenerate directions of $\tilde{h}^{(02)3}$. The $x^l_2$ are therefore not proper gauge modes. 

In complete analogy, after also performing the second step in figure \ref{bvpfig1}, i.e.\ after imposing the equations of motion at $n=3$ and solving for $x_3(x_4,x_2)$, the $(R_2)_r$ likewise generally no longer define degenerate directions of the (new) effective Hessian at $n=2$. The originally first class post--constraints ${}^+C^2_r$ can become {\bf second class} on account of the new holonomic constraints $H^2_l$ which arise at $n=2$ on--shell (see table \ref{conal}). At this stage, the situation is as in figure \ref{bvpfig2}. It should be noted that further integrating out the $x_2$ and solving for $x_2(x_4,x_0)$ may produce new holonomic constraints $\tilde{H}$ at $n=0,4$ which, in general, are independent of the previous holonomic constraints at $n=0,4$. 

In conclusion, on solutions to the equations of motion at $n=1,3$, the classification and numbers of the null vectors and, correspondingly of the different types of constraints and degrees of freedom, at $n=2$ generally changes. %Vectors $(Z_2)_z$ of type (4) generally become new vectors $(\tilde{V}_2)_{\tilde{\gamma}}$ of type (5), vectors $(L_2)_l$ of type (2)(a) generally become new vectors $(\tilde{L}_2)_{\tilde{\lambda}}$ of type (2)(b), vectors $(R_2)_r$ of type (3)(a) generally become new vectors $(\tilde{R}_2)_{\tilde{\rho}}$ of type (3)(b), but also vectors $(V_2)_\gamma$ of type (5) may become new null vectors of some type, and so on. 
The only types of vectors, constraints and degrees of freedom which remain unaffected by all the equations of motion are type (1)(A) and (B): (\ref{effhessdeg}) shows that the $(Y_2)_I$ remain degenerate directions of the effective Hessian and since these vectors are both left and right null vectors at $n=2$ they will also always remain left and right null vectors of any effective Lagrange two--forms at $n=2$. This is a consequence of theorems 5.2 and 5.3 in \cite{Dittrich:2013jaa} which imply, firstly, that the coinciding type (1)(A) pre-- and post--constraints $C^2_I={}^+C^2_I={}^-C^2_I$ at $n=2$ are always first class generators of symmetries of any `effective' action (involving step $n=2$ but otherwise arbitrary numbers of other time steps), and, secondly, that the $x^I_2$ are genuine gauge modes that never enter any equations of motion. On the other hand, type (1)(B) vectors $(Y_2)_H$ are both left and right null vectors at $n=2$ and must thus always remain left and right null vectors of any effective Lagrangian two--form at $n=2$. The corresponding second class constraint pairs ${}^+C^2_H,{}^-C^2_H$ must always remain second class and therefore the $(Y_2)_H$ can never additionally become null vectors of any effective Hessian. Thus, the $x^H_2$ remain fixed, yet non--propagating degrees of freedom and the classification as type (1)(B) is also preserved.

In the present section we have only considered the situation for a boundary value problem. But one can analogously demonstrate the same state of affairs for the situation of an initial value problem where one solves  for $x_2(x_1,x_0)$, $x_3(x_1,x_0)$,... . Also for an initial value problem the classification of the null vectors, constraints and degrees of freedom changes in the same way under the inclusion of additional time steps.

That is, the decomposition of the transformation matrix $(T_2)_\Gamma{}^i$ and the degrees of freedom (\ref{lindecomp}) is no longer suitable for effective actions and we have to choose a new decomposition $(\tilde{T}_2)_{\tilde{\Gamma}}{}^i$ on--shell---according to the prescription in section \ref{sec_classdof}. This is not a fundamental problem of the formalism. It merely reflects the fact that, as generally discussed in \cite{Dittrich:2013jaa}, the notion of observables as propagating degrees of freedom---and even of symmetries and the reduced phase space---depends on the time steps between which one is evolving. Consequently, also the classification must become evolution move dependent. Coarse graining/shrinking or refining/growing the discretization changes the dynamics non-trivially.

%One therefore has to expect that for generic discrete systems with evolving phase spaces, the classification of null vectors, constraints and degrees of freedom is step dependent. This is precisely what we see in the present section. 

\begin{Example}
\emph{Since $\tilde{c}^{02}\equiv0$, all 12 vectors in (\ref{T2}) are trivially right null vectors of it, while $(T_2)_9,\ldots,(T_2)_{12}$ were not right null vectors of $c^2$. However, it turns out that none of the 12 vectors in (\ref{T2}) are (left) null vectors at $n=2$ of $c^3$ or $\tilde{h}^{(02)3}$ of the move $2\rightarrow3$ in figure \ref{fig_exp}. Hence, these vectors become type (3)(B) on solutions to the equations of motion at $n=1$.}
\end{Example}

\section{Quantum theory}\label{sec_QT}

In order to quantize systems governed by quadratic discrete actions, we shall follow the general quantum formalism for global moves developed in \cite{Hoehn:2014fka} (see also \cite{Hoehn:2014wwa} for the quantum formalism for local evolution moves). This formalism employs the Dirac algorithm \cite{Dirac,Henneaux:1992ig} for quantizing constrained systems. In the sequel, we restrict the exposition to systems with flat Euclidean (extended) configuration spaces $\cq_n\simeq\mathbb{R}^{Q}$ and $x_n,p^n\in\mathbb{R}$. We emphasize that, as explained in section \ref{sec_quadact}, we work on {\it extended} configuration spaces $\cq_n$ if the underlying discretization is temporally varying. That is, the $\cq_n$ are the same at each time step $n$, despite a possibly temporally varying number of dynamically relevant degrees of freedom (the unextended configuration spaces may vary in discrete time). The subsequent discussion therefore encompasses, in particular, temporally varying discretizations. We shall employ the classification of constraints and degrees of freedom of the previous sections since it directly carries over to the quantum theory.

\subsection{Imposition of the constraints}\label{sec_impose} 
 
The pre-- and post--constraints at $n$ are promoted to self-adjoint operators on the kinematical Hilbert space $\ch^{\rm kin}_n=L^2(\mathbb{R}^{Q},dx_n)$. For instance, the quantum versions of the pre-- and post--constraints (\ref{kincon}) at $n=1$ read
\ba\label{qcons}
{}^+\hat{C}^1_R&=&(R_1)_R{}^i\left(\hat{p}^1_i-\widehat{\f{\p S_1}{\p x_1^i}}\right)=(R_1)_R{}^i\left(\hat{p}^1_i-b^1_{ij}\hat{x}^j_1\right),\q\q\q R=I,H,r,\rho\nn\\
{}^-\hat{C}^1_L&=&(L_1)_L{}^i\left(\hat{p}^1_i+\widehat{\f{\p S_2}{\p x_1^i}}\right)\,\,=(L_1)_L{}^i\left(\hat{p}^1_i+a^2_{ij}\hat{x}^j_1\right),\q\q\q L=I,H,l,\lambda.
\ea
Since there is only one kinematical Hilbert space $\ch^{\rm kin}_1$ at $n=1$ for both moves $0\rightarrow1$ and $1\rightarrow2$ we do not distinguish between pre-- and post--momenta at the operator level and just have $\hat{p}^1_i=-i\hbar\f{\p}{\p x^i_1}$ and $[\hat{x}^i_1,\hat{p}^1_j]=i\hbar\,\delta^i_j$ (see also \cite{Hoehn:2014fka} on this and how momentum matching translates into the quantum theory). The spectra of these quantum pre-- and post--constraints are absolutely continuous. The orbits $\cg^-_1$ associated to the pre--constraints and called {\it pre--orbits}, as well as $\cg^+_1$ associated to the post--constraints and called {\it post--orbits}, are non-compact.

Next, the quantum constraints (\ref{qcons}) must be implemented. At $n=1$ we distinguish between
\begin{itemize}
\item the {\it post--physical Hilbert space} ${}^+\ch^{\rm phys}_1$ containing the {\it post--physical states} ${}^+\psi^{\rm phys}_1$ which are annihilated by all ${}^+\hat{C}^1_R$, and
\item the {\it pre--physical Hilbert space} ${}^-\ch^{\rm phys}_1$ containing the {\it pre--physical states} ${}^-\phi^{\rm phys}_1$ which are annihilated by all ${}^-\hat{C}^1_L$. 
\end{itemize}
The post-- and pre--physical states are constructed from kinematical states $\psi^{\rm kin}_1\in\ch^{\rm kin}_1$ by means of an improper projection \cite{Hoehn:2014fka}
\ba
{}^+\psi^{\rm phys}_1={}^+\mathbb{P}_1\,\psi^{\rm kin}_1,\q\q\q\q{}^-\phi^{\rm phys}_1={}^-\mathbb{P}_1\,\phi^{\rm kin}_1.\label{proj}
\ea
The {\it post--} and {\it pre--projector}
\ba
{}^+\mathbb{P}_1:=\prod_R\delta({}^+\hat{C}^1_R),\q\q\q\q {}^-\mathbb{P}_1:=\prod_L\delta({}^-\hat{C}^1_L),
\ea
are group averaging projectors \cite{Marolf:1995cn,Marolf:2000iq,Thiemann:2007zz} with 
\ba
\delta(\hat{C})=\f{1}{2\pi\hbar}\int_\mathbb{R}ds\,e^{is\hat{C}/\hbar}\nn.
\ea
We note that, since the pre-- and post--constraints each form an abelian set, i.e.\ $[{}^-\hat{C}^1_L,{}^-\hat{C}^1_{L'}]=0=[{}^+\hat{C}^1_R,{}^+\hat{C}^1_{R'}]$, the pre-- and post--projectors are equivalent to 
\ba
{}^+\mathbb{P}_1&=&\f{1}{(2\pi\hbar)^{N^1_R}}\int_{\mathbb{R}^{N^1_R}}\prod_{R}\,ds_1^R\,e^{i/\hbar\sum_R\,s_1^R{}^+\hat{C}^1_R},\nn\\
{}^-\mathbb{P}_1&=&\f{1}{(2\pi\hbar)^{N^1_L}}\int_{\mathbb{R}^{N^1_L}}\prod_{L}\,ds_1^L\,e^{i/\hbar\sum_L\,s_1^L{}^-\hat{C}^1_L}
\ea
where $N^1_R=N^1_I+N^1_H+N^1_r+N^1_\rho$ and $N^1_L=N^1_I+N^1_H+N^1_l+N^1_\lambda$.

Consider the post--physical states. Since
\ba
{}^+\hat{C}^1_R\,\psi_1^{\rm kin}=(R_1)_R{}^i\,e^{iS_1/\hbar}\,\hat{p}^1_i\,e^{-iS_1/\hbar}\,\psi_1^{\rm kin}\underset{(\ref{lindecomp})}{=}e^{i\,b^1_{ij}x^i_1x^j_1/2\hbar}\,\hat{p}^1_R\,e^{-i\,b^1_{ij}x^i_1x^j_1/2\hbar}\,\psi_1^{\rm kin}\nn
\ea
%Hence,
%\ba
%\left(\hat{C}_I\right)^n=\left(e^{iS(\lambda^I,x^\alpha)/\hbar}\,\hat{p}_I\,e^{-iS(\lambda^I,x^\alpha)/\hbar}\right)^n=e^{iS(\lambda^I,x^\alpha)/\hbar}\,\left(\hat{p}_I\right)^n\,e^{-iS(\lambda^I,x^\alpha)/\hbar}\nn
%\ea
%and the (improper) projectors take the form (the spectra of the $\hat{C}_I$ are absolutely continuous)
%\ba
%\delta(\hat{C}_I)=\f{1}{2\pi\hbar}\int dt\,e^{is\hat{C}_I/\hbar}=\f{1}{2\pi\hbar}\int dt\,e^{iS/\hbar}\,e^{it\hat{p}_I/\hbar}\,e^{-iS/\hbar}.
%\ea
%Since the constraints are abelian, $[\hat{C}_I,\hat{C}_J]=0$, the different (improper) projectors $\delta(\hat{C}_I)$ commute and no factor ordering ambiguity in the definition of the physical states arises,
one finds, using the splitting of variables as in section \ref{sec_classdof}, for ${}^+\psi^{\rm phys}_1={}^+\mathbb{P}_1\,\psi^{\rm kin}_1$
\ba
{}^+\psi^{\rm phys}_1(x^R_1,x^B_1)&=&\f{1}{(2\pi\hbar)^{N^1_R}}\int_{\mathbb{R}^{N^1_R}} \prod_R\left(ds_1^R\,e^{i\,b^1_{ij}x^i_1x^j_1/2\hbar}\,e^{is^R_1\hat{p}^1_R/\hbar}\,e^{-i\,b^1_{ij}x^i_1x^j_1/2\hbar}\right)\psi^{\rm kin}_1(x^i_1)\nn\\
&=&\f{1}{(2\pi\hbar)^{N^1_R}}\int_{\mathbb{R}^{N^1_R}} e^{i\,b^1_{ij}x^i_1x^j_1/2\hbar}\prod_R\left(ds^R_1\,e^{is^R_1\hat{p}^1_R/\hbar}\right)\phi^{\rm kin}_1(x^R_1,x^B_1)\nn\\
&=&\f{1}{(2\pi\hbar)^{N^1_R}}\,e^{i\,b^1_{ij}x^i_1x^j_1/2\hbar}\int_{\mathbb{R}^{N^1_R}} \prod_Rds_1^R\,\phi^{\rm kin}_1(x^R_1+s^R_1,x^B_1)\nn\\
&=&e^{i\,b^1_{ij}x^i_1x^j_1/2\hbar}\,{}^+{\Psi}^{\rm phys}_1(x^B_1),\label{lemphys}
\ea
where ${}^+{\Psi}^{\rm phys}_1$ must be a square integrable function in the post--observables $x^B_1$ (see below). In the second line we made use of the definition $\phi^{\rm kin}_1:=e^{-i\,b^1_{ij}x^i_1x^j_1/2\hbar}\,\psi^{\rm kin}_1$. The post--physical state ${}^+\psi_1^{\rm phys}(x^R_1,x^B_1)$ therefore only depends on the {\it a priori free} $x^R_1$ through the factor $e^{i\,b^1_{ij}x^i_1x^j_1/2\hbar}$, where $b^1_{ij}x^i_1x^j_1=b^1_{RR'}x^R_1x^{R'}_1+b^1_{BB'}x^B_1x^{B'}_1+2b^1_{RB}x^R_1x^B_1$. 

(\ref{lemphys}) is clearly annihilated by the ${}^+\hat{C}^1_R$ in (\ref{qcons}). This shows that the post--projector ${}^+\mathbb{P}_1$ is improper: a double action of the post--projector ${}^+\mathbb{P}_1\cdot{}^+\mathbb{P}_1\,\psi^{\rm kin}_1={}^+\mathbb{P}_1{}^+\psi^{\rm phys}_1=``\infty"\cdot{}^+\psi^{\rm phys}_1$ yields a divergence because the second action leads to an integration of a constant function over the non-compact post--orbit $\cg^+_1$.

Similarly, one finds
\ba
{}^-\psi^{\rm phys}_1(x^L_1,x^A_1)=e^{-ia^2_{ij}x^i_1x^j_1/2\hbar}\,{}^-{\Psi}^{\rm phys}_1(x^A_1)\label{lemphys2}
\ea
with ${}^-{\Psi}^{\rm phys}_1$ a square integrable function in the pre--observables $x^A_1$ (see below). This state is evidently annihilated by ${}^-\hat{C}^1_L$ in (\ref{qcons}). Again, ${}^-\mathbb{P}_1$ is an improper projector and leads to a divergence if doubly employed.

Notice that the form of the solutions (\ref{lemphys}, \ref{lemphys2}) implies that a totally constrained move, as appearing, e.g., in a discrete version of the `no-boundary proposal' \cite{Hoehn:2014fka}, yields a unique pre-- and post--physical state. We shall see this in the scalar field example below.

Before moving on to the dynamics, we need to introduce the physical inner products for the two physical Hilbert spaces at $n=1$. We just summarize the results from \cite{Hoehn:2014fka}. In particular, from lemma 4.2 in \cite{Hoehn:2014fka}, which applies only to constraints linear in the momenta (and thus to the systems in this article), it follows that in the present case the {\it post--physical inner product} in ${}^+\ch^{\rm phys}_1$ is given by
\ba
\langle{}^+\psi^{\rm phys}_1|{}^+\xi^{\rm phys}_1\rangle_{\rm phys+}&=&\langle\psi^{\rm kin}_1|{}^+\mathbb{P}_1|\xi^{\rm kin}_1\rangle_{\ch^{\rm kin}_1}\nn\\
&=&\int_{\mathbb{R}^{N^1_B}}\prod_B\,dx^B_1({}^+\psi^{\rm phys}_1(x^R_1,x^B_1))^*{}^+\xi^{\rm phys}_1(x^R_1,x^B_1)\nn\\
&\underset{(\ref{lemphys})}{=}&\int_{\mathbb{R}^{N^1_B}}\prod_B\,dx^B_1({}^+{\Psi}^{\rm phys}_1(x^B_1))^*{}^+{\Xi}^{\rm phys}_1(x^B_1)
\ea
where $N_B^1=N^1_l+N^1_\lambda+N^1_z+N^1_\gamma$. Thus, ${}^+{\Psi}^{\rm phys}_1,{}^+{\Xi}^{\rm phys}_1$ must be square integrable in the post--observables $x^B_1$. Similarly, the {\it pre--physical inner product} in ${}^-\ch^{\rm phys}_1$ reads  \cite{Hoehn:2014fka}
\ba
\langle{}^-\xi^{\rm phys}_1|{}^-\psi^{\rm phys}_1\rangle_{\rm phys-}&=&\langle\xi^{\rm kin}_1|{}^-\mathbb{P}_1|\psi^{\rm kin}_1\rangle_{\ch^{\rm kin}_1}\nn\\
&=&\int_{\mathbb{R}^{N^1_A}}\prod_A\,dx^A_1({}^-\xi^{\rm phys}_1(x^L_1,x^A_1))^*{}^-\psi^{\rm phys}_1(x^L_1,x^A_1)\nn\\
&\underset{(\ref{lemphys2})}{=}&\int_{\mathbb{R}^{N^1_A}}\prod_A\,dx^A_1({}^-{\Xi}^{\rm phys}_1(x^A_1))^*{}^-{\Psi}^{\rm phys}_1(x^A_1)
\ea
with $N^1_A=N^1_r+N^1_\rho+N^1_z+N^1_\gamma$. Consequently, ${}^-{\Xi}^{\rm phys}_1,{}^-{\Psi}^{\rm phys}_1$ must be square integrable in the pre--observables $x^A_1$. Again, the last two lines only hold for systems with constraints linear in the momenta.

\begin{Example}
\emph{We return to the toy model of the free scalar field on the expanding square lattice. The pre--physical state at $n=0$ and the post--physical state at $n=1$ are unique and given by
\ba\label{expstates}
{}^-\psi^{\rm phys}_0=1,\q\q\q{}^+\psi^{\rm phys}_1=e^{i\sum_{i=1}^4\left((1+\f{1}{4}m^2)(\phi_1^i)^2-\phi_1^i\phi_1^{i+1}\right)/\hbar}.
\ea
There is no square integrable part in the states because the move $0\rightarrow1$ is fully constrained and thus devoid of propagating degrees of freedom. We thus have ${}^-\ch^{\rm phys}_0\simeq{}^+\ch^{\rm phys}_1\simeq\mathbb{C}$.}

\emph{Moreover, the pre-- and post--physical states at $n=1,2$, respectively, read
\ba
{}^-\psi^{\rm phys}_1&=&e^{-i\sum_{i=1}^4\left((3+\f{3}{4}m^2)(\phi_1^i)^2-\phi_1^i\phi_1^{i+1}\right)/\hbar}\,{}^-{\Psi}^{\rm phys}_1(\phi_1^1,\dots,\phi_1^4) ,\nn\\
{}^+\psi^{\rm phys}_2&=& e^{ib^2_{ij}\phi^i_2\phi_2^j/\hbar}\,{}^+{\Psi}^{\rm phys}_2(\Phi_2^9,\ldots,\Phi_2^{12}),\label{expstates2}
\ea
where $b^2$ is given in (\ref{b2}), ${}^-{\Psi}^{\rm phys}_1$ is square integrable in the pre--observables $\phi_1^1,\ldots,\phi_1^4$ and ${}^+{\Psi}^{\rm phys}_2$ is square integrable in the post--observables $\Phi_2^9,\ldots,\Phi_2^{12}$. The physical Hilbert spaces of the move $1\rightarrow2$ are therefore ${}^-\ch^{\rm phys}_1\simeq{}^+\ch^{\rm phys}_2\simeq L^2(\mathbb{R}^4)$.}
\end{Example}

\subsection{Dynamics and propagators}

Despite the absence of a Hamiltonian, we have all the ingredients to construct the quantum dynamics. In particular, the action $S_{n+1}$ contains the entire information about the dynamics and we shall employ it, in analogy to the classical case, to generate a quantum time evolution map for a global move $n\rightarrow n+1$.

The general idea, prior to explicitly using the action, is to construct a suitable propagator $K_{0\rightarrow 1}$, associated to the move $0\rightarrow 1$, to define a map from $\ch^{\rm kin}_0$ to ${}^+\ch^{\rm phys}_{1}$,
\ba
{}^+\psi^{\rm phys}_{1}=\int dx_0\,K_{0\rightarrow 1}(x_0,x_{1})\,\psi^{\rm kin}_0,\label{propmap}
\ea
using that the dynamics should project onto the constraints \cite{Hoehn:2014fka}. This implies non-trivial consistency requirements: the propagator has to satisfy both pre-- and post--constraints \cite{Hoehn:2014fka}:
\ba
{}^+\hat{C}^{1}_R\,K_{0\rightarrow 1}=0={}^-\hat{C}^0_L(K_{0\rightarrow 1})^*,\label{cond}
\ea
where $(K_{0\rightarrow 1})^*=K_{1\rightarrow 0}$ is the reverse propagator and ${}^*$ denotes complex conjugation. This suggests to establish the notion of a {\it kinematical propagator} $\kappa_{0\rightarrow 1}(x_0,x_{1})$ as a square integrable function on $\cq_0\times\cq_{1}$ which does not satisfy any constraints --- in analogy to kinematical states $\psi^{\rm kin}_0\in\ch^{\rm kin}_0$. The {\it physical propagator} may then be reformulated as \cite{Hoehn:2014fka}
\ba
K_{0\rightarrow 1}={}^+\mathbb{P}_{1}\,({}^-\mathbb{P}_0)^*\,\kappa_{0\rightarrow 1}.\label{kinprop}
\ea
Given the analysis around (\ref{lemphys}, \ref{lemphys2}), the physical propagator must take the form
\ba
K_{0\rightarrow 1}=e^{i\,(a^1_{ij}x^i_0x^j_0+b^{1}_{ij}x^i_1x^j_1)/2\hbar}\,\mathfrak{K}_{0\rightarrow 1}(x^A_0,x^B_1),\label{physprop2}
\ea
where $\mathfrak{K}_{0\rightarrow 1}$ is the propagator of the propagating pre--observables $x^A_0$ at $n=0$ and the propagating post--observables $x^B_1$ at $n=1$. It must be square integrable in these variables (see below). The propagator can thus only depend on the {\it a priori free} $x^R_1$ and the {\it a posteriori free} $x^L_0$ through the phase-prefactor. 

Using (\ref{kinprop}) and square integrability, one can rewrite (\ref{propmap}) as
\ba
{}^+\psi^{\rm phys}_{1}&=&\int dx_0\,{}^+\mathbb{P}_{1}\,({}^-\mathbb{P}_0)^*\,\kappa_{0\rightarrow 1}\,\psi^{\rm kin}_0=\int dx_0\,{}^+\mathbb{P}_{1}\,\kappa_{0\rightarrow 1}\,{}^-\mathbb{P}_0\,\psi^{\rm kin}_0\nn\\
&\underset{(\ref{proj})}{=}&\int dx_0\,{}^+\mathbb{P}_{1}\,\kappa_{0\rightarrow 1}\,{}^-\psi^{\rm phys}_0,\label{physprop3}
\ea
such that
\ba
U_{0\rightarrow 1}:=\int dx_0{}^+\mathbb{P}_0\,\kappa_{0\rightarrow 1}\label{unitary}
\ea
defines a map from ${}^-\ch^{\rm phys}_0$ to ${}^+\ch^{\rm phys}_{1}$. We define the {\it pre--} and {\it post--fixed propagators}
\ba
K^{f_+}_{0\rightarrow1}={}^+\mathbb{P}_1\,\kappa_{0\rightarrow1},\q\q\q\q K^{f_-}_{0\rightarrow1}=({}^-\mathbb{P}_0)^*\,\kappa_{0\rightarrow1},\nn
\ea
respectively. The integral map $U_{0\rightarrow1}$ and its time reverse $U_{1\rightarrow0}:{}^+\ch^{\rm phys}_{1}\rightarrow {}^-\ch^{\rm phys}_0$ then read
\ba
U_{0\rightarrow 1}=\int dx_0\,K^{f_+}_{0\rightarrow1}\,\q\q\q U_{1\rightarrow0}=\int dx_1\,\left(K^{f_-}_{0\rightarrow1}\right)^*.\nn
\ea
For quadratic discrete actions, we can rewrite the pre-- and post--fixed propagators in a more suggestive form which makes the fact that they are (`gauge') fixed explicit. Namely, according to lemma 4.1 in \cite{Hoehn:2014fka}, for systems with constraints linear in the momenta, the following holds:
\ba
(2\pi\hbar)^{N^0_L}\,{}^-\mathbb{P}_0\,\prod_{L=1}^{N^0_L}\delta(x^L_0-x'^L_0)\,{}^-\mathbb{P}_0\,\psi^{\rm kin}_0={}^-\mathbb{P}_0\,\psi^{\rm kin}_0.\label{identity}
\ea 
The analogous expression is true for post--physical states and post--projectors. Inserting this into (\ref{physprop3}) and performing the same for the time reverse map (and, moreover, absorbing the powers of $2\pi\hbar$ into the physical states) entails that
\ba
K^{f_+}_{0\rightarrow1}=K_{0\rightarrow1}\,\prod_{L=1}^{N^0_L}\delta(x^L_0-x'^L_0),\q\q\q K^{f_-}_{0\rightarrow1}= K_{0\rightarrow1}\,\prod_{R=1}^{N^1_R}\delta(x^R_1-x'^R_1).\label{fixedprops}
\ea
The pre--fixed propagator is thus the physical propagator multiplied with a Faddeev-Popov fixing condition on the pre--orbit $\cg^-_0$, while the post--fixed propagator is the physical propagator multiplied with a Faddeev-Popov fixing condition on the post--orbit $\cg^+_1$.

This permits us to require invertibility of $U_{0\rightarrow1}$ on the pre-- and post--physical Hilbert spaces in the following form:
\ba\label{coninvert}
\int\,dx_0\,K^{f_+}_{0\rightarrow1}(x_1,x_0)\left(K^{f_-}_{0\rightarrow1}(x'_1,x_0)\right)^*&=&\delta^{(Q)}(x'_1-x_1),\nn\\
\int dx_1\left(K^{f_-}_{0\rightarrow1}(x_1,x_0)\right)^*K^{f_+}_{0\rightarrow1}(x_1,x'_0)&=&\delta^{(Q)}(x'_0-x_0).
\ea 
The right hand sides feature delta functions involving {\it all} $Q$ configuration variables, i.e.\ {\it both} free parameters and observables. This is a consequence of the form of the fixed propagators (\ref{fixedprops}) which involve the Faddeev-Popov fixing conditions on the free parameters. $U_{0\rightarrow1}$ is then a unitary isomorphism and constitutes the quantum analogue of $\mathfrak{H}_n$ \cite{Hoehn:2014fka}. 

In fact, (\ref{physprop3}, \ref{unitary}) can be rewritten in a more useful way: lemma 4.3 of \cite{Hoehn:2014fka}, which applies to systems with constraints linear in the momenta as in the current situation, implies that (up to a prefactor, given by a power of $2\pi\hbar$, which can be absorbed into the states)
\ba
{}^+\psi^{\rm phys}_{1}&=&\int_{\mathbb{R}^{N_A^0}} \prod_{A}dx^A_0\, K_{0\rightarrow1}\,{}^-\psi^{\rm phys}_0\nn\\
&\underset{(\ref{lemphys2}, \ref{physprop2})}{=}&e^{i\,b^{1}_{ij}x^i_1x^j_1/2\hbar}\int_{\mathbb{R}^{N_A^0}} \prod_{A}dx^A_0\,\mathfrak{K}_{0\rightarrow 1}(x^A_0,x^B_1)\,{}^-{\Psi}^{\rm phys}_0(x^A_0).\label{abcd}
\ea
From this we can see that the propagator part $\mathfrak{K}_{0\rightarrow 1}$, corresponding to the actually propagating degrees of freedom, must be square integrable in the pre--observables $x^A_0$ (and by time reversed arguments also in the post--observables $x^B_1$).

The previous conditions are not sufficient to uniquely specify the propagator $K_{0\rightarrow1}$. In particular, apart from the condition of square integrability, the form of the physical part of the propagator $\mathfrak{K}_{0\rightarrow 1}$ in (\ref{physprop2}), which ultimately contains the dynamics of the propagating degrees of freedom, is thus far left undetermined. At this stage, we can finally make use of the action $S_1$ associated to the move $0\rightarrow 1$ to uniquely single out a propagator. More precisely, we make the propagator ansatz \cite{Hoehn:2014fka}
\ba
K_{0\rightarrow 1}(x_0,x_{1})=M_{0\rightarrow 1}(x_0,x_{1})\,e^{iS_{1}(x_0,x_{1})/\hbar}.\nn
\ea
For quadratic actions and, in particular, semiclassical limit approximations, the propagator usually takes this form of a measure times a phase factor containing the classical action \cite{miller1,miller2}. Specifically, comparing with (\ref{physprop2}) and using the form of the action (\ref{Sk}) and the variable splitting of section \ref{sec_classdof}, the physical part of the propagator will be given by 
\ba
\mathfrak{K}_{0\rightarrow 1}(x^A_0,x^B_1)=M_{0\rightarrow 1}\,e^{ic^1_{AB}x^A_0x^B_1/\hbar}.\nn
\ea
Finally, thanks to (\ref{fixedprops}), the fixed propagators take the form
\ba
K^{f_+}_{0\rightarrow1}=M^{f_+}_{0\rightarrow1}\,e^{iS_1/\hbar},\q\q\q K^{f_-}_{0\rightarrow1}=M^{f_-}_{0\rightarrow1}\,e^{iS_1/\hbar}\nn
\ea
with {\it pre--} and {\it post--fixed measures}
\ba
M^{f_+}_{0\rightarrow1}:=M_{0\rightarrow1}\,\prod_{L=1}^{N^0_L}\delta(x^L_0-x'^L_0),\q\q\q M^{f_-}_{0\rightarrow1}:=M_{0\rightarrow1}\,\prod_{R=1}^{N^1_R}\delta(x^R_1-x'^R_1),\nn
\ea
respectively.

\subsection{Determining the propagator measure}

It remains to determine the measure $M_{0\rightarrow1}$. To this end, we shall employ the invertibility (or unitarity) conditions (\ref{coninvert}). The second invertibility condition in (\ref{coninvert}) is equivalent to
\ba
\delta^{(Q)}(x^i_0-x'^i_0)=\int dx_1\,M^{f_+}_{0\rightarrow1}(x_0,x_1)(M^{f_-}_{0\rightarrow1}(x_0',x_1))^*\,e^{ic^1_{ij}x^j_1(x_0^i-x_0'^i)/\hbar},\label{quadinv}
\ea
Using the variable splitting of section \ref{sec_classdof}, this condition takes the form
\ba
\delta^{(Q)}(x^i_0-x'^i_0)=\int \prod_\Gamma dx^\Gamma_1\,\big|\det (T_1)\big|\,M^{f_+}_{0\rightarrow1}(x_0,x_1)(M^{f_-}_{0\rightarrow1}(x_0',x_1))^*\,e^{ic^1_{AB}x^B_1(x_0^A-x_0'^A)/\hbar}.\nn
\ea
Noting from (\ref{h0}) that the pre--momenta canonically conjugate to the $x_0^A$ are given by ${}^-\pi^0_A:=-c^1_{AB}x^B_1$, the last equation can be transformed into
\ba
\delta^{(Q)}(x^i_0-x'^i_0)=\int \prod_Rdx^R_1\prod_Ad{}^-\pi^0_A\,\f{\big|\det (T_1)\big|}{\big|\det(c^1_{AB}/\hbar)\big|}\,M^{f_+}_{0\rightarrow1}(x_0,x_1)(M^{f_-}_{0\rightarrow1}(x_0',x_1))^*\,e^{-i{}^-\pi^0_A(x_0^A-x_0'^A)}.\nn
\ea
The Jacobian of $(T_0)$ translates between the delta functions with arguments $x_0^i$ and $x_0^\Gamma$. Hence,
\ba
M^{f_+}_{0\rightarrow1}(x_0,x_1)(M^{f_-}_{0\rightarrow1}(x_0',x_1))^*&=&\f{1}{(2\pi)^{N^0_A}}\big|\det(T^{-1}_0)\det(c^1_{AB}/\hbar)\det(T^{-1}_1)\big|\nn\\
&&\q\q\q\q\q\times\prod_R \delta(x_1^R-x'^R_1)\prod_L\delta(x_0^L-x'^L_0).\nn
\ea
(Recall from section \ref{sec_classdof} that $N^0_A=N^1_B$.) The analogous procedure can be done for the first condition in (\ref{coninvert}). The final solution (up to unitary representation changes) for the (unfixed) propagator is therefore
\ba\label{propquad}
K_{0\rightarrow1}(x_0,x_1)=\sqrt{\left(-2\pi i\hbar\right)^{-N^0_A}\big|\det(T^{-1}_0)\det(c^1_{AB})\det(T^{-1}_1)\big|}\,\,e^{iS_1(x_0,x_1)/\hbar}
\ea
where $S_1$ is given by (\ref{Sk}). (This is the generalization of equation (3.5) in \cite{Hoehn:2014fka}, which is valid only for regular moves, to constrained moves for quadratic discrete actions.) Consequently, for quadratic discrete actions, the measure $M_{0\rightarrow1}$ is constant and unique (up to unitary transformations), despite the absence of an evolution equation, such as a Schr\"odinger equation, for the discrete propagator.

Clearly, the propagator (\ref{propquad}) satisfies all quantum post--constraints at $n=1$ in (\ref{qcons}) and all quantum pre--constraints at $n=0$, defined similarly, in the manner of (\ref{cond}).

\begin{Example}
\emph{We briefly state the propagators for the example of the scalar field on a growing square lattice. The propagator associated to the move $0\rightarrow1$ must satisfy the pre--constraints at $n=0$ and post--constraints at $n=1$. Since the solutions to the latter are unique, the propagator factorizes into the pre-- and post--physical state at $n=0,1$, respectively
\ba
K_{0\rightarrow1}={}^+\psi^{\rm phys}_1\,({}^-\psi^{\rm phys}_0)^*=e^{i\sum_{i=1}^4\left((1+\f{1}{4}m^2)(\phi_1^i)^2-\phi_1^i\phi_1^{i+1}\right)/\hbar}.
\ea
The factorization reflects the absence of a coupling and hence of propagating degrees of freedom between steps $n=0,1$. This propagator and the pre-- and post--physical states (\ref{expstates}) trivially satisfy the quantum evolution equation (\ref{abcd}). }

\emph{Evaluating (\ref{propquad}) for the propagator of the non-trivial move $1\rightarrow2$ yields
\ba
K_{1\rightarrow2}=\f{1}{(\pi i\hbar)^2}\,e^{iS_2(\{\phi_1\},\{\phi_2\})/\hbar}\nn,
\ea
where $S_2$ is given in (\ref{exampleaction}). As can be easily checked, this propagator and the pre-- and post--physical states in (\ref{expstates2}) fulfill the quantum evolution equation (\ref{abcd}).}
\end{Example}

\subsection{Composition of propagators}

Next, let us compose the propagators of the moves $0\rightarrow1$ and $1\rightarrow2$ to an `effective' propagator for the `effective' move $0\rightarrow2$. As seen in the classical part, it is the composition of moves where non-trivialities arise.

To begin with, we note that the projectors $\mathbb{P}^I_1=\prod_I\delta(\hat{C}^1_I)$ onto the type (1)(A) gauge symmetry generating constraints $\hat{C}^1_I$ appear in both ${}^+\mathbb{P}_1$ and ${}^-\mathbb{P}_1$ and thus in both $K_{0\rightarrow1}$ and $K_{1\rightarrow2}$ because type (1)(A) constraints are both pre-- and post--constraints. We have seen in section \ref{sec_impose} that a double action of a projector leads to divergences because the projectors are improper. Hence, the type (1)(A) constraints lead to divergences in state sums resulting from a spurious integration over non-compact gauge orbits (in contrast to other constraints which are only implemented once and thereby do not involve a second integration over non-compact gauge orbits). These gauge divergences can be easily regularized by using (\ref{identity}), i.e.\ by introducing a Faddeev-Popov gauge fixing factor $\prod_I\delta(x_1^I-x'^I_1)$. In this special quadratic action case, with the gauge fixing condition conjugate to a constraint linear in the momenta, the Faddeev-Popov determinant is just one \cite{Hoehn:2014fka} (for general actions one would obtain a non-trivial Faddeev-Popov determinant):
\ba
K_{0\rightarrow2}&=&\int dx_1\prod_I\delta(x_1^I-x'^I_1) \,K_{1\rightarrow2}\,K_{0\rightarrow1}\nn\\
&=&M_{1\rightarrow2}\,M_{0\rightarrow1}\,e^{\f{i}{2\hbar}(a^1_{ij}x_0^ix_0^j+b^2_{ij}x_2^ix_2^j)}\int dx_1\,\prod_I\delta(x_1^I-x'^I_1)\,e^{\f{i}{\hbar}(\f{1}{2}h^{12}_{ij}x_1^ix_1^j+J^1_jx_1^j)},\nn
\ea
where we have defined the `source vector'
\ba
J^1_j=c^1_{ij}x^i_0+c^2_{ji}x_2^i.\nn
\ea
Transforming the variables as in (\ref{lindecomp}), where $\nu$ runs over the null vectors of the Hessian $h^{12}$, $I,l,r,z$,
\ba
K_{0\rightarrow2}&=&M_{1\rightarrow2}\,M_{0\rightarrow1}\,e^{\f{i}{2\hbar}(a^1_{ij}x_0^ix_0^j+b^2_{ij}x_2^ix_2^j)}\nn\\
&&\times\int \prod_\alpha dx^\alpha_1\prod_\nu dx^\nu_1\,|\det(T_1)|\,\prod_I\delta(x_1^I-x'^I_1)\,e^{\f{i}{\hbar}(\f{1}{2}h^{12}_{\alpha\beta}x_1^\alpha x_1^\beta+J^1_\alpha x_1^\alpha+J_\nu^1x_1^\nu)}.\nn
\ea
Using Gaussian integration, comparing to (\ref{effcoeff}) and defining
\ba
S_{02}:=\frac{1}{2}\tilde{a}^{02}_{ij}x^i_0x^j_0+\frac{1}{2}\tilde{b}^{02}_{ij}x^i_2x^j_2+\tilde{c}^{02}_{ij}x^i_0x^j_2\nn
\ea
(this is the effective action (\ref{effact}) without holonomic and boundary data constraints) yields
\ba
K_{0\rightarrow2}&=&M_{1\rightarrow2}\,M_{0\rightarrow1}\sqrt{\f{(2\pi i\hbar)^{N^1_\alpha}}{\det(h^{12}_{\alpha\beta})}}\,|\det(T_1)|\,e^{\f{i}{2\hbar}(a^1_{ij}x_0^ix_0^j+b^2_{ij}x_2^ix_2^j-J^1_\alpha h^{\alpha\beta}_{12}J^1_\beta)}\label{stepbla}\\
&&\q\q\q\q\q\q\q\q\q\q\q\q\q\q\q\q\q\q\times\int \prod_\nu dx^\nu_1\,\prod_I\delta(x_1^I-x'^I_1)\,e^{\f{i}{\hbar}J_\nu^1x_1^\nu}\nn\\
&\underset{\text{(\ref{effact})}}{=}&M_{1\rightarrow2}\,M_{0\rightarrow1}\sqrt{\f{(2\pi i\hbar)^{N^1_\alpha}}{\det(h^{12}_{\alpha\beta})}}\,|\det(T_1)|\,e^{\f{i}{\hbar}{S}_{02}(x_0,x_2)}\int \prod_\nu dx^\nu_1\,\prod_I\delta(x_1^I-x'^I_1)\,e^{\f{i}{\hbar}J_\nu^1x_1^\nu},\nn
\ea
where $N^1_\alpha=N^1_H+N^1_\lambda+N^1_\rho+N^1_\gamma$. Noting that $J^1_I=0$ and that $J^1_\nu$, $\nu=l,r,z$ are precisely the holonomic constraints (\ref{holcon2}, \ref{holcon3}) and the boundary data constraints (\ref{2stepcon}), one finds
\ba\label{effpropquad}
K_{0\rightarrow2}=M_{0\rightarrow2}\,e^{\f{i}{\hbar}{S}_{02}(x_0,x_2)}\prod_l\delta(H^0_l)\prod_r\delta(H^2_r)\prod_z\delta\left(B^{02}_z\right),
\ea
where the measure finally reads (up to an overall constant phase)
\ba
M_{0\rightarrow2}:=\sqrt{(2\pi i\hbar)^{N^1_H+N^1_l+N^1_r-N^1_\gamma}\f{|\det(T_0^{-1})\det(c^1_{AB})\det(c^2_{A'B'})\det(T_2^{-1})|}{\det(h^{12}_{\alpha\beta})}}\,\,.\label{effmeasure}
\ea
A number of comments are necessary:
\begin{itemize}
\item Just as in the continuum, for quadratic actions the integration over step $n=1$ in the quantum theory is thus equivalent to solving the classical equations of motion at $n=1$. In particular, the holonomic constraints at $n=0,2$ (\ref{holcon1}, \ref{holcon2}) and the boundary data constraints (\ref{2stepcon})---which are secondary constraints and thus neither pre-- nor post--constraints---are now implemented by delta-functions.
\item One easily verifies that, if no holonomic or boundary data constraints $H^0_l,H^2_r,B^{02}_z$ arise, the `effective' propagator $K_{0\rightarrow2}$ (\ref{effpropquad}) is annihilated --- in the sense of (\ref{cond}) --- by the quantum versions of the `effective' pre-- and post--constraints given in (\ref{effcons}). That is, new `effective' constraints at $n=0,2$ can also arise in the quantum theory, such that the number of quantum pre-- and post--constraints is likewise move dependent. In particular, this happens if:
\begin{itemize}
\item[(a)] First class coarse graining pre--constraints ${}^-\hat{C}^1_\lambda$ of type (2)(B) arise at $n=1$. These `propagate' under the dynamics back to $n=0$ (they involve post--observables of the move $0\rightarrow1$) to ensure that the finer discretization of the move $0\rightarrow1$ does not carry more dynamical information than the coarser discretization of the move $1\rightarrow2$ can support.
\item[(b)] First class refining post--constraints ${}^+\hat{C}^1_\rho$ of type (3)(B) arise at $n=1$. These `propagate' under the dynamics to $n=2$ (they involve pre--observables of the move $1\rightarrow2$) to ensure that the physical states on the finer discretization of the  move $1\rightarrow2$ do not carry more dynamical information than the coarser move $0\rightarrow1$ transports.
\end{itemize}
\item On the other hand, if holonomic or boundary data constraints $H^0_l,H^2_r,B^{02}_z$ arise, the `effective' propagator $K_{0\rightarrow2}$ (\ref{effpropquad}) is 
\begin{itemize}
\item[(a)] annihilated by the primary quantum pre--constraints at $n=0$ and the primary quantum post--constraints at $n=2$, contained in the `effective' set (\ref{effcons}), which have already existed prior to integrating out step $n=1$. This is easily verified by noting that the $H^0_l,H^2_r,B^{02}_z$ only involve data that propagate in the moves $0\rightarrow1$ and $1\rightarrow2$, while the primary pre-- and post--constraints at $n=0,2$ only involve free data.
\item[(b)] generally {\it not} annihilated by the remaining secondary `effective' quantum constraints corresponding to (\ref{effcons}). The reason is that the `effective' constraints (\ref{effcons}) are defined {\it prior} to implementing the holonomic and boundary data constraints. By contrast, in (\ref{effpropquad}) the latter are already implemented. (In fact, it is straightforward to convince oneself that, had one not carried out the integrals over the Lagrange multipliers  $x^\nu_1$, $\nu=l,r,z$ in (\ref{stepbla}), which ultimately implement the holonomic and boundary data constraints, the resulting expression would be annihilated by all `effective' quantum constraints corresponding to (\ref{effcons}).) Instead, one has to integrate over the delta functions in (\ref{effpropquad}) in order to solve the holonomic and boundary data constraints. The resulting expression would be a new `effective' propagator with a new `effective' action in its phase which coincides with the action one would get from (\ref{effact}) by solving the holonomic and boundary data constraints contained in it. This object would then solve a new set of `effective' quantum constraints. We shall not carry out these steps explicitly as they depend on one's choice of independent variables (see also the discussion in sections \ref{sec_unique} and \ref{sec_varcon}).
\end{itemize}
\item The measure (\ref{effmeasure}) of the `effective' propagator is not in general of precisely the normalized form (\ref{propquad}). This is, however, not a problem because the propagator $K_{0\rightarrow2}$ in (\ref{effpropquad}) is not a {\it fixed} propagator. The fixed measure, obtained via the invertibility conditions (\ref{coninvert}) for the move $0\rightarrow2$, would be of a slightly different form in order to ensure normalization. We emphasize that the change of the measure does not arise if only degrees of freedom of type (1)(A) and (5) arise. In that case, no new constraints are produced and (\ref{effmeasure}) becomes
\ba
M_{0\rightarrow2}:=\sqrt{(2\pi i\hbar)^{-N^1_\gamma}{|\det(T_0^{-1})\det(\tilde{c}^{02}_{AB})\det(T_2^{-1})|}}\,\,,\nn
\ea
which is precisely of the normalized form (\ref{propquad}). In any case, what really matters is the functional form of the propagator, rather than the precise form of its constant measure. 
\item If new quantum pre-- and post--constraints arise at $n=0,2$, as a consequence of integrating out step $n=1$, the pre--physical Hilbert space at $n=0$ and the post--physical Hilbert space at $n=2$ will change to a new pair ${}^-\widetilde{\ch}^{\rm phys}_0,{}^+\widetilde{\ch}^{\rm phys}_2$. The latter contain states which satisfy {\it all} `effective' quantum pre-- and post--constraints at $n=0,2$, respectively. In fact, ${}^-\widetilde{\ch}^{\rm phys}_0:={}^-\widetilde{\mathbb{P}}_0^{\rm new}({}^-\ch^{\rm phys}_0)$ and ${}^+\widetilde{\ch}^{\rm phys}_2:={}^+\widetilde{\mathbb{P}}_2^{\rm new}({}^+\ch^{\rm phys}_2)$, where ${}^-\widetilde{\mathbb{P}}_0^{\rm new}$ and ${}^+\widetilde{\mathbb{P}}_2^{\rm new}$ are the projectors corresponding to the newly produced pre-- and post--constraints. Since such projectors are improper, ${}^-\widetilde{\ch}^{\rm phys}_0$ and ${}^+\widetilde{\ch}^{\rm phys}_2$ will no longer be contained in the original ${}^-\ch^{\rm phys}_0$ and ${}^+\ch^{\rm phys}_2$, respectively. Coarse graining thus leads to non-unitary projections of physical Hilbert spaces. This does not come as a surprise, given that coarse graining removes dynamical degrees of freedom. We shall not further elaborate on this, since this has been extensively discussed in \cite{Hoehn:2014fka,Hoehn:2014wwa}.
\item Related to this, the fact that the number of constraints can increase at a given time step has severe repercussions for observables. In particular, not all quantum pre--observables at $n=0$ of the move $0\rightarrow1$ survive as `effective' pre--observables of the move $0\rightarrow2$ because the number of constraints which they have to commute with can increase. The analogous statement holds for quantum post--observables at $n=2$. In this sense, coarse graining in either time direction projects out observables non-unitarily. If the propagating constraints can be interpreted as coarse graining conditions, the observables which are projected out can be interpreted as corresponding to `too finely grained' degrees of freedom (see also \cite{Hoehn:2014fka,Hoehn:2014wwa}).
\item Finally, a full state sum would be generated by composing more and more evolution moves in the manner just discussed. 
\end{itemize}

\begin{Example}
\emph{As seen at the end of section \ref{sec_unique}, the effective action for the effective move $0\rightarrow2$ of the scalar field on the growing square lattice is merely a boundary term with $\tilde{a}^{02}=\tilde{c}^{02}\equiv0$ and a somewhat convoluted expression for $\tilde{b}^{02}$ which we abstain from showcasing explicitly. The move $0\rightarrow2$ is evidently fully constrained as appropriate for a `creation from nothing'. }

\emph{After integrating out step $n=1$, both the pre--physical state and the post--physical state at $n=0,2$, respectively are unique,
\ba
{}^-\tilde{\psi}^{\rm phys}_0=1,\q\q\q {}^+\tilde{\psi}^{\rm phys}_2=e^{i\tilde{b}^{02}_{ij}\phi^i_2\phi_2^j/\hbar},\nn
\ea
Accordingly, the effective propagator factorizes
\ba
K_{0\rightarrow2}={}^+\tilde{\psi}^{\rm phys}_2\,({}^-\tilde{\psi}^{\rm phys}_0)^*=e^{i\tilde{b}^{02}_{ij}\phi^i_2\phi_2^j/\hbar}.
\ea
Again, these pre-- and post--physical states and the effective propagator trivially satisfy the evolution equation (\ref{abcd}).}

\emph{As a result, the effective physical Hilbert spaces of the move $0\rightarrow2$ become one-dimensional ${}^-\tilde{\ch}^{\rm phys}_0\simeq{}^+\tilde{\ch}^{\rm phys}_2\simeq\mathbb{C}$. Since previously ${}^+\ch^{\rm phys}_2\simeq L^2(\mathbb{R}^4)$ for the move $1\rightarrow2$, a non-unitary projection of the post--physical Hilbert space at $n=2$ has taken place. The composition with the move $0\rightarrow1$ is a non-trivial lattice shrinking which renders all degrees of freedom at $n=2$ non-dynamical for the move $0\rightarrow2$. This result highlights the move or spacetime region dependence of physical Hilbert spaces and propagating degrees of freedom.  }
\end{Example}

\section{Conclusions and outlook}\label{sec_summary}

Quadratic discrete actions permit to explicitly solve all equations of motion and the corresponding quantum dynamics. In the present manuscript we have taken advantage of this and, firstly, classified constraints, equations of motion and degrees of freedom, arising from quadratic discrete actions, into eight types which distinguish between gauge modes and different kinds of propagating degrees of freedom. Secondly, we performed the analogous discussion in the quantum theory. This analysis applies to variational discrete systems with both temporally varying or constant discretization. 

This article confirms the general results of \cite{Dittrich:2013jaa,Hoehn:2014fka,Hoehn:2014wwa} and shows explicitly how changes in the discretization, deriving from dynamical coarse graining and refining operations, or a shrinking and growing lattice, lead to non-trivial changes in the dynamical content of the system. The classification of constraints and degrees of freedom becomes evolution move, and thus, in a spacetime context, spacetime region dependent. Constraints, admitting the interpretation as coarse graining or refining conditions \cite{Dittrich:2013xwa}, can propagate under time evolution as consistency conditions to other time steps, for instance, to ensure that a finer discretization does not carry more dynamical information than a coarser discretization can support at another time step. Composing evolution moves on temporally varying discretizations changes the reduced phase space and physical Hilbert space at a given time step; in particular, coarse graining or shrinking the spatial lattice leads to non-unitary projections of physical Hilbert spaces. 

Nevertheless, the formalism and this classification are fully consistent. In fact, this formalism constitutes a canonical complement to the covariant coarse graining techniques developed in \cite{Dittrich:2012jq,Bahr:2009qc,Bahr:2011uj,Dittrich:2012qb,Bahr:2010cq,Dittrich:2011zh}. It can be applied to arbitrary variational discrete systems governed by quadratic actions such as, for example, linearized 4D Regge Calculus \cite{Dittrich:2009fb,dh4,Bahr:2009ku,Rocek:1982fr}. Furthermore, while the present work focuses on technical aspects of the formalism, a forthcoming article \cite{biancameted}, using Fourier decomposition, will provide a more detailed physical discussion and interpretation specific to a scalar field on an evolving lattice. Lastly, it is hoped that the results of this research may contribute to a better understanding of the discretization changing dynamics inherent to several quantum gravity approaches \cite{Thiemann:1996ay,Thiemann:1996aw,Alesci:2010gb,Dittrich:2011ke,Dittrich:2013xwa,Bonzom:2011hm}.

\section*{Acknowledgements}
The author thanks Bianca Dittrich for discussion. Most of the classical part of this research was performed while the author was still at the Institute for Theoretical Physics of Universiteit Utrecht. Research at Perimeter Institute is supported by the Government of Canada through Industry Canada and by the Province of Ontario through the Ministry of Research and Innovation.

\providecommand{\href}[2]{#2}\begingroup\raggedright\endgroup

\end{document}